\documentclass{aa}  

\usepackage[breaklinks=true,colorlinks,citecolor=blue]{hyperref}
\usepackage{graphicx}
\usepackage{enumerate}
\usepackage{epsfig}
\usepackage{color}
\usepackage{txfonts}
\usepackage{natbib}
\usepackage{multirow}
\usepackage{ulem} 
\usepackage{amssymb}
\usepackage{xcolor}

\definecolor{ao}{rgb}{0.0, 0.5, 0.0}

\newcommand{\kmps}{\rm km~s\ensuremath{^{-1} }\,}
\newcommand{\kmskpc}{km~s\ensuremath{^{-1}}~kpc\ensuremath{^{-1} }\,}
\newcommand{\Msun}{M\ensuremath{_\odot}}

\newcommand{\Gaia}{{\it Gaia}\,}

\newcommand{\XY}{$(X,Y)$\,}
\newcommand{\XgYg}{$(X_g,Y_g)$\,}
\newcommand{\Rgphi}{$(R_g,\phi)$\,}

\newcommand{\RVphi}{(R,v\ensuremath{_\phi})\,}
\newcommand{\RgVphi}{(R\ensuremath{_g},v\ensuremath{_\phi})\,}
\newcommand{\RdashVphi}{$R-$v\ensuremath{_\phi}}
\newcommand{\RgdashVphi}{$R_g$-v\ensuremath{_\phi}}

\newcommand{\VV}{v\ensuremath{_\phi}\,}

\newcommand{\uv}{v\ensuremath{_R}-v\ensuremath{_\phi}\,}
\newcommand{\UV}{(v\ensuremath{_R},v\ensuremath{_\phi})\,}

\newcommand{\her}{Hercules\,}
\newcommand{\arc}{Arcturus\,}
\newcommand{\ple}{Pleiades\,}
\newcommand{\hya}{Hyades\,}
\newcommand{\hor}{Horn\,}
\newcommand{\ha}{Hat\,}
\newcommand{\sir}{Sirius\,}
\newcommand{\cb}{Coma Berenices\,}

\newcommand{\aFe}{\ensuremath{\rm [\alpha/Fe]}\,}

\newcommand{\FeH}{\ensuremath{\rm [Fe/H]}\,}

\newcommand{\RRg}{\ensuremath{\rm R-R_g}\,}

\begin{document} 

\title{Chemo-kinematics of the Milky Way spiral arms \\ and bar resonances:  connection to ridges and \\ moving groups in the Solar vicinity}
\titlerunning{Chemo-kinematics of the Milky Way spiral arms}

\author{Sergey Khoperskov$^{1,2,3,4}$\thanks{E-mail: sergey.khoperskov@gmail.com} and Ortwin Gerhard$^{2}$}
\authorrunning{Sergey Khoperskov \& Ortwin Gerhard}

\institute{$^1$ Leibniz-Institut für Astrophysik Potsdam (AIP), An der Sternwarte 16, 14482 Potsdam, Germany\\ $^2$ Max-Planck-Institut f\"{u}r extraterrestrische Physik, Gie{\ss}enbachstrasse 1, 85748 Garching, Germany \\ 
 $^3$ Institute of Astronomy, Russian Academy of Sciences, 48 Pyatnitskya St., Moscow, 119017, Russia \\ $^4$ GEPI, Observatoire de Paris, Université PSL, CNRS, 5 Place Jules Janssen, 92190 Meudon, France }

\abstract{

Making use of a new high-resolution spiral galaxy simulation, as well as \Gaia~DR2 and EDR3 data complemented by chemical abundances from the Galah~DR3, APOGEE~DR16, LAMOST~DR5 surveys, we explore the possible link between the Milky Way~(MW) spiral arms, \RVphi-ridges and moving groups in local \uv space. 
We show that the tightly wound main spiral arms in the $N$-body simulation can be successfully identified using overdensities in angular momentum (AM) or guiding space, as well as in the distribution of dynamically cold stars close to their guiding centers. Stars in the AM overdensities that travel over many kpc in radius trace extended density ridges in \RVphi space and overdensities in the \uv plane of a Solar neighbourhood (SNd)-like region, similar to those observed in the \Gaia data.
Similarly, the AM space of the MW contains several overdensities which correlate with a wave-like radial velocity pattern; this pattern is also reproduced by stars well beyond the SNd. 
 We find that the fraction of \Gaia stars located near their guiding centers shows three large-scale structures that approximately coincide with the MW spiral arms traced by distributions of maser sources in the Sagittarius, Local, and Perseus arms. This approach does not work for the Scutum arm near the end of the bar. Similar to the simulation, the stars in the AM overdensities follow the main \RVphi density ridges with nearly constant angular momentum. When these ridges cross the SNd, they can be matched with the main \uv features. Thus we suggest that the Hat is the inner tail of the Perseus arm, one of the Hercules components is the Sagittarius arm and the Arcturus stream is likely to be the outermost tail of the Scutum-Centaurus arm. Based on previous work, the bar corotation is suggested to coincide with the second, $v_\phi\approx -55$ km/s Hercules stream ridge, and the OLR with the Sirius stream. 
The latter is supported by a sharp decrease of the mean metallicity beyond the Sirius stream, which is an expected behaviour of the OLR, limiting migration of the metal-rich stars from the inner MW. 
In various phase-space coordinates the AM overdensities stars have systematically higher mean metallicity by about $0.05$~dex compared to surrounding stars which is a predicted behaviour of the spiral arms.
We show that the wave-like metallicity pattern can be traced at least up to $\rm |z|\approx 1$~kpc, linked to radial velocity variations seen even further~($|z| \approx 2$~kpc) from the Galactic mid-plane.

}


\maketitle

\section{Introduction}\label{sec::intro}
The MW disk has a complex kinematic
structure~\citep[see, e.g.,][]{2011MNRAS.412.2026S,2013MNRAS.436..101W} which is likely affected by the central bar and the Galactic spiral arms~\citep{2012MNRAS.425.2335S,2015MNRAS.453.1867G}. Moreover, the Galactic disk is also perturbed externally by its satellites, particularly the Sagittarius dwarf galaxy~\citep[see, e.g.,][]{2011Natur.477..301P,2013MNRAS.429..159G}. Both non-axisymmetric density structures in the disk and external interactions leave signatures in the kinematics of the MW stars, consistent with the presence of multiple substructures in the Solar neighbourhood~(SNd) \citep[e.g.,][]{1998AJ....115.2384D,2006A&A...449..533A,2008A&A...483..453F,2008A&A...490..135A}.

The recent data releases~DR2 and EDR3 of ESA's \Gaia mission~\citep{2018A&A...616A..11G,2021A&A...649A...1G} have made it possible to deepen our understanding of the known velocity and phase-space features, and also to discover many new ones. In the extended SNd the \Gaia~data have provided more evidence for the lack of equilibrium of the MW disk. For instance, previously known vertical wave-like motions~\citep{2012ApJ...750L..41W} were detected in $z-v_z$ space in the form of a snail-like structure~\citep{2018Natur.561..360A}. The precise measurements from \Gaia also provided a more comprehensive view on the in-plane phase-space structure of the MW~\citep{2018A&A...616A..11G,2018MNRAS.479L.108K,2018Natur.561..360A}, allowing a number of radially extended features and short kinematic arches to be identified in \uv velocity space~\citep{2018A&A...619A..72R,2019MNRAS.489.4962K,2019A&A...631A..47K}. These complex stellar motions are not easily interpreted considering a single mechanism. However, several attempts have been made to explain the origin of various features, involving spiral arms~\citep{2019MNRAS.484.3154S,2018MNRAS.481.3794H,2019MNRAS.490.1026H,2020ApJ...888...75B}, bar resonances~\citep{2019MNRAS.488.3324F,2019A&A...632A.107M,2019A&A...626A..41M} and external intruders~\citep{2019MNRAS.485.3134L,2021MNRAS.504.3168B}.

Over the decades, much attention was attracted by the agglomeration of stars in \uv kinematic space dubbed as the ``Hercules'' stream~(or the ``u-anomaly''), with an asymmetric drift of about $-45$~\kmps and negative radial velocity. After \Gaia DR2 it became clear that the Hercules stream consists of at least two elongated features~\citep{2018A&A...616A..11G}. For a long time numerical works and analytic calculations had argued that this stream can be explained as the effect of the MW bar if the Sun is placed just outside its Outer Lindblad Resonance~(OLR) - then Hercules arises naturally from resonant interaction between the disk stars and a short ($\approx \sim3$~kpc), fast-rotating bar~\citep{2000AJ....119..800D,2001A&A...373..511F,2009ApJ...700L..78A,2014A&A...563A..60A,2017MNRAS.466L.113M}. 
The alternative idea, supported by more recent work on the structure and dynamics of the bulge and bar \citep{2015MNRAS.450.4050W,2017MNRAS.465.1621P} as well as by subsequent chemo-kinematical measurements~\citep{2019MNRAS.490.4740B,2019A&A...632A.121W}, suggests that the Hercules stream is related to the corotation resonance \citep{2017MNRAS.464L..80P,2019A&A...626A..41M,2020MNRAS.495..895B} which requires a slowly rotating long bar~\citep[$\approx 4.5$~kpc, see, e.g.,][]{2019MNRAS.489.3519C,2019MNRAS.488.4552S,2019MNRAS.490.4740B}.

The spiral structure is also believed to perturb the kinematics in the SNd, creating multiple overdensities in \uv space~\citep{2011MNRAS.417..762Q, 2011MNRAS.418.1423A,2016MNRAS.457.2569M}. \cite{2018MNRAS.480.3132Q} suggested that the ridges and arcs seen in the local velocity distributions are consistent with the presence of multiple spiral arms~\citep[see also][]{2018MNRAS.481.3794H,2018A&A...615A..10M,2020ApJ...888...75B} which may be transient structures~\citep{2019MNRAS.490.1026H}. The properties of the MW spiral structure are less constrained than those of the Galactic bar because most of our knowledge about the spiral arms is based on HII regions~\citep{1976A&A....49...57G,2003A&A...397..133R}, giant molecular clouds~\citep[see, e.g.,][]{1986ApJS...60..695C,1986ApJ...305..892D,1988ApJ...331..181G,2009A&A...499..473H}, pulsar dispersion measures~\citep[see, e.g.,][]{1993ApJ...411..674T}, the distribution of young star-forming regions~\citep[see, e.g.,][]{2014A&A...569A.125H,2014ApJ...783..130R,2019ApJ...885..131R} and analysis of extinction maps~\citep[see, e.g.,][]{2006A&A...453..635M,2019A&A...625A.135L,2020A&A...641A..79H}.
Then the number of arms, their precise locations and strength can only be constrained indirectly by modeling the phase-space features in the disk~\citep[see, e.g.,][]{2012MNRAS.425.2335S,2015MNRAS.453.1867G}. 

Interactions with satellites often trigger the formation of spiral arms~\citep[see, e.g.,][]{1986ApJ...309..472Q,2010MNRAS.403..625D,2018MNRAS.474.5645P} and could perhaps even stimulate the formation of bars~\citep[see, e.g.,][]{1990A&A...230...37G,2004MNRAS.347..277M}. Therefore, it is natural to assume that the external perturbations affect the kinematics of the disk, including the local SNd kinematics~\citep[see, e.g.,][]{2009MNRAS.396L..56M,2012MNRAS.419.2163G}, via the impact of tidally-induced large-scale stellar density structures. Note, however, that the properties of the self-excited and tidally-induced spirals~(or bar) could be different also implying a different impact on the kinematics of the stellar populations in the disk~\citep{2021MNRAS.504.3168B}.

Thanks to the \Gaia RVS sample~(DR2), the largest ever available sample of stars with 6D phase-space information, it is possible to identify stellar overdensities corresponding to the spiral arms in angular momentum~(or guiding) space~\citep{2020A&A...634L...8K}. This makes it possible to constrain the physical locations of these stellar density structures in the Galactic plane. The angular momentum space analysis sharpens the stellar overdensities by ``squizzing'' the stellar orbits towards their guiding centres, reducing the blurring by their epicyclic motions. The success of this idea is confirmed by {\it (i)} a good correspondence of the recovered large-scale stellar density structure with the location of spiral arms obtained with high-mass tar formation regions ~\citep{2014ApJ...783..130R} and {\it (ii)} comparison with $N$-body/hydrodynamical simulations showing a similar relation between stellar angular momentum overdensities and their spirals arms.

The aim of this work is to explore the link between the MW spiral arms, the radially extended ridges in the \RVphi plane, and the numerous local moving groups in the SNd \uv plane. We illustrate our findings in the MW with a parallel, comprehensive analysis of a new high-resolution $N$-body simulation of a MW-type spiral galaxy. The structure of the paper is the following. In Section~\ref{sec::model} we explore various manifestations of the tightly-wound spiral structure in the new simulations: in the galactic plane, in \RVphi space, and in a SNd-like region. In Section~\ref{sec::data} we define our sample of stars from \Gaia~DR2 and EDR3, and provide new evidence of the MW stellar spiral arms' presence in the angular momentum~(or guiding) space, and their contribution to various phase-space features. Section~\ref{sec::feh} describes chemical abundance information from Galah~(DR~3), APOGEE~(DR~16) and LAMOST~(DR~5) used to highlight signatures of the bar and spiral arms in these data. Finally, we discuss our results and summarize our findings in Section \ref{sec::concl}.

\section{A simulated galactic disk with tightly wound spiral arms}\label{sec::model}
In order to investigate \Gaia-like phase-space features in a MW-type disk, in this section we focus on the analysis of a new, high-resolution $N$-body simulation of an isolated disk galaxy, with tightly-wound, multi-arm spiral structure. 

\subsection{Model description}\label{sec::nbody_description}
 We performed a single high-resolution, $N$-body simulation of a disk galaxy. The initial disk is represented by a Miyamoto-Nagai density profile~\citep{1975PASJ...27..533M} with characteristic scale length of $4$~kpc, vertical thicknesses of $0.15$~kpc, and mass of $6 \times 10^{10}$~\Msun. The simulation includes a live dark matter halo whose density distribution follows a Plummer sphere~\citep{1911MNRAS..71..460P} with a total mass of $6.2 \times 10^{11}$~\Msun\ and scale radius of $21$~kpc. This choice of parameters leads to a galaxy mass model with a circular velocity of $\approx 220$~\kmps, but for an easier comparison with MW kinematics we re-scaled our galaxy to have a circular velocity of $240$~\kmps near the Solar radius at $8.2$~kpc \citep{2016ARA&A..54..529B}. The initial setup was generated using the iterative method provided in the AGAMA software~\citep{2019MNRAS.482.1525V}. 

In order to resolve substructures in the phase-space, we used $120 \times 10^6$ particles to model the stellar disk component and $61 451 200$ particles to model the dark matter halo, making this model one of the highest resolution $N$-body simulations of an isolated disk galaxy~\citep[see, e.g.,][]{2019MNRAS.482.1983F,2019A&A...622L...6K,2020MNRAS.499.2416A}. Note however, that the number of particles in a SNd-like region in this high-resolution model is still by a factor of $\sim\!5$ lower than the number of stars already available in the \Gaia~RVS sample.

For the $N$-body system integration, we used our parallel version of the TREE-GRAPE code~\citep[][Sect. 4.1]{2005PASJ...57.1009F} with multithread usage under the SSE and AVX instructions. In recent years we already used and extensively tested the hardware-accelerator-based gravity calculation routine in several galaxy dynamics studies, obtaining accurate results with a good performance~\citep[see, e.g.,][]{2017MNRAS.470...20S,2018A&A...620A.154K,2019A&A...628A..11D}. In the simulation we adopted the standard opening angle $\theta = 0.5$ and a gravitational softening parameter equal to $4$~pc. For the time integration, we used a leapfrog integrator with a fixed step size of $0.1$~Myr.

\begin{figure}[t!]
\begin{center}
\includegraphics[width=1\hsize]{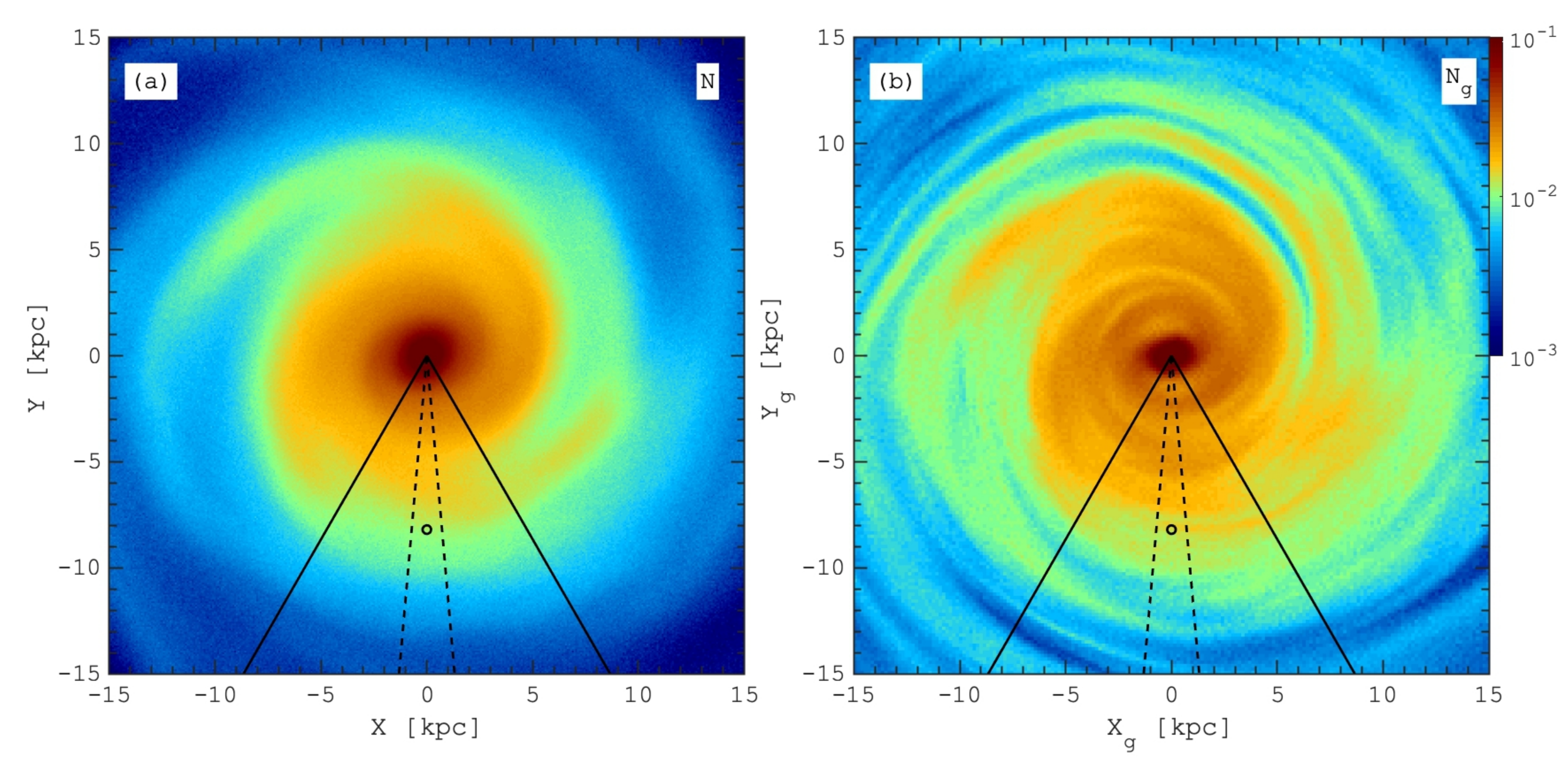}
\caption{Stellar density maps of the $N$-body model in \XY~(left) and \XgYg~coordinates (right, see  Eq.~\ref{eq::xg_yg} for the guiding coordinates transformation). The solid lines highlight a $60^\circ$-region in azimuth, comparable to the coverage in the MW of the \Gaia RVS sample, dashed lines limit the region we adopt in the analysis of \RdashVphi-ridges, and a SNd-like region~($0.2$~kpc solid line circle) is placed at $Y=-8.0$~kpc.}\label{fig::model_density_maps}
\end{center}
\end{figure}

\begin{figure*}[t!]
\begin{center}
\includegraphics[width=0.9\hsize]{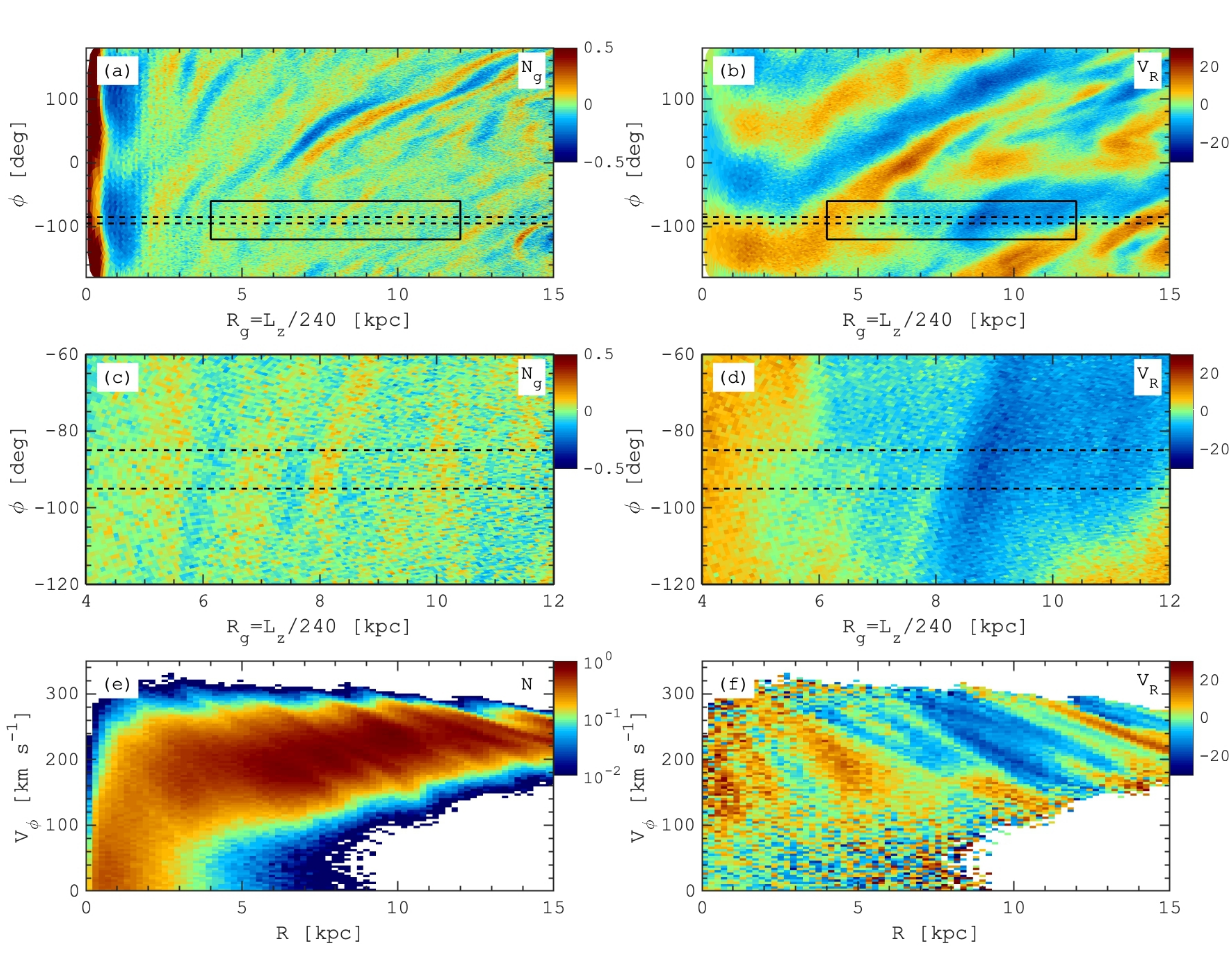}
\caption{{\it Top row:} unsharp-masked stellar density distribution in the N-body simulation {\it(a)} and the mean radial velocity map {\it(b)} in the \Rgphi-plane. {\it The middle row} shows zoomed maps of the rectangular region indicated in the top row: density perturbation~{\it(c)}, and the mean radial velocity map~{\it(d)}. {\it The bottom row} shows the density~{\it(e)} and the mean radial velocity~{\it(f)} in the \RVphi-coordinates for all stars in a narrow azimuthal strip, indicated by the dashed lines in the top and middle rows.}\label{fig::model_guiding_maps}
\end{center}
\end{figure*}

\begin{figure*}[t!]
\begin{center}
\includegraphics[width=1\hsize]{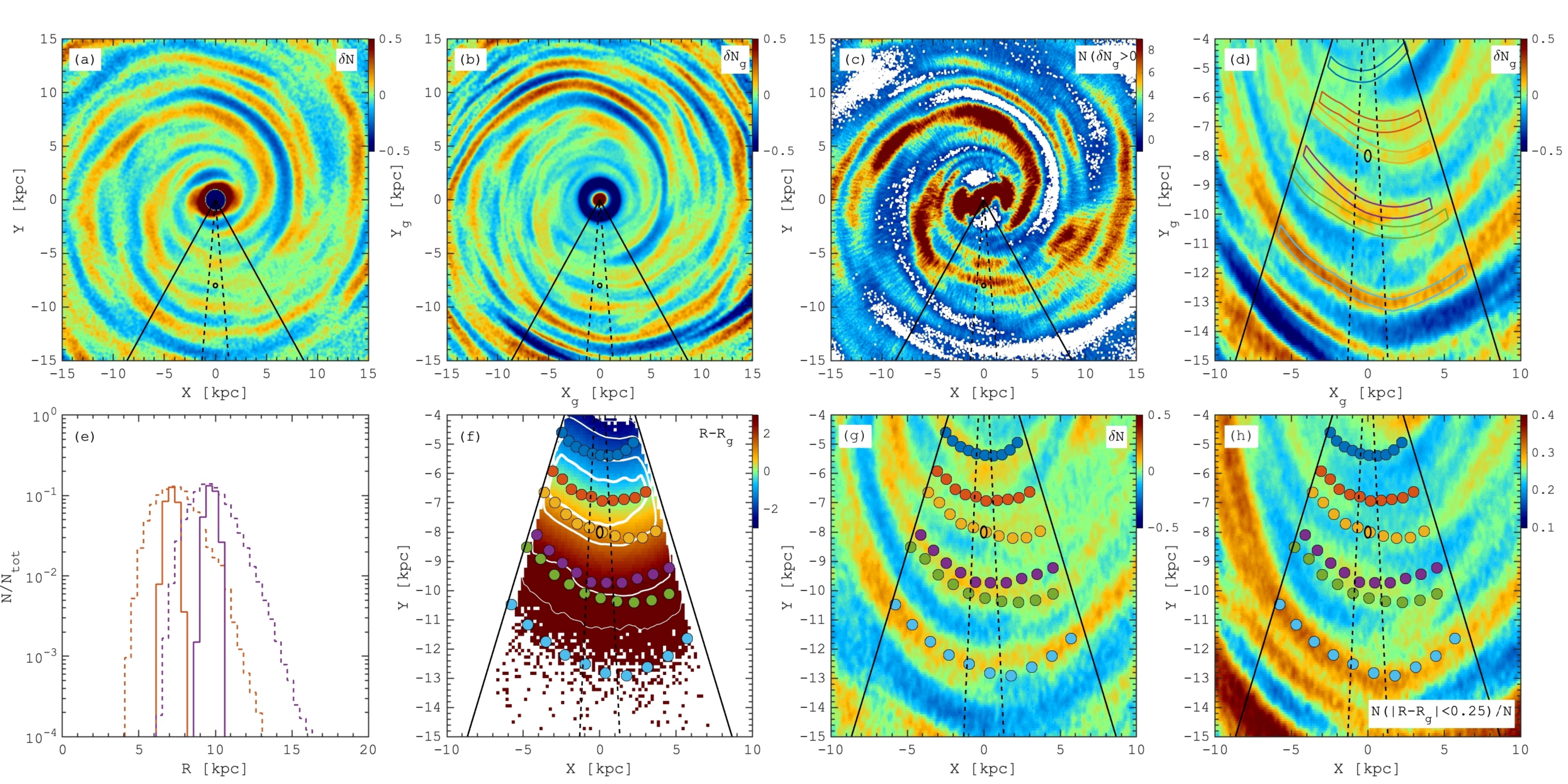}
\caption{Properties of spiral arms in the $N$-body simulation. {\it (a):} unsharp masking of stellar density in the galaxy's \XY-plane, showing a tightly wound spiral arm density perturbation.  {\it (b):}  unsharp-masked stellar density distribution in guiding \XgYg coordinates.  {\it (c):} density distribution of stars in \XY, weighted by the positive density perturbation at their corresponding \XgYg coordinates. {\it (d):} background map is a zoom-in of {\it (b)}; coloured lines indicate $\pm 200$~pc-wide radial selection regions along the peaks of the distribution in this map. {\it (e):} distributions of galactocentric radii for all stars in two \XgYg selection regions (second overdensity -- red, fourth overdensity -- purple; dashed lines). Solid lines correspond to the subset of stars located close to their guiding radii, $\rm |R-R_g|<0.25$~kpc. {\it (f):} background scatter plot shows the \XY positions of all particles in the second~(red) \XgYg overdensity region, where the white contours correspond to $0.8$, $0.5$, $0.1$ and $0.01$ density levels. Filled circles show the locations of density maxima as illustrated in panel {\it (e)} for all overdensities marked in frame {\it (d)}. {\it (g):} correspondence between spiral arm density perturbation (zoom-in from panel {\it (a)}) and location of such density maxima corresponding to all overdensities selected in \XgYg (panel {\it d}). {\it (h):} background map represents the fraction of stars in \XY coordinates near their guiding centers~(dynamically cold population with $|R-R_g|<0.25$), while the filled circles show the location of the same density maxima as in frames {\it (f)} and {\it (g)}.}
\label{fig::model_spirals_selection}
\end{center}
\end{figure*}

\begin{figure*}[t!]
\begin{center}
\includegraphics[width=0.85\hsize]{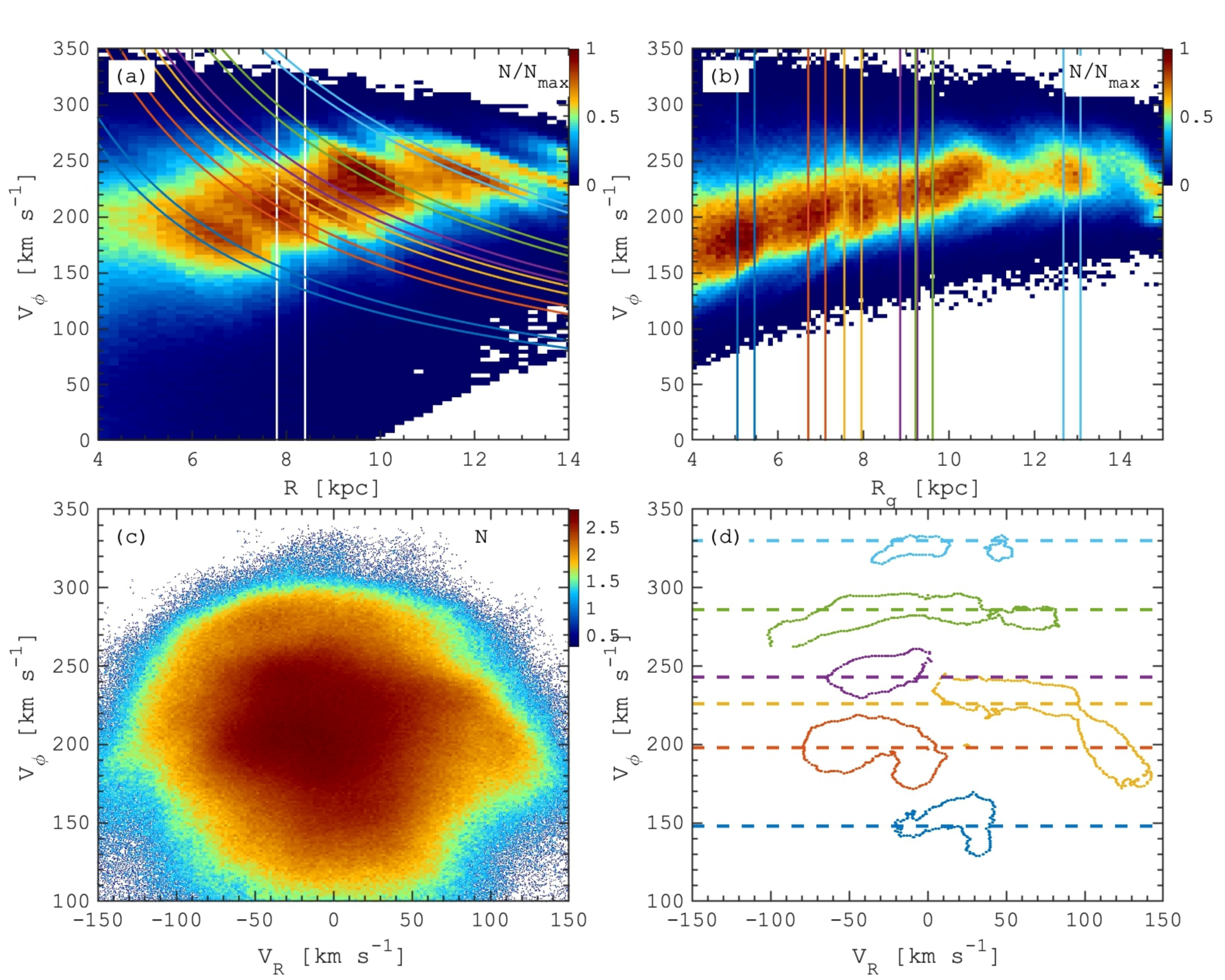}
\caption{Relation between diagonal ridges and the moving groups in $N$-body simulation. {\it (a):} density distribution in the \RVphi plane for stars selected in a $10^\circ$ azimuthal cone~(see Fig.~\ref{fig::model_spirals_selection}). Coloured lines represent the selected spiral arm regions in guiding coordinates~(see Fig.~\ref{fig::model_spirals_selection}(d)). White vertical lines limit the SNd-like region at $8\pm0.2$~kpc. {\it (b):} same as in (a) but transformed to \RgdashVphi ~coordinates. {\it (c):} phase-space \UV density distribution for stars in the SNd-like region between the white lines in {\it (a)}. {\it (d):} Coloured density contours highlight the positive overdensities in \UV coordinates after unsharp masking of the density distribution in {\it (c)}. The colours of the \UV-overdensities are chosen to match the angular momentum~(guiding radius) of the large-scale selections made in \XgYg~(see Fig.~\ref{fig::model_spirals_selection}(d)) and thus providing correspondence between the local SNd-like \UV-features and large-scale \RVphi-ridges shown in {\it (a)}.}\label{fig::model_ridges_uv}
\end{center}
\end{figure*}

\subsection{Angular momentum (guiding) space}\label{sec::nbody_analysis}
The $N$-body model presented here is unstable towards the formation of a multi-arm, tightly-wound spiral structure in the early phases of evolution while a strong stellar bar, as in many MW models, forms at later times. To focus our analysis on the spiral arms, we consider a single snapshot at $880$~Myr for which we extract the particle positions and velocities that are used in the phase-space plots below. Fig.~\ref{fig::model_density_maps}~(a) shows the stellar density map of this snapshot. The superposed solid lines highlight a $60^\circ$ cone region roughly comparable to the \Gaia radial velocities sample coverage~(with $<10\%$ parallax error), the dashed lines the region we adopt in the analysis of \RgVphi ridges, and a $0.2$~kpc radius SNd-like region is placed at $8.0$~kpc from the centre, within which we explore the substructures~(so-called moving groups) in simulated \uv-space.

Similar to our previous study~\citep[][]{2020A&A...634L...8K} in Fig.~\ref{fig::model_density_maps}~(b) we present the global structure of the angular momentum space in Cartesian ``guiding coordinates'' defined as
\begin{equation}
    X_g = R_g \, \cos(\phi), \; Y_g = R_g \, \sin(\phi), \label{eq::xg_yg},
\end{equation}
where $R_g = L_z / (240 {\rm km s}^{-1})$ is the guiding radius of the particles, $L_z = R \times v_\phi$ is the angular momentum, $R$ is the instantaneous value of the galactocentric distance, $\phi$ is the azimuthal angle in cylindrical coordinates and $v_\phi$ is the instantaneous value of the azimuthal velocity. In this notation the guiding radii of stars correspond to the angular momentum normalized by the circular velocity beyond $>4$~kpc in the disk. For near-circular orbits in an axisymmetric potential, in which $L_z$ is conserved, $R_g$ is the true time-independent guiding centre radius. For more realistic potentials with non-axisymmetric perturbations, $R_g$ can be thought of an approximate orbit-averaged mean radius. 

As shown in \cite{2020A&A...634L...8K}, this approach allows us to decrease the effects of the radial oscillations of stars and detect phase-space overdensities in the disk that would otherwise be too blurred to see. The approximation of a constant circular velocity is reasonable for galaxies with a nearly flat rotation curve. E.g., recent \Gaia data analysis shows a gently declining rotation curve in the Milky Way, with gradient $-1.7\pm0.1$~\kmskpc in \cite{2019ApJ...871..120E} and $-1.41\pm0.2$~\kmskpc in ~\cite{2019ApJ...870L..10M}. This corresponds to a $\approx 2\%$ change in the derived guiding radii at $3$~kpc from the Sun, which we neglect here for simplicity.

By transforming the in-plane coordinates of stars \XY to \XgYg we use a simpler description than the full transformation to the action-angle variables of a nearby axisymmetric potential, whereby also the azimuthal blurring can be removed by transforming the star to the $\theta_\phi$ conjugate to $R_g$.  Such a procedure requires knowledge of the full gravitational potential which in view of the later application to the MW in Section~\ref{sec::data} we wish to avoid. However, for the N-body model particles we used the  action-angle transformation in the epicyclic approximation \citep[e.g.,]{2016MNRAS.457.2569M} to construct a map similar to the \XgYg map in Fig.~\ref{fig::model_density_maps}. However, differences between both map are small. Moreover, we use the particle distribution in \XgYg only for identification, and transfer back to \XY before comparing with other \XY information or data. Therefore the exact transformation used to reduce the epicyclic blurring is not important.

The map of the particle distribution resulting from the coordinate transformation of Eq.~\ref{eq::xg_yg} is compared to the original density distribution in Fig.~\ref{fig::model_density_maps}. In the original density map the spiral structure is represented by $2-3$ main large-scale and broad arms. In contrast, the density distribution in the guiding coordinates shows multiple narrow and tightly-wound density structures. Some of these overdensities in Fig.~\ref{fig::model_density_maps}(b) evidently correspond to the main spiral arms in Fig.~\ref{fig::model_density_maps}(a), however, as expected, more concentrated in radius. A few less prominent overdensities in Fig.~\ref{fig::model_density_maps}(b) are faintly seen in the left panel, while others do not seem to match any structures in \XY space. Note that a direct match of the density structures in \XY and \XgYg coordinates is not expected because the spiral pattern while tightly wound is not axisymmetric,  nor is the gravitational potential. However, an exact match is not needed; similar as in  \citep[][]{2020A&A...634L...8K}, once stars in the \XgYg overdensities are identified, we back-transform, determine the corresponding features in \XY space, and compare with the physical spiral arms.

\subsection{$N$-body model: angular momentum~(guiding) space overdensities}

We start our analysis by providing a more detailed description of the angular momentum space. Fig.~\ref{fig::model_guiding_maps}~(a, c) shows unsharp-masked angular momentum density distributions in polar $(R_g,\phi)$ coordinates defined as:
\begin{equation}
\rm \delta N_g(R_g,\phi) = \left[ N_g(R_g,\phi) - \langle N_g(R_g,\phi) \rangle \right] / \langle N_g(R_g,\phi) \rangle\,, \label{eq::sharping} 
\end{equation}
where $\rm \langle N_g \rangle$ denotes the mean number density map constructed by convolving the density distribution with a 2D Gaussian kernel of width $500$~pc. The large scale map in Fig.~\ref{fig::model_guiding_maps}(a) shows a large number of radially extended overdensities with an amplitude of up to $20\%$. In a local, \Gaia-like region illustrated in frame (c) these overdensities are visible as a more nearly vertically extended~(along azimuth) wave-like pattern. These maps and Fig.~\ref{fig::model_guiding_maps}(b,d), which show corresponding mean radial velocity distributions, can be compared to similar maps based on the \Gaia radial velocity sample presented in \cite{2019MNRAS.490.5414F} and \cite{2019A&A...626A..41M}. On large scales, see Fig.~\ref{fig::model_guiding_maps}(b), we find a near $180^\circ$ symmetry of the $v_R(R_g,\phi)$ velocity distribution~\citep[see also][]{2021MNRAS.500.4710C}, likely caused by the central oval~(proto-bar in the model). On top of the large-scale pattern we observe multiple sharp velocity wave-like modulations~\citep[see also][]{2019MNRAS.490.5414F}.
Interestingly, these radial velocity oscillations are tightly connected to the density variations~(see Fig.~\ref{fig::model_guiding_maps}(a, c)). For instance, comparison of frames (c) and (d) implies that overdensities~($\delta N_g>0$) generally correspond to the more negative radial velocity regions, which can be explained when the stellar overdensities in the guiding space contain stars associated with tightly-wound spiral arms~\citep[][]{2012MNRAS.425.2335S,2014MNRAS.440.2564F,2016A&A...589A..13A}.

Next, Fig.~\ref{fig::model_guiding_maps}(e) shows the \RVphi plane for particles selected in a narrow $10^\circ$ region. Similar to a number of previous studies we observe so-called ridges, i.e., radially-extended~($3-5$~kpc long) diagonal stellar overdensities roughly corresponding to constant angular momentum. The map in Fig.~\ref{fig::model_guiding_maps}(f) depicts the radial velocity variations associated with these density ridges. The global structure and the amplitude of the velocity pattern ($15-30$~\kmps) is similar to that found in the MW with \Gaia-RVS~\citep[see, e.g.,][]{2019MNRAS.488.3324F,2019MNRAS.485.3134L}.

Having seen that our model is qualitatively similar to the MW disk kinematics seen in \Gaia-RVS, we now explore further the origin of the \XgYg overdensities and their connection to the model's spiral arms. Due to the exponential radial density profile in the model the contrast of the spiral arms in a colour map such as Fig.~\ref{fig::model_density_maps}(a) is not large; for a more precise description we therefore use an unsharp-masked map of the \XY stellar density distribution; see Fig.~\ref{fig::model_spirals_selection}(a). This map indeed depicts a multi-armed, tightly-wound spiral pattern, where some of the arms (overdensities with $\delta N>0$) can be traced over $360^\circ$ in azimuth. The amplitudes of the spiral arms do not exceed $10-20\%$ and also vary along the individual arms. 

Next to this figure we show the unsharp-masked density map in the guiding~\XgYg coordinates~(Fig.~\ref{fig::model_spirals_selection}(b)). Interestingly, the global morphology of the density perturbation is similar in both maps, however, the guiding space overdensities are more radially concentrated and thus can have larger amplitudes~(up to $\delta N_g \approx 0.4$), and several of them apparently can map to one, broader, spatial spiral arm. The larger concentrations and amplitudes are easily understood, because the coordinate transformation of Eq.~\ref{eq::xg_yg} moves stars close the their mean radial location, thus enhancing the amplitude of angular momentum structures in a given azimuthal range. Note that the peak angular momentum corresponding to a certain overdensity varies with azimuth, visible clearly over large angular scales in Fig.~\ref{fig::model_guiding_maps}(a) but to a lesser degree also on smaller scales (Fig.~\ref{fig::model_guiding_maps}(c)).

Without making explicit connections between particular structures in \XgYg and \XY, we can establish a global correspondence between the overdensities in the guiding coordinates and the real space stellar overdensities:  Fig.~\ref{fig::model_spirals_selection}(c) shows the \XY density map obtained by summing the contributions of all back-transformed particles from the \XgYg-distribution, but weighted by the amplitude of the $\delta N_g$ perturbation in Fig.~\ref{fig::model_spirals_selection}(b). The multiple structures in frame (c) nicely trace the large-scale locations of the spiral arm overdensities in frame (a), showing that the peaks in the angular momentum distribution of this model are related to the positive spiral density perturbation in \XY space, even if one apparent density arm may contain more than one blurred angular momentum peak.

Next we focus on a smaller region in the simulated galaxy, similar to that covered by the \Gaia data, to illustrate more explicitly the connection between particular guiding space overdensities and the corresponding spiral arms in Fig.~\ref{fig::model_spirals_selection}(a). Fig.~\ref{fig::model_spirals_selection}(d) zooms in on the lower part of frame (b), highlighting the guiding space overdensities by the coloured bands. These bands contain stars located within $R_g \pm 200$~pc regions near the peaks of the overdensities. In Fig.~\ref{fig::model_spirals_selection}(e), the dashed lines show the number of stars as a function of galactocentric distance for the second~(red) and the fourth~(purple) angular momentum overdensities. Stars from both overdensities span a very large radial range, $\sim\!6-7$~kpc, with significant overlap between them. Clearly, stars from several other \XgYg overdensities will also overlap with these two; this is illustrated in Fig.~\ref{fig::model_spirals_selection}(e,f). 

To characterize the amplitude of the orbital oscillations we calculate the difference between current galactocentric distance and the guiding radius $R-R_g$. This parameter represents an average radial excursion of stars because even though stars with large radial oscillations can be found near their guiding centers, the probability of this is small, while dynamically cold stars always stay near their guiding centers. From Fig.~\ref{fig::model_spirals_selection}(c) a guiding space overdensity, near its peak, should mainly contain stars with low $\rm |R-R_g|$ while the tails of the $\rm |R-R_g|$ distribution should be populated by hotter stars. This is illustrated in Fig.~\ref{fig::model_spirals_selection}(e) where the solid lines represent the radial distribution of stars with low mean $\rm |R-R_g|<0.25$~kpc. For two~(red and purple) overdensities, these distributions are narrow and located near the maxima of the density distributions for all $\rm |R-R_g|$~(dashed lines). In frame (f) of Fig.~\ref{fig::model_spirals_selection} the scatter plot depicts the \XY distribution of stars from one of the aforementioned overdensities~(red one), colour-coded by the value of $\rm |R-R_g|$. Large filled circles correspond to the location of density maxima similar to those illustrated in frame (e)) for all overdensities marked in frame (d). Note that although the selection stripes in frame (d) have similar shapes to those traced out by the fitted density maxima~(circles in frame (f)), these structures are not the same because the latter represent the mean locations of stars in \XY coordinates for a given overdensity in \XgYg.

Finally, having identified the \XgYg locations of the stellar density structures for the angular momentum overdensities, we compare them in Fig.~\ref{fig::model_spirals_selection}(g) with the spiral arm density perturbations (zoom in of map in frame (a)) where the filled circles are the same as in frame (f). It is nicely seen that the locus of stars selected near the peaks of the guiding center map correspond well to the real spiral arms. This agreement parallels our previous finding~\citep{2020A&A...634L...8K} with the same procedure as described here, that the density maxima of MW stars selected in the angular momentum~(or guiding \XgYg) space overdensities match the locations of high-mass star-forming regions used as tracers of the MW spiral arms.  Fig.~\ref{fig::model_spirals_selection} shows that in our high-resolution simulation single guiding space overdensities dominate individual spiral arms. However, this does not rule out
the possibility that in other simulations, depending on the dynamical nature of the spiral arms, several such overdensities contribute substantially to a single arm~\citep[see, e.g,][]{2020MNRAS.497..818H}. In this case, the XY-locus of the guiding space overdensities would overlap, which we have not observed in the available \Gaia data~\citep{2020A&A...634L...8K}, at least for the most prominent features. This issue could be further tested with forthcoming \Gaia releases.

\begin{figure*}[t!]
\begin{center}
\includegraphics[width=0.85\hsize]{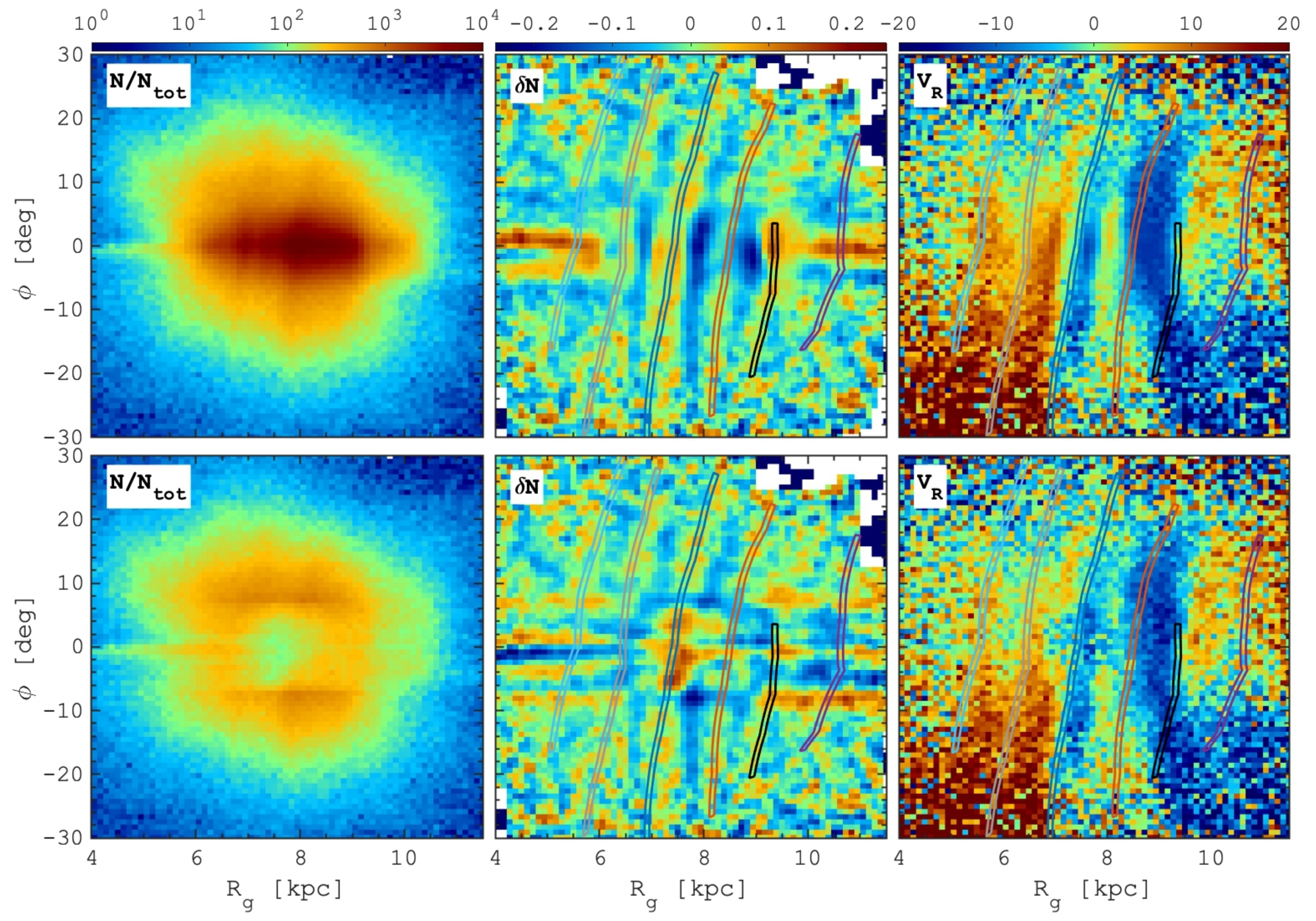}
\caption{Angular momentum~(or guiding space) structure of the G3RV2 stars near the Galactic plane~($|z|<200$~pc): number of stars~(left), unsharp masking of the density distribution~(center) and the mean radial velocity~(right). Bottom row shows the same but for stars outside the SNd~(beyond $1$~kpc cylindrical distance from the Sun). Coloured lines depict the angular momentum selections used in \XgYg coordinates to identify overdensities in \cite{2020A&A...634L...8K}. The hole in the density distribution~($R_g \approx 7.6$~kpc) in the bottom left frame is shifted towards the Galactic center because the SNd contains more stars from the inner regions.}\label{fig::GaiaPlane}
\end{center}
\end{figure*}

Another approach for locating the spiral arms in this model is shown in frame (h) of Fig.~\ref{fig::model_spirals_selection} (see also frame (e)), where we see that the density maxima of the spiral arms also correspond to maxima in the fraction of stars near their guiding centers. Specifically, a map of the fraction of stars with $|R-R_g|<0.25$ is shown together with the same filled circles in panels (f) and (g) which represent the mean locations of the spiral arms in the \XY. We see a good correspondence of the spiral arms with a higher fraction of dynamically cold stars. This suggests that also in the \Gaia data we can trace the mean locations of the MW spiral arms by finding the regions with a larger fraction of stars with low $|R-R_g|$~(see Section~\ref{sec::mw_spirals}). Note that a larger fraction of stars with cold orbits near the spiral arms does not necessarily suggest that these stars formed locally in the arms. Likely these dynamically cold disk populations react more efficiently, compared to kinematically hot stars, to gravitational perturbations of the spiral arms~\citep[see, e.g.,][]{2018A&A...611L...2K}. 

\subsection{Linking the spiral arms with \RdashVphi-ridges and the \uv space in the $N$-body simulation}

In the previous section we identified the location of tightly-wound spiral arms in our $N$-body model by using the overdensities in the guiding~(or angular momentum) coordinates. We believe that one of the most exciting properties of the spirals is a large radial spread of the stars associated with individual arms~(see frames (e) and (f) in Fig.~\ref{fig::model_spirals_selection}). In this section we discuss the manifestations of such behaviour in different phase-space projections well-studied in the MW. In particular, we quantify how much the spiral arms contribute to the \RVphi-structures across the galactic disk, and in the local \uv space as intensively explored in the SNd over the last decades~(see references in Introduction).

In Fig.~\ref{fig::model_ridges_uv} we present the analysis of \RVphi space taken in a narrow $10^\circ$ azimuthal cone selection~(see dashed lines in Figs.~\ref{fig::model_density_maps}-\ref{fig::model_spirals_selection}) and the \uv space selected in between $R = 8\pm0.2$~kpc in this region. As we already noticed in the previous section the \RVphi space depicts a large number of diagonal overdensities -- ridges, spanning over $3-7$~kpc across the galaxy. To investigate the relation with the spiral arms as in the previous section we overplot the contours reflecting our angular momentum selections~(see frame (d) in Fig.~\ref{fig::model_spirals_selection}). It is easy to see that  that these coincide well with the \RVphi density-ridges. In Fig.~\ref{fig::model_ridges_uv}(b) we also show the \RgVphi space where the diagonal ridges are transformed into nearly vertically-aligned  overdensities which, in fact, depict the azimuthal velocity structure of the \XgYg overdensities. Similar to the previous frame by contours we highlight our angular momentum selections showing a nice agreement between the radially-extended kinematical structures and the spiral arms in the N-body simulation. Note also that, in contrast to the \RVphi space, here we do not see an overlap between the different structures demonstrating why it is convenient to identify the underlying structures in angular momentum space.

\begin{figure*}[t]
\begin{center}
\includegraphics[width=0.9\hsize]{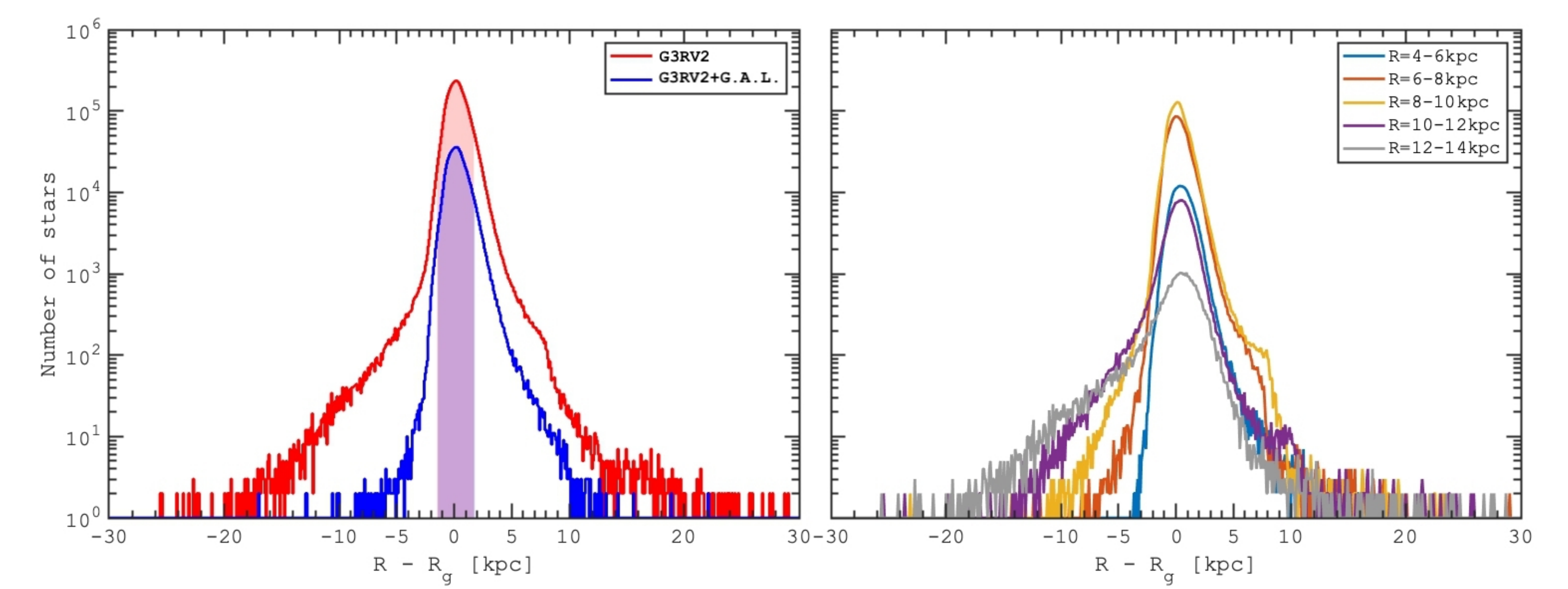}
\caption{{\it Left:} distribution of the difference between the current galactocentric position of stars and their guiding centers for the G3RV2 sample~(red) and G3RV2 all stars in common with Galah, APOGEE and LAMOST~(G.A.L., blue, see Sect.~\ref{sec::feh}). {\it Right:} distribution of $R-R_g$ for G3RV2 stars at different galactocentric radii. Shaded areas in the left frame correspond to 95\% of the distribution around its maxima. The distributions suggest that most stars are found less than $2$~kpc away from their guiding centers. Visible asymmetry of the distribution is caused by the contribution of accreted into the MW stars~(see Sec.~\ref{sec::mw_spirals} and Appendix~\ref{sec::app1} for details).}\label{fig::r_rg}
\end{center}
\end{figure*}

In the MW the diagonal density ridges discovered with GRV2 pass through the solar vicinity and thus, as shown in several recent works~\citep{2018A&A...619A..72R,2018MNRAS.480.3132Q,2019MNRAS.485L.104M,2019MNRAS.488.3324F}, may contribute to the known moving groups in the \uv space. In our simulation, we arbitrarily set up a SNd-like region at $8\pm0.2$~kpc shown with white lines in Fig.~\ref{fig::model_ridges_uv}(a).  In Fig.~\ref{fig::model_ridges_uv}(c) we show the corresponding density distribution in the \uv coordinates. A clearer picture of the velocity structures is shown in Fig.~\ref{fig::model_ridges_uv}(d) where we show zero-levels of the main overdensities derived by applying unsharp masking to the number star counts~(frame (c)). In this plot we can clearly see a number of features with an arch-like shape, similar to ones found in the SNd~\citep{2018A&A...616A..11G}.  A match between the \uv overdensities and particular \RVphi ridges~(and thus underlying spirals) can be done by comparing their mean azimuthal  velocities. For the \RVphi ridges in the SNd-like region these are shown by horizontal dashed lines. The colour of the \uv overdensities is taken to be same as the colour of the nearby angular momentum selection.  This demonstrates a perfect match between the ridges and the main \uv overdensities, in particular, we can identify the contribution of all six stellar density structures. Keeping in mind that these ridges correspond to the spiral arms, we suggest that stars tracing the inner disk spirals can be found in the bottom part of the \uv plane~(below LSR) while the outer disk spiral arms are imprinted in the top of the distribution. For, example, the highest-$v_\phi$ velocity \uv-features~(green and light blue) correspond to spiral arms with locus well beyond $8$~kpc.

We demonstrated that in our $N$-body simulation tightly-wound spiral arms correspond to radially-extended stellar density structures which have higher contrast in the guiding radii distribution. The radial extension of the spiral arms leads to their appearance at different galactocentric radii, causing prominent features in the \RVphi plane, which, in turn, contribute to the substructures in the \uv space at a certain galactic radius.  The many similarities of the disk kinematics in this simulation with the Gaia data then suggest that a number of kinematic arches~(or moving groups) in the SNd \uv plane may similarly be associated with the MW spiral arms.

\section{MW phase-space}\label{sec::data}
In this section, using \Gaia data we explore the structure and kinematics of stars in the SNd \uv space, \RVphi coordinates, and in the Galactic plane, indicating how the most prominent phase-space features are linked to the MW stellar density structures~(spiral arms and main bar resonances). We will use the dynamical structure found in our $N$-body simulation as a guide to find the correlations between in-plane overdensities in the guiding coordinates and the diagonal ridges and moving groups.   

\Gaia~(EDR3+DR2) is the largest available 6D phase-space dataset for $\approx7.2$ million stars brighter than $G_{\rm RVS} = 12$~mag which is making possible precise studies of the MW structure on large scales and trace the kinematics of the disk across multiple kpcs around the Sun.  For this work, we first took all sources with an available $5$~parameter astrometric solution~(sky positions, parallaxes, and proper motions) from \Gaia EDR3~\citep[][]{2021A&A...649A...4L} and radial velocities from \Gaia DR2~\citep[][]{2018A&A...616A...1G}. From this sample (hereafter G3RV2), we selected stars with positive parallaxes, and relative errors on parallaxes less than $10\%$. We estimated the parallax zero-point as a function of magnitude, colour and ecliptic latitude by using the routine that is available as part of the \Gaia EDR3 access facilities~\citep{2021A&A...649A...2L}\footnote{\href{https://gitlab.com/icc-ub/public/gaiadr3_zeropoint}{https://gitlab.com/icc-ub/public/gaiadr3zeropoint}}. Additionally we took into account the median parallax of the quasars of $\rm -17 \mu as$~\citep{2021A&A...649A...2L}. Distances were computed by inverting parallaxes. For calculating positions and velocities in the galactocentric rest-frame, we assumed an in-plane distance of the Sun from the Galactic centre of $8.19$~kpc~\citep{2018A&A...615L..15G}, a velocity of the Local Standard of Rest, $v_{LSR}=240~\kmps$~\citep{2014ApJ...783..130R}, and a peculiar velocity of the Sun with respect to the LSR, $U_\odot = 11.1$~\kmps, $v_\odot =12.24$~\kmps, $W_\odot=7.25$~\kmps~\citep{2010MNRAS.403.1829S}.

\subsection{MW spiral arms in the G3RV2}\label{sec::mw_spirals}

First we present the angular momentum~(or guiding space) distribution for the G3RV2 stars. Different from our previous work we do not apply the homogenization of the density distribution~\citep[see Fig.1 in][]{2020A&A...634L...8K}. We assume a constant value of the circular velocity $v_c=240 {\rm km s}^-1$ for computing the guiding radii of the stars; as discussed in Sec.~\ref{sec::nbody_analysis}, this is a good approximation. In Fig.~\ref{fig::GaiaPlane} we show the density distribution and its unsharp masking in the guiding polar~($R_g,\phi$) coordinates for better comparison with some recent works~\citep{2019A&A...626A..41M,2019MNRAS.490.5414F,2021MNRAS.500.4710C}. The density distribution reveals the presence of  several notable structures seen as azimuthally-extended  tightly-wound trailing overdensities. The unsharp masking allows us to highlight six main structures which are prominent across the entire G3RV2 sample. These overdensities overlap with the radial velocity pattern~(right frames), first discovered in \cite{2018MNRAS.478.3809S} and discussed also in \cite{2019MNRAS.490.5414F} who interpreted the complex velocity pattern as the manifestation of the spiral structure. More recently~\cite{2021arXiv210505263W} suggested that some of the angular momentum overdensities can be associated with the main resonances of the MW bar~\citep[see, also,][]{2019A&A...632A.107M}.

Since the G3RV2 dataset, while covering a large area in the MW, is still dominated by the stars in the extended Solar vicinity, it is important to test to what extent the density-velocity patterns we observe are caused by the local stars. In the bottom frames of Fig.~\ref{fig::GaiaPlane} we show exactly the same maps as in the top but for stars located outside $1$~kpc from the Sun. Obviously, the density distribution and its unsharp masking are now significantly impacted by our selection but still reveal the main stellar density structures in the unsharp mask. In particular, the radial velocity map is almost identical to that based on the entire G3RV2, implying that the stellar density structures in the angular momentum space can not be explained by local stars~(or selection/extinction biases). This does not support the possibility discussed in the recent study by \cite{2020MNRAS.497..818H}, that the overdensities in $\rm (X_g,Y_g)$ we found in ~\cite{2020A&A...634L...8K} are generated by the local \uv moving groups because the G3RV2 sample is dominated by stars in the SNd. Although the large majority of stars in the G3RV2 sample indeed can be found close to the Sun, the distant~($D>1$~kpc) stars show very similar patterns in Fig.~\ref{fig::GaiaPlane} as the full sample.

From Fig.~\ref{fig::GaiaPlane} we conclude that, similar to our $N$-body simulation, the MW angular momentum space is not featureless but contains the imprint of a number of overdensities which in the simulation represent the tightly-wound spiral arms. Although, in \cite{2020A&A...634L...8K} we already showed that the \Gaia DR2~(RVS) angular momentum overdensities are made of stars in the spiral arms and the main bar resonances. Here we use another approach; we recall that in the model~(see Fig.~\ref{fig::model_spirals_selection}(h)) the fraction of stars near their guiding centers~(cold orbits stars) clearly highlight the location of the spiral arms. Using the same idea for the MW, in Fig.~\ref{fig::r_rg} we show the global distribution of \RRg for all stars in G3RV2 where the shaded areas in left show that the $95\%$ of stars are located within $\approx 2$~kpc away from their guiding centers. The visible asymmetry of the distributions is caused by the contribution of accreted into the MW stars~(see end of this Section and Appendix~\ref{sec::app1} for details).

\begin{figure}[t!]
\begin{center}
\includegraphics[width=0.9\hsize]{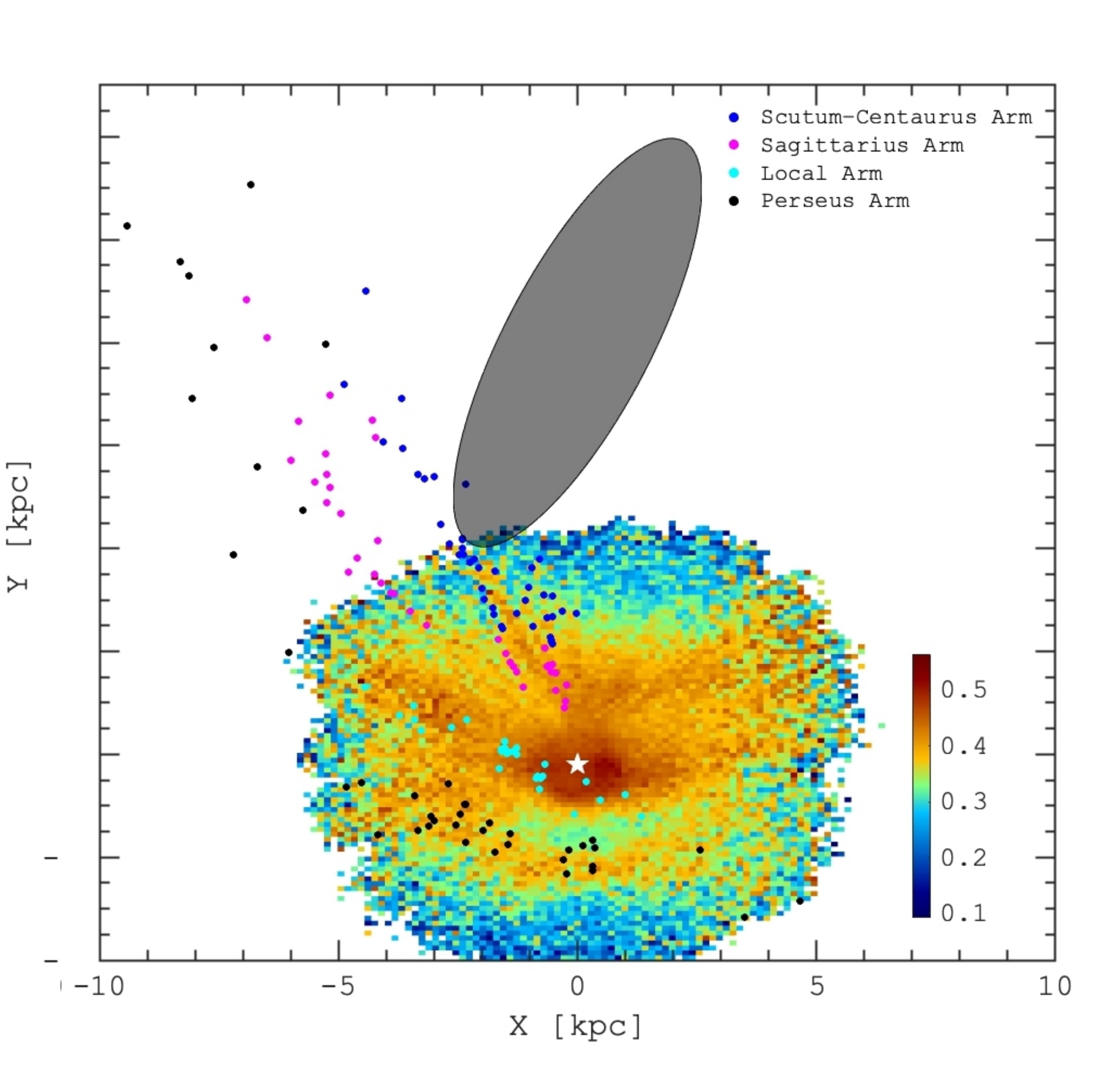}
\caption{Fraction of G3RV2 stars with $|R-R_g|<0.25$~kpc. The coloured points correspond to the location of the high-mass star-forming regions from \cite{2019ApJ...885..131R} associated with Scutum-Centaurus~(blue), Sagittarius~(magenta), Local~(cyan) and Perseus~(black) arms. A star symbol corresponds to the location of the Sun, and the grey oval represents the orientation of $\approx 4.6$~kpc long Milky Way bar~\citep{2015MNRAS.450.4050W}.}\label{fig::Gaia_vs_masers}
\end{center}
\end{figure}

\begin{figure}[t!]
\begin{center}
\includegraphics[width=1\hsize]{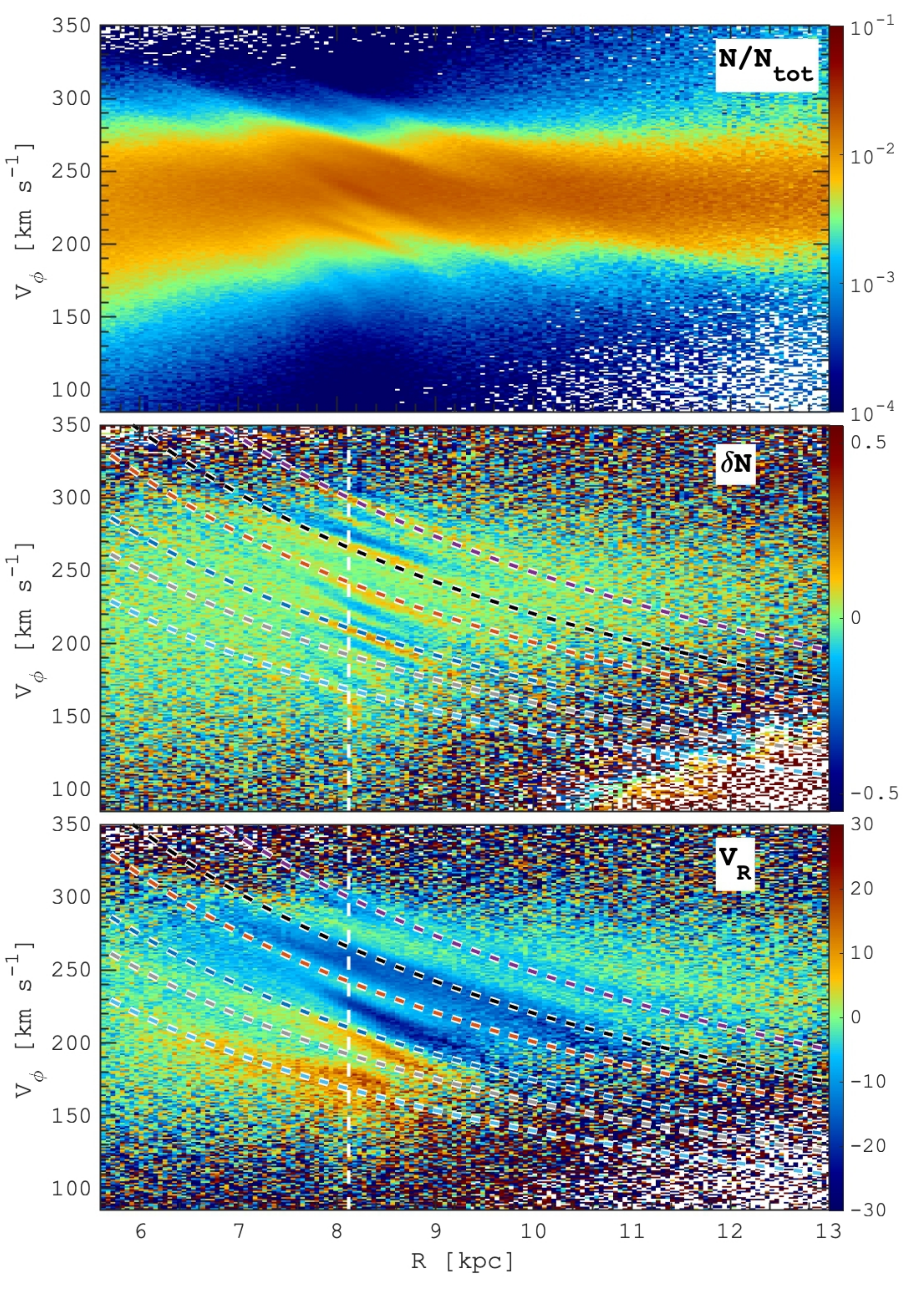}
\caption{Azimuthal velocity component structure in a narrow region $\pm 5^\circ$ along the Galactic radius. {\it Top frame} shows the density of stars normalized at each galactocentric distance. {\it Middle frame} shows unsharp masking of the density distribution. {\it Bottom frame} shows the same distribution but colour-coded by the mean radial velocity. Coloured diagonal lines depict the angular momentum selections in a given azimuthal range, using \XgYg coordinates to identify overdensities in \cite{2020A&A...634L...8K}~(see also Fig.~\ref{fig::GaiaPlane}).}\label{fig::GaiaRidges}
\end{center}
\end{figure}

In Fig.~\ref{fig::Gaia_vs_masers}~we present the fraction of stars located near their guiding centres~($|R-R_g|<0.25$~kpc) with an approximate location of the bar~(grey oval) and the maser sources with $<10\%$ distance error ~\citep{2019ApJ...885..131R} that are associated with the four MW spiral arms~(Scutum-Centaurus~(blue), Sagittarius~(magenta), Local~(cyan) and Perseus~(black)). Note that here we do not involve the coordinate transformation and show the distributions in the real physical space. The distribution of stars near their guiding centres show three large-scale regions which match well the locations of the  maser sources in spiral arms. In particular, we find good agreement for the Perseus and Local arms while the Sagittarius arm is less evident but still visible in G3RV2. Our new approach fails only for the innermost Scutum-Sagittarius arm where, in fact, the stellar overdensity also overlaps with the masers but it does not represent a continued structure. Nevertheless, the approach of identifying of the spiral arms through the distribution of the dynamically cold stars provides good results in both the $N$-body model of the tightly-wound spiral galaxy and in the MW~\citep[see recent finding by][]{2021A&A...651A.104P}.

\begin{figure*}[t!]
\begin{center}
\includegraphics[width=1\hsize]{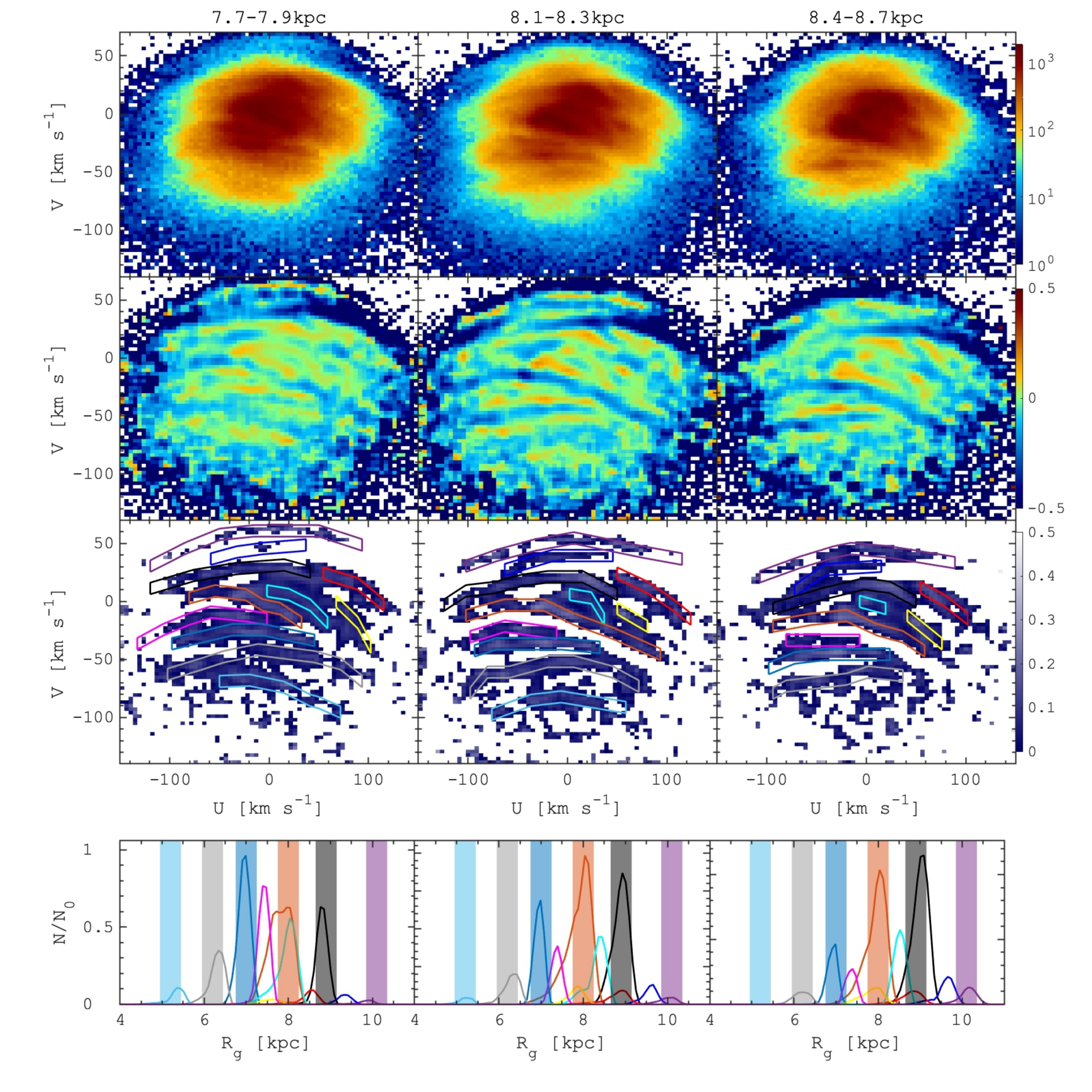}
\caption{ {\it First row.} Number density of stars in kinematic \uv space for the G3RV2 stars selected in three regions along the galactocentric radius: $R = 7.8$~kpc~(left), $R = 8.2$~kpc~(center), $R = 8.55~kpc$~(right). Colour depicts a number of stars in $\log$-scale. {\it Second row.} Unsharp masking maps~(see Eq.~\ref{eq::sharping}) for corresponded density distributions. {\it Third row.} Positive part of the unsharp masking maps~(overdensities) where coloured boxes highlight different isolated kinematic overdensities. {\it Forth row.} Number of stars from each \uv-plane selected feature as a function of their angular momentum~(or guiding radius). Coloured boxes depict the angular momentum selections used in \XgYg coordinates to identify the main spiral arms of the MW, which coincide with known star-forming regions, and the bar resonances from~\cite{2020A&A...634L...8K}. In particular, the Arcturus stream~(light-blue) is a low-velocity tail of Scutum-Centaurus spiral arm, part of the Hercules stream~(grey) corresponds to the bar corotation while its top part is likely to be the Sagittarius arm~(blue), the Sirius stream represents the bar OLR~(black), the Hat is likely the high-velocity tail of the Perseus arm~(purple).}\label{fig::GaiaUV}
\end{center}
\end{figure*}

An interesting feature is a second component in the distributions at $R-R_g\gtrsim 5$~kpc in Fig.~\ref{fig::r_rg}. Obviously, such stars exhibit rotation in the opposite direction compared to the MW disk. By checking the kinematics and spatial distribution of these stars we found that the vast majority of them show halo-like kinematics, therefore, representing both accreted and heated populations~\citep[see, e.g.,][]{2018MNRAS.478..611B,2020MNRAS.494.3880B,2019A&A...632A...4D}. Although beyond the scope of the paper, in Appendix~1 we confirm this finding using chemical abundance information in the $\aFe-\FeH$ plane.

\subsection{Connection of \RdashVphi-ridges with the SNd \uv structure in the MW}\label{sec::gaiaUV}

Next we focus on the radial structure of the MW spiral arms directly presented in Fig.~\ref{fig::Gaia_vs_masers} and their connection to the stellar overdensities in Fig.~\ref{fig::GaiaPlane}. Similar to the simulation analysis, in Fig.~\ref{fig::GaiaRidges} we show the radial distribution of azimuthal velocities for the G3RV2 stars selected in a narrow region~($10^\circ$) along the galactocentric distance vector through the Solar position. In order to highlight the stellar overdensities we normalize the stellar density distribution at each radius separately~(Fig.~\ref{fig::GaiaRidges}~top) and also present an unsharp masking residual map~(Fig.~\ref{fig::GaiaRidges}~middle) where several diagonal \RVphi-ridges are now even more prominent~\citep[see, e.g., also,][]{2018Natur.561..360A,2019MNRAS.488.3324F,2019MNRAS.486.1167B}. Despite the density normalization the ridges are still most prominent close to the SNd. Also once we colour-code the \RVphi map by the mean radial velocity, the density ridges~(see unsharp masking map) coexist with the wave-like velocity pattern extending across the whole radial range available with G3RV2. This suggests that the real extensions of the density ridges are much larger then we observe in the density map. We expect that \RVphi-ridges will be more prominent when the MW disk coverage will be increased in the forthcoming \Gaia releases~(DR3 and DR4; see also recent EDR3 results for the outer MW disk~\citep{2021A&A...649A...8G}). 

To match the \RVphi ridges and overdensities in \XgYg we show the lines of a constant angular momentum corresponding to the mean stellar density structures in the guiding coordinates~\citep[see Fig.~\ref{fig::GaiaPlane} and in][]{2020A&A...634L...8K} or colour bands in Fig.~\ref{fig::GaiaPlane}. In Fig.~\ref{fig::GaiaRidges} the location of the density ridges agrees with the mean angular momentum of stars in \Rgphi overdensities calculated in the same azimuthal range ($<10^\circ$). Similar to the $N$-body model discussed above we see a significant radial extension of the main stellar density structures across the large area in the MW which can be traced along \RVphi with nearly constant angular momentum. 


Finally, we take an in-depth look at the \uv space in three regions along the galactic radius with high coverage by G3RV2. In Fig.~\ref{fig::GaiaUV}~(top) we show the number density of stars at $R\approx 7.45, 8.2$ and $8.55$~kpc from the Galactic centre. Many substructures appear in this coordinate spaces showing how complex is the velocity distribution in the SNd. The \uv structure is nearly self-similar in three regions along radius with a clear vertical shift of the main features suggesting approximate conservation of the mean angular momentum of underlying density structures~\citep[see, also,][]{2018A&A...619A..72R}. 

We briefly describe the most notable features of the local velocity distribution. The Hercules stream indeed consists of two (or even three) features which are horizontally aligned in the SNd but being slightly tilted in the inner~(negative inclination) disk and the outer~(positive inclination). These several low-\VV features are separated from a central blob~(near LSR) by a diagonal gap which gradually decreases its mean azimuthal velocity with larger galactocentric distance. There is a superposition of several features previously identified as the \hor, \hya, \ple, \cb beyond the gap. Next, beyond \sir, there is another sharp break in the density distribution separating a very high-velocity feature -- the \ha. Almost all the features can be traced out to $\approx 0.5-0.8$~kpc from the Solar radius, and we can see that the groups of stars associated with different SNd moving groups can be also found at different Galactic radii. Although for a long time it was believed that the \uv moving groups are the local group of stars the \Gaia data show their significant radial extent, already noticed by \cite{2018A&A...619A..72R} who found that many of the SNd \uv kinematic substructures can be traced over $2-3$ kpc along radius implying their non-local nature.

To identify more precisely the stars contributing to the different features a wavelength analysis has been used~\citep[see, e.g.,][]{2017A&A...608A..73K,2018A&A...619A..72R}; however, roughly same results can be obtained by applying unsharp masking~(see Eq.~\ref{eq::sharping}) to the density maps presented in~Fig.~\ref{fig::GaiaUV}~(top). The result is presented in the second row of Fig.~\ref{fig::GaiaUV} where numerous features appear to be more prominent and isolated from each other, especially at the outskirts of the density distribution. To sharpen the features even more we show only the positive part of the residual maps and highlight the overdensities with different colour boxes~(third row). Of course, the boundaries of the selections are somewhat arbitrary, however, we are confident about the robust measurement of the main components because we obtain a similar sample of features as presented in~\cite{2018A&A...619A..72R}, however, in our analysis, their A11-A12 features seem to represent a single component.

Once stars are selected in the \UV plane their angular momenta can be determined, and similar as for the \RVphi-ridges, we can check how well they can be associated with the stellar density structures~(spiral arms, main bar resonance). In other words, what is the locus of the orbits of stars in different \uv features?  The bottom row of Fig.~\ref{fig::GaiaUV} shows the distributions of guiding radii $R_g$~(or angular momenta) of stars found in different \uv-features where we also marked the location of the MW spiral arms and the bar resonances from~\cite{2020A&A...634L...8K}~(corresponding colour bands in Fig.~\ref{fig::GaiaPlane} and colour ridges in Fig.~\ref{fig::GaiaRidges}). First, we note that the \uv features, being identified at different galactocentric radii, coincide well with each other in the angular momentum space. Second, the agreement between the \uv features and various in-plane stellar density structures~(both spiral arms and main bar resonances) in the angular momentum space implies that the so-called local moving groups are in fact pieces of the large-scale stellar density structures in the MW disk. In particular, the match between these ``local'' \uv features and the MW structures suggests, e.g. that the Arcturus stream is a low-velocity tail of the Scutum-Centaurus spiral arm, part of the Hercules stream corresponds to the bar corotation~\citep[see, also,][]{2020MNRAS.495..895B} while its top part is likely to be the Sagittarius arm, the Sirius stream represents the bar OLR~\citep[see, also,][]{2020arXiv201101233T,2020MNRAS.495..895B}, the Hat is likely to be a high-velocity tail of the Perseus arm. Note that we can not properly associate three tiny \uv features but we expect that they are likely to be either high-order resonances of the bar~\citep[see, e.g.][]{2019A&A...626A..41M,2020ApJ...888...75B,2020MNRAS.499.2416A} or previously unknown weak arms or spurs/featherings of the main MW spiral arms.

\subsection{Summary on the MW phase-space}

In this Section we started the analysis of the density structure and kinematics in the angular momentum space depicting several large-scale azumithally-extended~($\sim 60^\circ$) overdensities. The locus of these overdensities coincides with the MW spiral arms identified by the location of high-mass star-forming regions implying that the angular momentum peaks contain the stars of the spiral structure. For the first time we also demonstrated that in \XY coordinates the spiral structure can be directly detected as regions in the disk with a higher fraction of cold-orbit stars~(near their guiding centers). Stars in the angular momentum overdensities~(spiral arms and the bar resonances) can be traced along the \RVphi density ridges. Once these radially-extended ridges pass through the SNd region, in the \uv space the main structures identified as previously-known moving groups~\citep[see also][]{2005AJ....130..576Q,2018MNRAS.480.3132Q}. The resonances of the bar show the same behaviour contributing to the Hercules and Sirius streams for corotation and the OLR respectively. Therefore we conclude that the most prominent moving groups in the \uv space are the pieces of the MW spiral arms and the bar resonances. In other words, {\it co-called local moving groups are not local}.

\subsection{Other interpretations of \uv plane substructures} \label{sec::alternatives1}

Because our analysis of the overdensities in the local \uv velocity plane is based on the properties of the large-scale angular momentum structures, it is interesting to discuss and compare with alternative interpretations of some of these features. For instance, using the kinematics of the \Gaia stars in a small annulus around the solar radius \cite{2019A&A...632A.107M} argued that the constant angular momentum with azimuth for the second ridge of Hercules is unexpected for the bar's co-rotation. However, the more recent determination of corotation for a decelerating bar by~\cite{2021MNRAS.505.2412C} appears to agree with our prediction of ${\rm v}_\phi$.

\cite{2019A&A...626A..41M} showed that a locally slightly declining rotation curve would displace the bar resonances at the top of Hercules for the corotation and at the location of the Hat for the OLR. We note here that assuming a slightly incorrect rotation curve in the guiding radius transformation changes the $R_g$ of the AM overdensities, but since the stars identified with them are back-transfered to \XY before comparing with the observed \uv diagram, this has no influence on their final ${\rm v}_\phi$ velocities. Similarly, the assumed value of the solar velocity causes the same shift for the observed stars and the subset of stars identified with an overdensity.

 As we discussed above, only the {\it  main} \uv features are here associated with the spiral arms and/or major~(corotation, OLR) bar resonances. However, we do not rule out scenarios where some  \uv features may represent high-order resonances of the bar~\citep[see, e.g.,][]{2019A&A...626A..41M,2020MNRAS.499.2416A}. According to some simulations of barred galaxies~\citep[see, e.g,][]{2002ApJ...569L..83A,2007MNRAS.379.1155C}, the number of stars trapped by the high-order bar resonances is substantially lower compared to the populations of corotation and the OLR. Thus the \uv features associated with high-order resonances are expected to be significantly weaker~(contain fewer stars) compared to the corotation/OLR streams.

For a long time dissolved star clusters were thought to cause the main moving groups in the \uv-space~\citep[][]{1965gast.book..111E,1997ESASP.402..525S,2007AJ....133..694D}. Recent models suggest that disrupted star clusters can be detected in the \uv space no longer than $\sim 1$~Gyr after their  formation~\citep{2019ApJ...884..173K} and thus their remnants are unlikely to be prominent over a wide range of galactocentric radii because of the rapid distortion, especially in the presence of the spiral arms.

Another scenario used for interpretation of some of the \uv moving groups assumes the presence of stars from accreted satellites in the MW disk~\citep[see, e.g.,][]{2009MNRAS.396L..56M}. In this case, the accreted populations can have a substantial phase-space overlap with in-situ populations~\citep{2017A&A...604A.106J}, thus the chemical abundances are key tools in the understanding of the impact of mergers on the structure of the MW phase-space. We focus in the next section on the stellar abundance patterns in phase-space across the MW disk. 

\begin{figure*}[t!]
\begin{center}
\includegraphics[width=0.85\hsize]{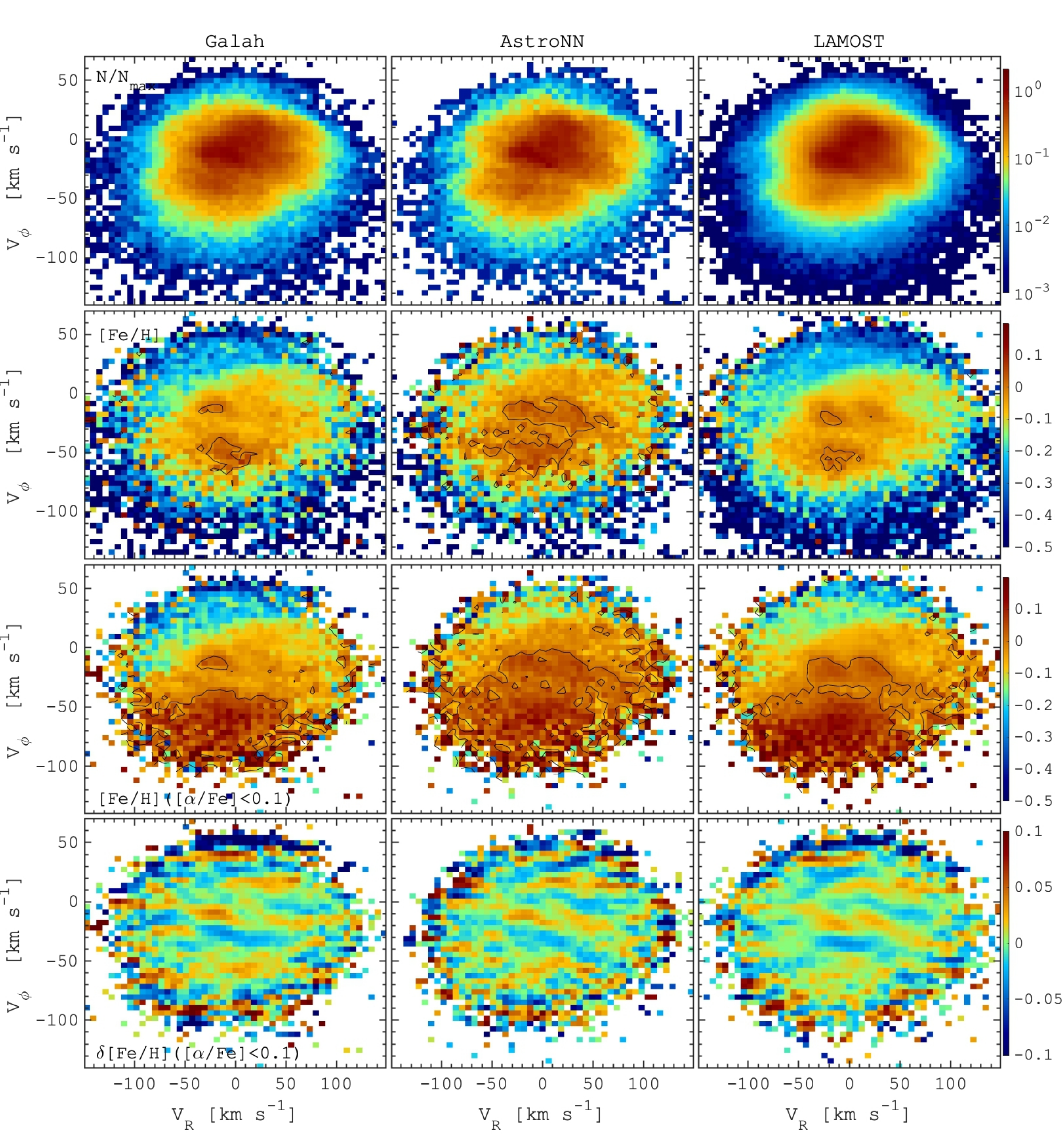}
\caption{Kinematic \uv space for the G3RV2 stars in the SNd cross-matched with Galah~(left), AstroNN~(center) and LAMOST~(right): density distribution~(first row), colour-coded by the mean \FeH~(second row),  the mean \FeH for thin-disk stars (low-$\alpha$ sequence, third row) and the residual \FeH distribution for low-$\alpha$ stars~(bottom row). The \uv features are well established with the higher mean metallicity~($\delta \FeH \approx0.05$~dex), and two groups of kinematic features are separated by low-\FeH gaps behind the corotation~(\her streams) and the OLR~(\sir stream). At a given galactocentric radius the metallicity gradient as a function of the azimuthal velocity reflects the metallicity gradient in the MW along the guiding radius~(or angular momentum).}\label{fig::GaiaFeH1}
\end{center}
\end{figure*}

\section{Gaia and spectroscopic surveys: chemical abundances in the SNd, \RVphi, \XgYg, and $R_g-z/V_z$}\label{sec::feh}

A combination of stellar kinematic and abundance information based on high-resolution spectroscopy is a valuable instrument to explore the present-day properties and evolution of the MW stellar populations. To study chemical abundances trends associated with the MW spiral arms and bar resonances, we cross-matched the G3RV2 catalogue with Galah DR3, APOGEE DR16~(astroNN) and LAMOST DR5 stars. 

We adopted the data from the public version of GALAH DR3~\citep{2021MNRAS.506..150B}\footnote{https://www.galah-survey.org/news/announcing-galah-dr3} where stellar abundances are derived using the Spectroscopy Made Easy code~\citep{2017A&A...597A..16P}. For our study we used only stars with SNR~$>20$, $\rm cannon\_flag=0$ and $\rm abundance\_flag <3$ which result in a precision of individual abundances of $\sim0.05$~dex. We used the data from the high-resolution ($R \approx 22500$) spectroscopic survey APOGEE DR~16~\citep{2017AJ....154...94M,2020ApJS..249....3A}, adopting the astroNN catalogue~\citep{2019MNRAS.483.3255L,2019MNRAS.490.4740B}\footnote{https://data.sdss.org/sas/dr16/apogee/vac}, from which we selected the stars with uncertainties $< 0.05$~dex in both \FeH and \aFe. Finally, we adopted the stellar abundance catalog of ~\cite{2019ApJS..245...34X}\footnote{http://dr5.lamost.org/doc/vac} derived from the LAMOST DR5 low-resolution spectra \citep[$R\sim1800$,][]{2015RAA....15.1095L,2017MNRAS.467.1890X} where abundances are measured using The DD–Payne~\citep{2019ApJ...879...69T, 2019ApJS..245...34X}.  Here we selected only stars with recommended labels and  SNR~$>30$ for which the chemical abundances have a typical precision $\sim 0.05- 0.07$~dex. 

For each survey we obtain a cross-matched catalogue of stars with precise \FeH and \aFe from the spectroscopic catalogue and positions and velocities from G3RV2, which we will refer to as 'G3RV2+G.A.L. samples' for Galah, APOGEE and LAMOST. For the stars in common between these spectroscopic samples for both \FeH and \aFe, we observe clear one-to-one relations, however, with some systematic shift and scatter. This is likely due to different spectral resolution and techniques for measuring the stellar parameters which does not allow us to merge the catalogues easily. Therefore in the following sections we look for chemical abundance features in MW with the three samples separately, taking advantage of the large number of stars in each. For additional exploration see also Appendix~\ref{sec::app1}.

\subsection{Chemical abundances in the SNd}\label{sec::Fe_UV}
 Metallicity variations of stars associated with the known \uv moving groups have been reported in a number of previous studies~\citep[see, e.g.,][]{2007ApJ...655L..89B,2009IAUS..254..139W,2014A&A...562A..71B,2014ApJ...787...31Z,2019A&A...631A..47K}. Broadly speaking most of the \uv-substructures do not show chemical homogeneity with the abundance patterns being almost indistinguishable from that of the background field stars of the Galactic disk~\citep[thick or thin, see, e.g.,][]{2004A&A...418..989N, 2006MNRAS.367.1181B,2012MNRAS.425.3188R, 2019A&A...631A..47K,2020A&A...638A.154K}.  Thus dynamical mechanisms must generally be responsible for the \uv substructures, such as resonances of the Galactic bar and/or the Milky Way's spiral arms~\citep[see, e.g.,][]{2000AJ....119..800D, 2017ApJ...840L...2P, 2018RNAAS...2...32M, 2018MNRAS.480.3132Q, 2020ApJ...888...75B}.

In Fig.~\ref{fig::GaiaFeH1} we show the density maps together with the mean metallicity distributions in the \uv plane for SNd stars which are common between G3RV2 and the aforementioned surveys. We note that the samples of stars in each cross-match~(Galah, AstroNN, LAMOST) with G3RV2 is large enough to represent well the \uv density structure which we already discussed in Section~\ref{sec::gaiaUV}~(see Fig.~\ref{fig::GaiaUV}) where several of the main kinematic substructures appear to be still rather prominent. Since the contribution of the thick disk stars is believed to be significant in the SNd region~\citep[see Section 5.1.3 in][]{2016ARA&A..54..529B}, to highlight the metallicity composition of the \uv phase-space we focus on the stars with $\rm [\alpha/Fe]<0.1$. This cut does not fully guarantee a clean thin-disk stars sample, however, it significantly reduces a contamination from chemically-defined thick disk which is likely to have a different formation history~\citep[see, e.g.,][]{2013A&A...560A.109H} and thus different kinematic properties ~\citep[see, e.g.,][]{2011ApJ...738..187L}. 

We firstly see in Fig.~\ref{fig::GaiaFeH1}  a monotonic decrease of the mean metallicity with increasing azimuthal velocity. Since in the local SNd azimuthal velocity is a proxy of the guiding radius~(or angular momentum), this metallicity trend represents the negative radial metallicity gradient in the MW's thin disk along guiding radius~\citep[see, e.g.,][]{2021arXiv210202082K}. Following an inside-out galaxy formation scenario~\citep[e.g.,][]{1997ApJ...477..765C}, stars formed in the inner galaxy have a larger metallicities but also larger radial excursions~(or asymmetric drift) in the SNd, thus mainly contributing in the lower part of the \uv plane~(third row in Fig.~\ref{fig::GaiaFeH1}). On the other hand, stars formed in the outer disk, beyond the Solar radius, have on average lower metallicities but they still can be found in the SNd at azimuthal velocities higher than the LSR value~(third row in Fig.~\ref{fig::GaiaFeH1}). 

\begin{figure}[t!]
\begin{center}
\includegraphics[width=1\hsize]{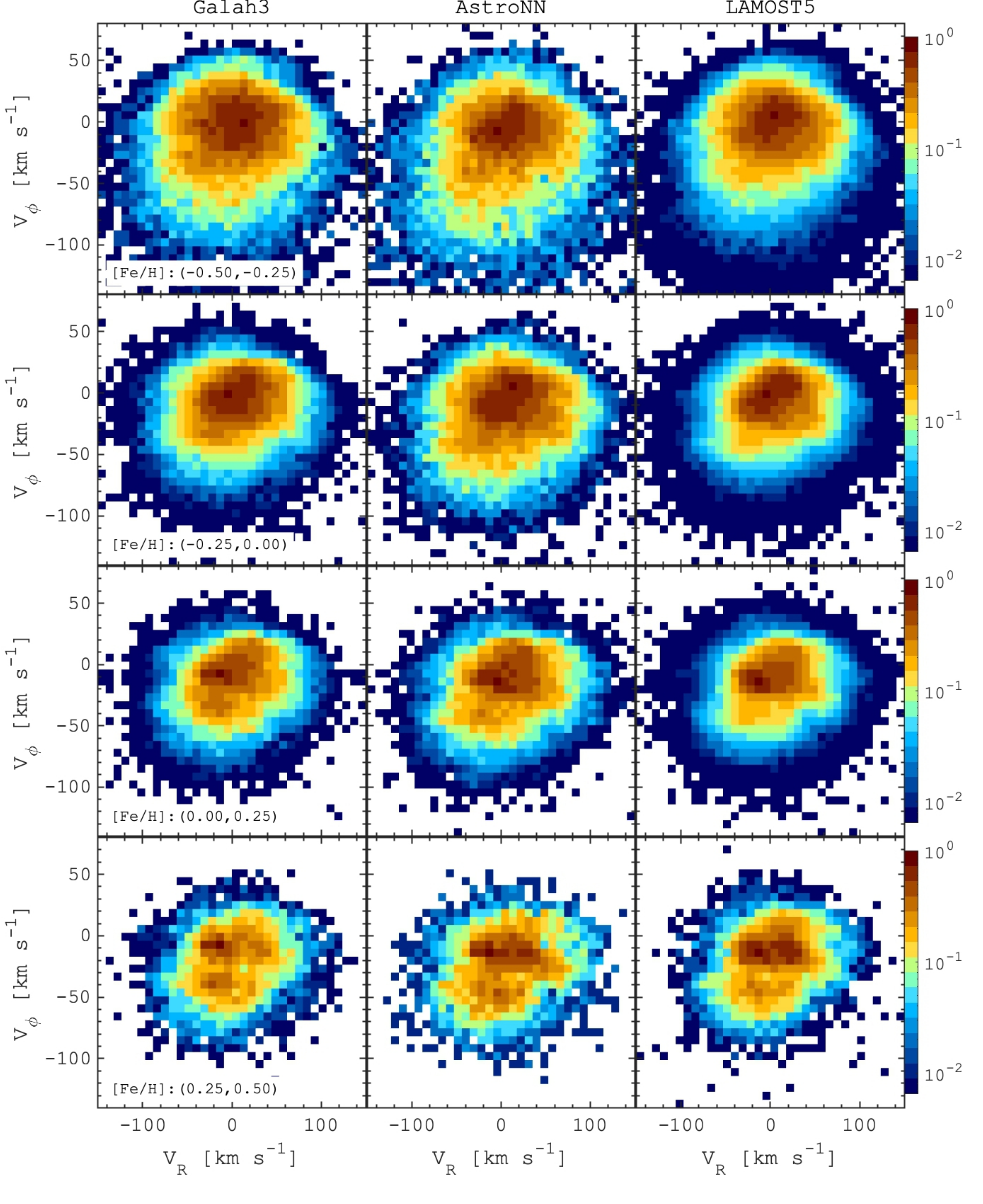}
\caption{Number density of stars in different metallicity bins in the \uv-coordinates for the G3RV2 sample cross-matched with Galah~(left), AstroNN~(center) and LAMOST~(right). }\label{fig::GaiaFeH2}
\end{center}
\end{figure}

Moving back to the mean metallicity maps in Fig.~\ref{fig::GaiaFeH1}, we see a very sharp, curved-shape decrease in the mean metallicity above $v_\phi \approx 20$ km/s. This prominent gradient implies that there is a mechanism limiting the outward motion of the high-\FeH stars beyond $R_g\approx9-10$~kpc. The most natural explanation is the presence of the OLR of the bar which does not allow the stars to pass through thus being a natural barrier between the inner $<9-10$~kpc and the outer disk~\citep{2019A&A...625A.105H}, as it has been suggested in a number of studies of barred galaxy dynamics~\citep{2015A&A...578A..58H,2020A&A...638A.144K,2020ApJ...889...81W}. It is worth mention that the orbits of stars can temporally cross the bar OLR~\citep[see][]{2016MNRAS.461.3835M} but the stars can not change their angular momentum to pass through the OLR from the inner galaxy. The azimuthal velocity ($\propto$ guiding radius in the SNd) of the OLR-caused metallicity pattern we observe correspond to the most recent direct estimates of its pattern speed of $\approx38-41$~\kmskpc~\citep{2019MNRAS.489.3519C,2019MNRAS.tmp.1855S,2019MNRAS.490.4740B}. Another argument in favour of the OLR-impact on the SNd metallicity distribution is based on the distribution function calculations by \cite{2020MNRAS.495..895B} where the MW bar OLR is placed in the \uv diagram in the same region we suggest here~(beyond the Sirius stream) being also in agreement with our previous estimations of large-scale structures location based on the guiding space analysis~\citep{2020A&A...634L...8K}. Once the OLR location is constrained in the \uv space, the bar corotation radius should be placed near the Hercules stream(s)~(which corresponds to $R\approx 6$~kpc from the Galactic center) thus also supporting a number of recent studies~\citep{2017ApJ...840L...2P,2019A&A...632A.107M,2020ApJ...890..117D,2020MNRAS.499.2416A}. 

\begin{figure*}[t!]
\begin{center}
\includegraphics[width=0.85\hsize]{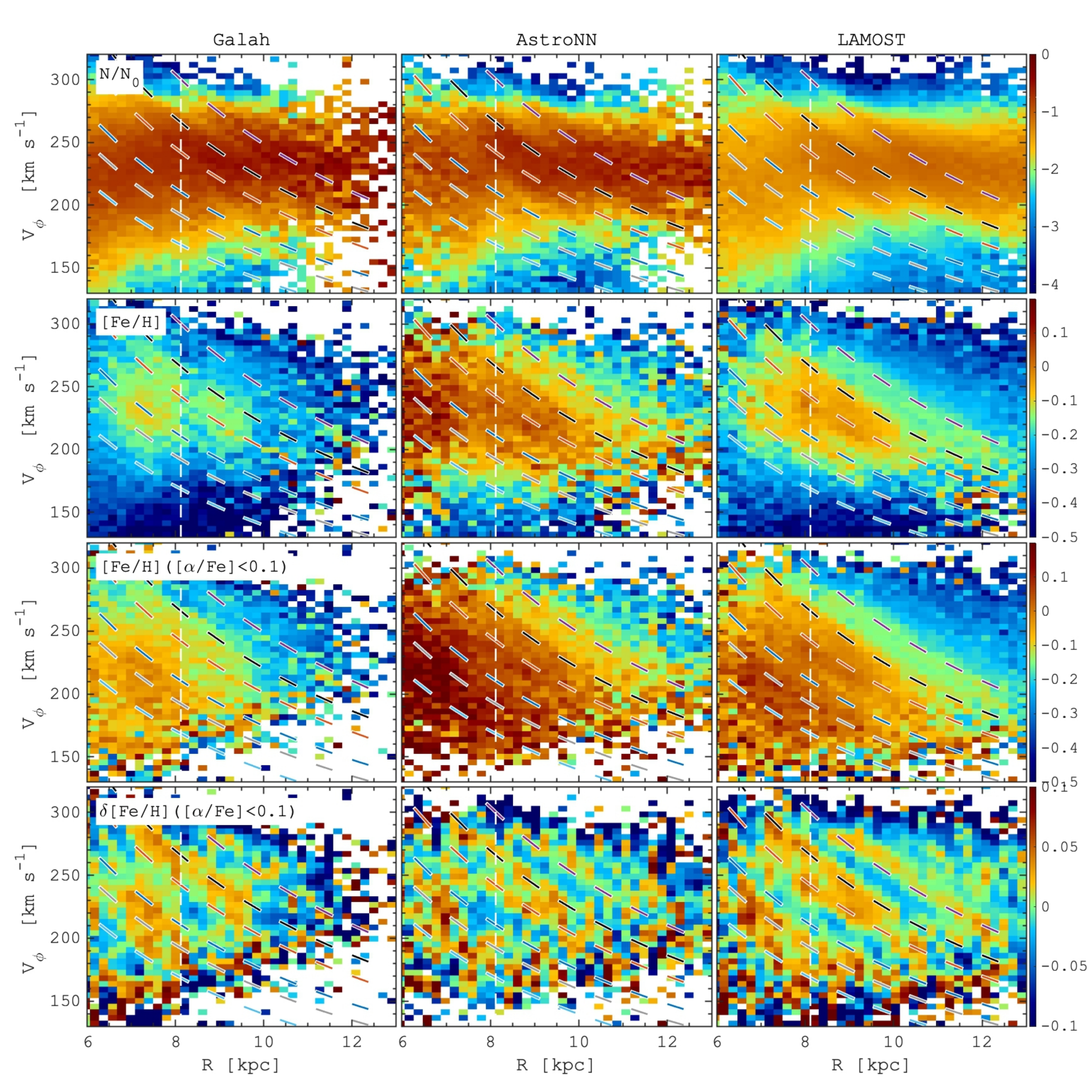}
\caption{Similar to Fig.~\ref{fig::GaiaFeH1} but in \RVphi space for the G3RV2 stars cross-matched with Galah~(left), AstroNN~(center) and LAMOST~(right): density distribution~(first row), colour-coded by the mean \FeH~(second row),  the mean \FeH for thin-disk stars (low-$\alpha$ sequence, third row) and the residual \FeH distribution for low-$\alpha$ stars~(bottom row). Coloured diagonal lines, same as in Fig.~\ref{fig::GaiaRidges}, depict the angular momentum selections in a given azimuthal range used in \XgYg coordinates to identify overdensities in \cite{2020A&A...634L...8K}~(see also Fig.~\ref{fig::GaiaPlane}). The \RVphi density ridges are well-established structures with the higher mean metallicity~($\delta \FeH \approx0.05$~dex). The impact of the OLR is seen as a sharp decrease of the metallicity beyond the black line.}\label{fig::GaiaFeH_Ridges}
\end{center}
\end{figure*}

Next we discuss the metallicity features focusing on the moving groups in the \uv space. In Fig.~\ref{fig::GaiaFeH1} most of the kinematic substructures are visible as groups of stars which tend to have a slightly higher metallicity compare to the local mean value. A typical difference of the \uv-plane features in the median metallicity is of $0.05$~dex which is somewhat similar to the numbers found by~\cite{2017A&A...601A..59A} by using RAVE data. As we discussed above the local moving groups are the pieces of the MW spiral arms and the bar resonances. Therefore the enhancement of the metallicity in the \uv space should correspond to one characterising the spiral arms on larger spatial scales. We return to this problem below once we present the metallicity distributions in other phase-space coordinates~(see Section~\ref{sec::feh_mechanism}).

\begin{figure*}[t!]
\begin{center}
\includegraphics[width=0.85\hsize]{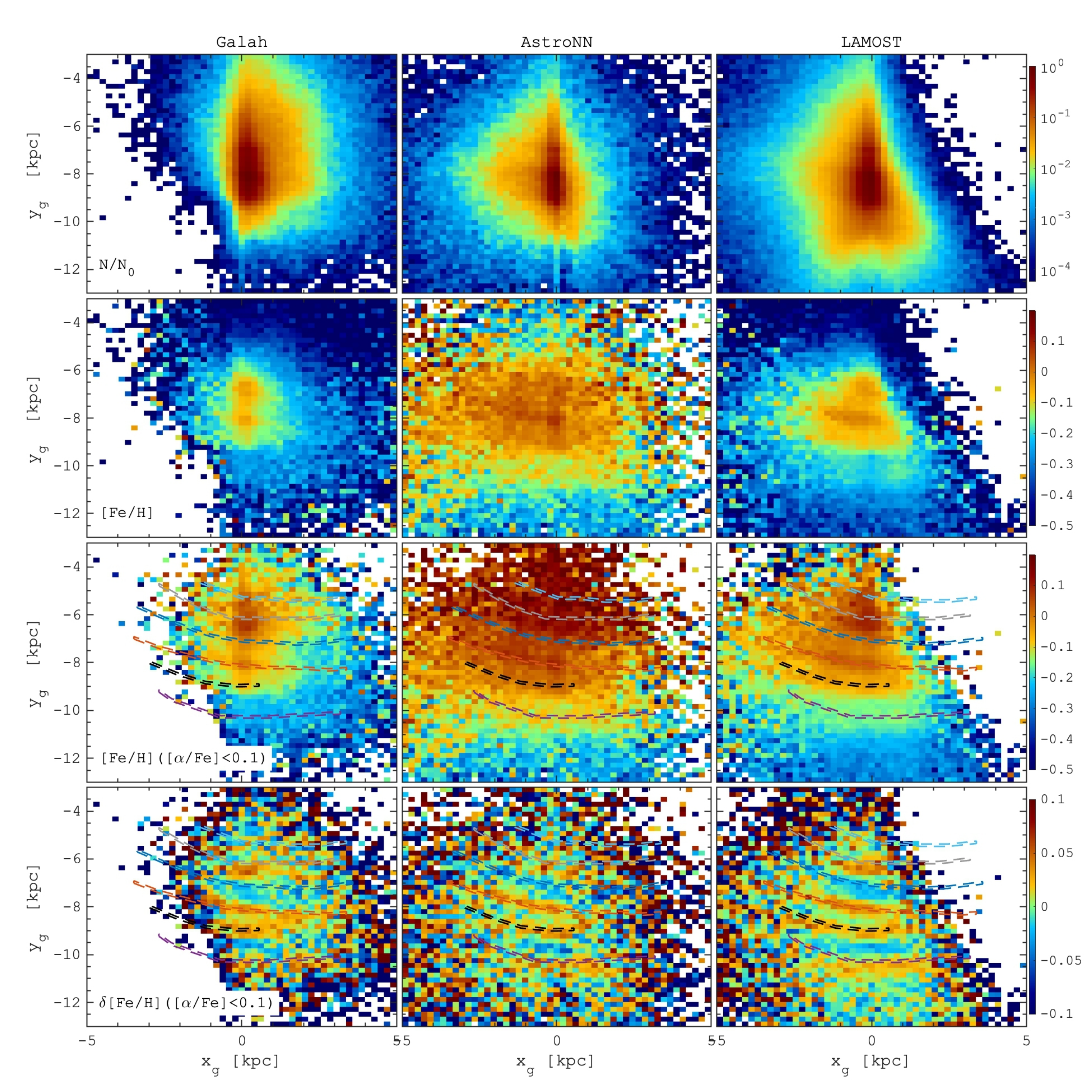}
\caption{Similar to Figs.~\ref{fig::GaiaFeH1} and \ref{fig::GaiaFeH_Ridges} but in the angular momentum~\XgYg space for the G3RV2 stars cross-matched with Galah~(left), AstroNN~(center) and LAMOST~(right): density distribution~(first row), colour-coded by the mean \FeH~(second row),  the mean \FeH for thin-disk stars (low-$\alpha$ sequence, third row) and the residual \FeH distribution for low-$\alpha$ stars~(bottom row). The locations of tightly-wound trailing regions of the higher mean metallicity~($\delta \FeH \approx0.05$~dex) correlate with the stellar overdensities associated with the main MW spiral arms.}\label{fig::GaiaFeH_Spirals}
\end{center}
\end{figure*}

Figure~\ref{fig::GaiaFeH2} shows the density distributions in the \uv plane in several \FeH bins. Variation of the distribution shape depicts a correlation between the stellar metallicity and spread of both radial and azimuthal velocity components~(velocity dispersion). In particular, the distribution is broader for low-\FeH stars while high-\FeH stars tend to have smaller peculiar velocities. Very high-\FeH stars~($>0.25$) show prominent overdensities in the area corresponded to the Hercules, \ple/\hya/\hor with a very sharp decrease of the density above the \sir stream. At the same time the lowest-\FeH distribution, being blurred, still shows asymmetric distribution reflecting the presence of the \her stream and a weak signal for \ple/\hya/\hor stars in this metallicity range. This picture suggests that the most prominent \UV-overdensities show a wide range in metallicity (from $-0.5$ to $+0.5$) implying their chemical homogeneity. An exception here is the \uv-streams with the very high \VV-velocity, above the LSR~(the \ha). This feature demonstrates a lack of high-metallicity stars~(\FeH>0.25) because these phase-space features are likely caused by the low-\FeH outer disk stellar density structures, beyond the OLR of the bar~(\sir stream).

Take another example from the opposite side of the \uv-plane. For a long time, the Arcturus stream~($\VV\sim - 100$~\kmps) has been associated with a relic of a disrupted extragalactic object~\citep{2004ApJ...601L..43N, 2006MNRAS.365.1309H}. Although the density distributions in Fig.~\ref{fig::GaiaFeH2} does not reveal the \arc stream as an overdensity~(but see Fig.~\ref{fig::GaiaUV} for the \Gaia stars) we can note the presence of high-\FeH stars~($0-0.25$~dex) near $\VV\sim - 100$~\kmps which indicates that the group of stars is likely to be a part of the stellar density structure in the inner disk of the MW~\citep[including thick disk stars][]{2009IAUS..254..139W, 2014ApJ...791..135H,2015ApJ...800L..32A}, that, according to our previous analysis~(see Fig.~\ref{fig::GaiaUV}), is associated with the innermost spiral arm~(Scutum-Centaurus), and its extragalactic origin can be ruled out. We believe that the presence of stars in a wide range of \FeH near the \arc stream is in agreement with a more detailed chemical abundance studies~\citep{2011ApJ...743..135R,2012MNRAS.425.3188R,2014A&A...562A..71B} suggesting the dynamical origin of the Arcturus stream being made of a mixture of thin and thick disk stars formed at different galactic radii.

\subsection{Chemical abundance patterns in the \RVphi coordinates}\label{sec::Fe_Rvphi}

Chemical abundance patterns associated with diagonal \RdashVphi-ridges have been presented in a number of studies~\citep{2019MNRAS.489.4962K,2019ApJ...887..193L,2020arXiv200108227W,2020MNRAS.491.2104W,2020ApJ...902...70W} where, for instance, \cite{2019MNRAS.489.4962K} suggested that the ridges are metal-rich because they made of younger stars that lie predominantly in the plane. Numerical simulations also demonstrate that, for instance, the OLR-related \RVphi-ridges contain younger, metal-rich and $\alpha$-poor stars compare to surrounding regions suggesting that this could be due to the fact that colder populations are more affected by the resonances caused by the bar~\citep[][]{2020MNRAS.494.5936F}. Alternatively, such a chemical abundance behaviour near the bar resonances can be explained by the outward migration of stars trapped at the bar resonances due to slowdown of the bar over time~\citep[see more details in][]{2020A&A...638A.144K}. In this case the stellar age distribution in the CR/OLR-ridges can be used to constrain the age of the bar and the epoch its slowdown.

In Fig.~\ref{fig::GaiaFeH_Ridges}, similar to the previous one in the \uv coordinates, we present both density and the mean metallicity maps in \RVphi plane for stars around $10^\circ$ along the line connecting Galactic center and the Sun. First, we note a sharp decrease of the metallicity above the ridge which we associated with the bar OLR and seen well in \uv space above the Sirius stream. Since the motions of thick disk stars are less affected by the structures in the galactic plane we again suggest that the \FeH-patterns of the ridges are more prominent once we remove stars of thick disk~(high-$\alpha$). In Fig.~\ref{fig::GaiaFeH_Ridges}~(third row) we show the mean metallicity maps for the low-$\alpha$~(or chemically-defined thin disk) stars. A large-scale diagonal shape of the metallicity gradient is obviously the result of larger radial excursion of the inner, metal-rich stars contributing to lower \VV-regions at large galactocentric radii. Interestingly, the tilt of the mean metallicity distribution implies that the metallicity gradient is more tightly connected to the angular momentum rather than to galactocentric distance~\citep{2021arXiv210202082K}. Once we extract the \FeH gradient~(bottom row of Fig.~\ref{fig::GaiaFeH_Ridges}) the density ridges appear to show a higher metallicity with an amplitude of $0.05$ dex thus confirming that the \uv metallicity features are the patches of large-scale structures spanning over the entire MW disk.

\begin{figure*}[t!]
\begin{center}
\includegraphics[width=1\hsize]{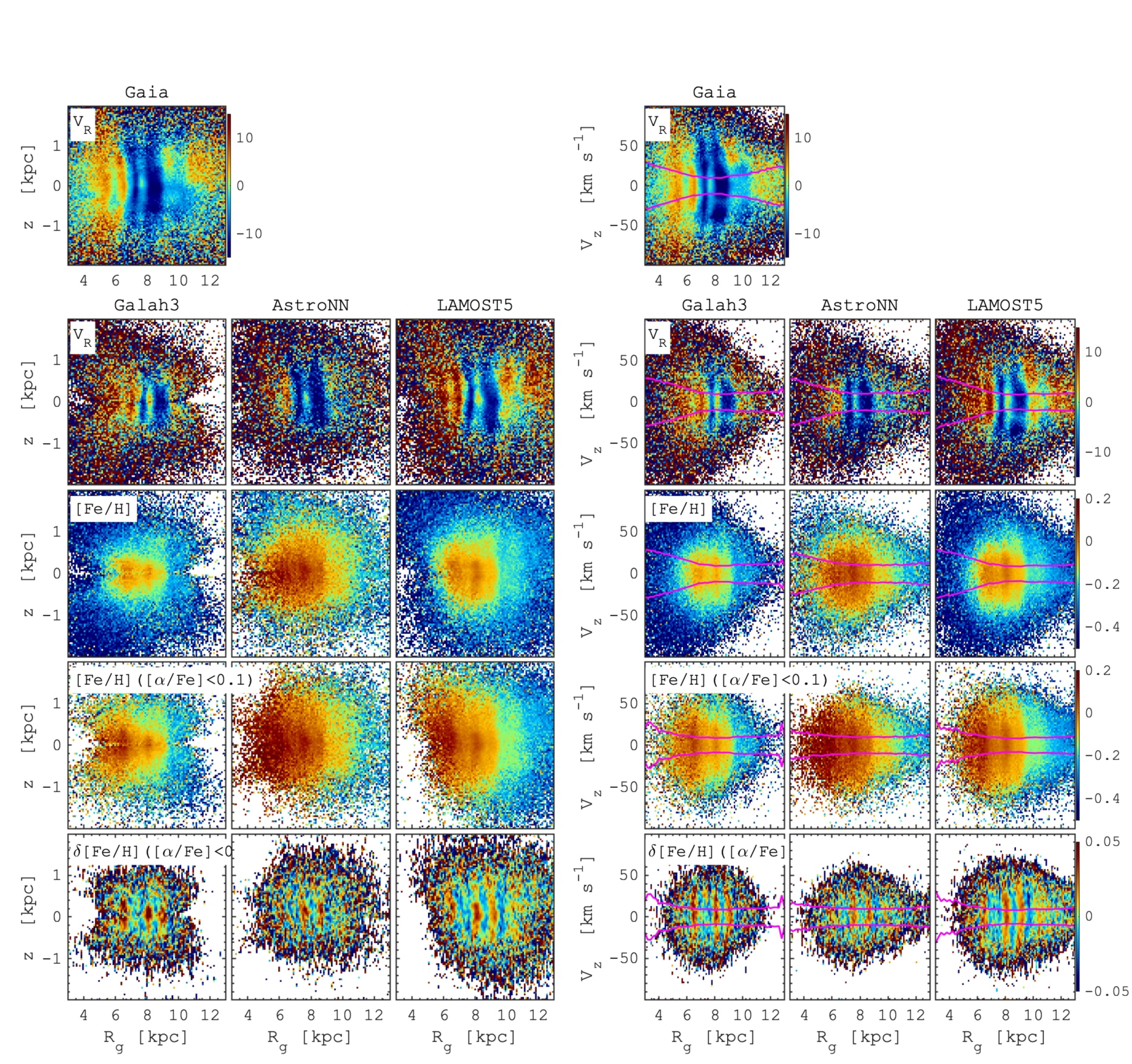}
\caption{Radial velocity and metallicity maps in $R_g-z$~(left) and $R_g-v_z$~(right) coordinates taken in a narrow $\pm 5^\circ$ azimuthal selection centred on the Sun-Galactic center line. The magenta contours in the right panels show 1-$\sigma$ levels of the vertical velocity distribution. Top frames show the mean radial velocity maps for the G3RV2. The following rows represent the mean radial velocity, the mean \FeH, the mean \FeH for $\aFe<0.1$ and the residual~(relative to the extracted 2D gradient) \FeH maps for the G3RV2 stars which are in common with Galah~(left), AstoNN~(center) and LAMOST~(right). The radial velocity wave-like pattern, correlating with the stellar overdensities in \XgYg and \RVphi coordinates~(sees Figs.~\ref{fig::GaiaPlane} and ~\ref{fig::GaiaRidges}), shows a large vertical extension. This suggests that the spiral arms and bar resonances can be traced far from the Galactic plane by stars with large vertical velocities. Similarly, the \FeH pattern detected significantly away from the Galactic plane, consistent with the radial velocity variations.}\label{fig::GaiaFeH_Vertical}
\end{center}
\end{figure*}

\subsection{Chemical abundance patterns in the \XgYg coordinates}

In Fig.~\ref{fig::GaiaFeH_Spirals} we present two-dimensional metallicity variations in the Galactic plane in \XgYg guiding coordinates. Similar to the previous figures of this section we show the mean maps of stellar density, \FeH, \FeH~($\rm [\alpha/Fe]<0.1$) and $\delta \FeH$ for Galah, APOGEE and LAMOST surveys cross-matched with G3RV2 separately. The density maps have different coverage in \XgYg plane due to different surveys footprints. However, all the data sets are large enough and show the existence of trailing angular momentum overdensities, which, as we showed above correspond to the MW spirals and the bar resonances~\citep[see, Fig.~\ref{fig::GaiaPlane}, and in][]{2020A&A...634L...8K}. The mean \FeH maps reveal a tight relation between these overdensities with narrow regions of enhanced~(relative amplitude $\delta\FeH \approx 0.05$) metallicity. Similar to Fig.~\ref{fig::GaiaFeH1} and \ref{fig::GaiaFeH_Ridges} we found a large-scale sharp decrease of the mean metallicity beyond the overdensity associated with the bar OLR which is even more prominent for the low-$\alpha$ stars. The most exciting result here, of course, is the presence of the azimuthally-extended metallicity-enhanced structures -- spiral arms implying that the chemical abundance peculiarities we discussed above is the large-scale phenomena which can be only caused by global mechanisms reshaping stellar populations across the entire MW disk.

\subsection{Chemical abundance patterns across the disk plane in $R_g-z$ and $R_g-v_z$ coordinates}

The last piece of our analysis involves the exploration of chemo-kinematic features perpendicular to the Galactic plane. Various velocity patterns perpendicular to the Galactic plane have been intensively discussed in a number of previous observational studies implying the importance of various factors~\citep[see, e.g.,][]{2013MNRAS.436..101W,2014MNRAS.440.1971W,2018A&A...616A..11G,2019MNRAS.490..797C}, including the impact of the spiral arms~\citep{2014MNRAS.440.2564F,2016MNRAS.457.2569M}. To link our previous analysis with the vertical manifestation of spirals in Fig.~\ref{fig::GaiaFeH_Vertical} we present the radial velocity and metallicity maps in $R_g-z$~(left) and $R_g-v_z$~(right) coordinates for the stars taken in a narrow azimuthal selection~(same as for \RVphi analysis). 

In Fig.~\ref{fig::GaiaFeH_Vertical} we see a prominent wave-like radial velocity waves along $R_g$, as we already described in Fig.~\ref{fig::GaiaPlane}, where the negative radial motions roughly correspond to the spiral arms location~\citep[see also][]{2012MNRAS.425.2335S,2014MNRAS.440.2564F,2016A&A...589A..13A}. The striking feature of such a pattern is its vertical extension. In particular, G3RV2 stars show the wave-like pattern out to $z\approx 2$~kpc and $v_z\approx 75$~\kmps. These numbers are slightly lower for the G3RV2+G.A.L. stars likely due to lower statistics. The vertical extension of the radial velocity pattern suggests that the spiral arms, in fact, contain relatively (dynamically) old stars with significant vertical excursion because they, being formed in a thin gaseous disk, already experienced some vertical heating. Therefore, the data suggest that the spiral arms of the MW can not be interpreted as a newly formed structure, visible only among young stars.

The vertical structure of the mean \FeH distribution shows a less prominent wave-like pattern compared to the radial velocity maps. However, once we removed the background distribution~(radial gradient), the residual distributions depict noticeable \FeH variations following the radial velocity pattern. The enhancement of the mean metallicity is close to the negative radial velocity regions, thus being associated with the spiral arms, similar to our analysis of \RVphi and \uv planes in previous sections. The amplitude of the metallicity variations is in the range of $0.05$~dex as in other phase-space coordinates. These variations can be traced up to $0.8-1$~kpc which is slightly lower compared to the ones for the radial velocity; however, we believe that this is likely the result of lower statistics and relatively large chemical abundance errors compared to the amplitude of the systematic \FeH pattern. Therefore, these results rule out the scenario proposed by \cite{2019MNRAS.489.4962K} to explain the \RVphi ridges by young stars only, because the main \RVphi-ridges correspond to the spiral arms which, as we see, can be traced far away from the Galactic plane. 

\subsection{On the mechanism of the \FeH patterns formation in the vicinity of spiral arms}\label{sec::feh_mechanism}

To summarize the results of this section, our analysis of the chemical abundances distribution demonstrates the existence of multiple large-scale stellar density structures with enhanced metallicity by $0.05$~dex. These \FeH-patterns follow the spiral arms derived as angular momentum overdensities traceable in the azimuthal direction across the entire G3RV2 footprint, in the radial direction in the form of the \RVphi ridges, as ``local'' \uv moving groups in the SNd and finally up to $0.8-1$~kpc away from the Galactic plane. Several different mechanisms can be involved in order to explain the origin of the chemical abundance patterns presented in Figs~\ref{fig::GaiaFeH1}-\ref{fig::GaiaFeH_Vertical}, such as {\it(i)} radial migration, {\it(ii)} a local enrichment and {\it (iii)} kinematic fractionation\footnote{originally proposed for the stellar populations separation in the presence of a bar in \cite{2017MNRAS.469.1587D}}. 

{\it (i)} In the presence of a radial metallicity gradient stellar radial migration results in the azimuthal variation in the residual metallicity characterized by a metal-rich trailing edge and a metal-poor leading edge of the spirals~\citep{2016MNRAS.460L..94G}. In this case the \FeH-pattern we discover would be the perfect manifestation of ongoing radial migration in the MW. However the radial migration is the most effective for stars on nearly-circular orbits in the vicinity of the corotation~\citep{2002MNRAS.336..785S} which for the slowly-rotating MW spiral arms~\citep[see][for the review]{2016ARA&A..54..529B} is located beyond the Solar radius. Therefore, ongoing stellar migration can not be responsible for the observed patterns in the inner disk. Note that some  $N$-body models suggest that rather open~(large pitch angles) spiral arms could be co-rotating structures where the corotation is located everywhere along the arms~\citep[see, e.g.,][]{2015MNRAS.453.1867G}. However, the MW spiral arms are usually assumed to be rigidly-rotating with a certain pattern speed, and there are no yet direct evidences in favour of the co-rotating nature of the MW spirals. Finally, there are no theoretical predictions concerning the vertical abundance patterns near the spirals caused by the radial migration.

{\it (ii)} Another process potentially responsible for the metallicity enhancement in spirals is discussed by ~\cite{2019A&A...628A..38S} where 2D chemical evolution models predict a local ISM enrichment leading to a faster chemical evolution~(and thus higher metallicity of stars) near the co-rotation radius. In this case, to explain a higher metallicity of the spirals in the MW, we again need to assume a co-rotating nature of the spirals, which should resemble their structure for a long-time~(enough to increase the metallicity of the ISM locally) which, however, contradicts the $N$-body simulations where spiral arms rapidly change the structure.

{\it (iii)} Finally, it is worth mentioning a mechanism independent of the nature of spiral arms, implying an enhancement of the mean metallicity near the spirals in the presence of multiple populations with different chemo-kinematical properties. In particular, assuming a decrease of the stellar velocity dispersion with metallicity (assuming realistic age-metallicity and age-velocity dispersion relations), it has been shown that a differential kinematical response of these stellar populations on the spiral arm gravitational potential leads to the azimuthal variations of the mean metallicity across the spiral arms~\citep{2018A&A...611L...2K}. 

The understanding of mechanisms driving the observed metallicity pattern in the vicinity of the spiral arms is yet beyond the scope of our work. We believe that more detailed chemical and dynamical models of the MW-type spiral galaxies are needed.

\section{Summary and Conclusions}\label{sec::concl}

We first studied a new high-resolution $N$-body simulation of a MW-like disk galaxy during the growth phase of a multi-arm, tightly wound spiral pattern, with the goal of investigating the phase-space structure that such a pattern imprints on the model's stellar disk. By shifting stars to their angular momentum-equivalent guiding radii $R_g$, much of the epicyclic blurring is removed, and a sharper view of the galaxy's substructure can be obtained~\citep[see, also, ][]{2020A&A...634L...8K} which is morphologically similar but not identical to the substructure in the density map (Figs.~\ref{fig::model_density_maps},\ref{fig::model_spirals_selection}). We showed that in this simulation the spiral structure can be recovered by analysing the disk stars in the angular momentum space. In the analysed snapshot, we demonstrated the following:
\begin{itemize}
    \item The spiral arms have amplitudes $10-15\%$. They are traced by stars located near the azimuthally extended overdensities in guiding (angular momentum) space~(see Fig.~\ref{fig::model_spirals_selection}), in particular by dynamically cold stars with small $R-R_g$.
    \item The guiding space overdensities both let us identify the peak densities of the real-space spiral arms, and they also contain a fraction of stars that travel far into the disk, leading to phase-space overdensities in disk regions kpcs away from the spiral arm with the same angular momentum.  
    \item In a region with a few kpc size and a few 10s of degrees angular extent such as the region surveyed by DR3V2, these overdensities appear as a more nearly vertically extended~(along azimuth), wave-like pattern in the $(\phi,L_z)$-plane~(Fig.~\ref{fig::model_guiding_maps}). 
    \item In the \RVphi-plane, these angular momentum overdensities give rise to radially extended~($3-5$~kpc) diagonal ridges with outwardly decreasing ${\rm v}_\phi$~(Figs.~\ref{fig::model_ridges_uv}). 
    \item At a typical SNd-like region, the ridges are visible as moving group-like overdensities in local velocity space with similar $10-15\%$ amplitudes~(Fig.~\ref{fig::model_ridges_uv}).     
\end{itemize}
In this simulation with a tightly-wound multi-arm spiral structure, the main angular momentum overdensities correspond to individual spiral arms. However, this seems not to be universal \citep{2020MNRAS.497..818H}; depending on the model and the nature of the spiral structure~(long-lived, transient, tidally-induced, grand design etc.), one may expect that a single spiral may correspond to several angular momentum overdensities. 

We chose this simulation because the patterns we observe in the $N$-body model are in many respects similar to the phase-space structure observed in the nearby Galactic disk by \Gaia. Therefore, we apply similar approaches to the Milky Way data. In our analysis of  \Gaia~(DR2+EDR3 we find:
\begin{itemize}
    \item The angular momentum overdensities in DR3V2 data are seen independently within and outside $1$~kpc from the Sun, just as in the homogenized \Gaia data subset in our earlier study~\citep{2020A&A...634L...8K}. They are large-scale, not local features in the surveyed Galactic disk region~(Fig.~\ref{fig::GaiaPlane}).
    
    \item Most of the \Gaia stars in the angular momentum overdensities are found close to their guiding radii~(e.g., $\approx 50\%$ within $|R-R_g|<0.4$~kpc, $95\%$ within $2$~kpc, see Fig.~\ref{fig::r_rg}). For several angular momentum overdensities, the stars within $250$~pc of their guiding radii are found in the Galactic plane in locations coinciding with the high-mass star formation regions from~\cite{2019ApJ...885..131R} - i.e. the Sagittarius, Local, and Perseus spiral arms~(see Fig.~\ref{fig::Gaia_vs_masers}). Just as in the $N$-body model, spiral overdensities can be identified as overdensities in the ``dynamically cold'' stars. The match does not work for the Scutum-Centaurus arm near the end of the bar.  
    
    \item Stars in the angular momentum overdensities follow the known \RVphi density ridges. When these radially-extended ridges pass through the SNd region, they result in \uv-velocity plane overdensities that match the well-known main moving groups~(see Fig.~\ref{fig::GaiaRidges},\ref{fig::GaiaUV}). In this interpretation, from low to high $V_\phi$, the Arcturus stream overlaps with low-velocity stars from the Scutum-Centaurus arm overdensity, part of Hercules corresponds to bar corotation, the top part overlaps with the Sagittarius arm overdensity, the Sirius stream lies on the bar OLR, and the high-velocity Hat on the Perseus arm overdensity~(see Figs.~\ref{fig::GaiaUV} and \citet{2020A&A...634L...8K} for the bar resonances).
\end{itemize}

 The overlap of the ``local'' kinematic features with the main spiral arm angular momentum overdensities suggests their genetic connection, however, alternative explanations can not be ruled out especially for weaker \uv-features that can not be yet associated with spirals and may represent higher-order bar resonances~\citep{2019A&A...626A..41M}.

Using additional data from the Galah, APOGEE, and LAMOST surveys, we finally analyzed the trends of mean metallicity across these structures.
\begin{itemize}
    \item Using low-$\alpha$ stars we find a uniform metallicity gradient in the \uv-plane from low- to high-${\rm v}_\phi$, and a corresponding gradient across the \RVphi-plane and in guiding space, reflecting the outward Galactic gradient~(Fig.~\ref{fig::GaiaFeH1}). The supersolar metallicity stars born in the inner disk contribute more strongly at low ${\rm v}_\phi$, while metal-poorer stars dominate at high ${\rm v}_\phi$.
    
    \item We find a sharp, curved-shape decrease in the mean metallicity beyond $R_g=9-10$~kpc which is naturally interpreted as the bar OLR barrier to the outer disk, as suggested by~\cite{2015A&A...578A..58H,2020A&A...638A.144K}~(see Figs.~\ref{fig::GaiaFeH1},~\ref{fig::GaiaFeH_Ridges} and \ref{fig::GaiaFeH_Spirals}). Although this OLR feature seems to be rather prominent, the impact of other high-order bar resonances on the chemical abundance variations in the disk is not fully explored in the literature. Using $N$-body models \cite{2021arXiv210505263W} recently predicted a weak \FeH signature of the main $4:1$ resonance, whose manifestation in the MW requires further investigation. 

    \item Stars of the main moving groups are seen to have slightly but systematically enhanced mean metallicity (by $\approx0.05$~dex) relative to the nearby average value~(Fig.~\ref{fig::GaiaFeH1}); this is also visible in the guiding~(angular momentum) space overdensities and \RVphi ridges~(Figs.~\ref{fig::GaiaFeH_Ridges}, \ref{fig::GaiaFeH_Spirals}). 

    \item We finally find noticeable periodic \FeH variations as a function of the angular momentum which can be traced up to $|z| \approx 0.8-1$ kpc and $|v_z| \approx 50$~\kmps (Fig.~\ref{fig::GaiaFeH_Vertical}). This suggests that the angular momentum overdensities also include stars that experienced substantial dynamical heating.

\end{itemize}
The enhancement of the metallicity near \uv-features and \RVphi-ridges provides additional arguments in favour of the link between the guiding space overdensities and the MW spiral arms, because such enhancements were predicted in a number of previous chemodynamical models~(see Sec.~\ref{sec::feh_mechanism}). 

In conclusion, these results consistently point to the interpretation that the most prominent phase-space features seen in the \Gaia data in SNd \uv space, the \RVphi plane, and the angular momentum diagram are linked to the large-scale MW spiral arms and main bar resonances. Future \Gaia data releases will have increased spatial coverage of the MW disk, making it possible to map the angular momentum structures at larger distances from the Sun and further test this interpretation. The physical parameters of the stars in the overdensities obtained by future spectroscopic surveys~(4MOST, WEAVE, MOONS) will allow us to better constrain the mechanisms for the discovered metallicity enhancements and in this way shed light on the true nature of the spiral structure in the MW.

\begin{acknowledgements}
 The authors thank the anonymous referee for a constructive report. They also wish to thank Jason Hunt for stimulating this work. SK thanks Dmitrii Byziaev, Jonny Clarke and Shola Wylie for their assistance with access to the APOGEE data; Evgeny Polyachenko for the help with the initial condition generation for the $N$-body model; Paola Di Matteo for the access to data storage facilities of the GEPI~(Paris Observatory); David Katz and Misha Haywood for several suggestions about the \Gaia and APOGEE data treatment respectively. SK also thanks Ivan Minchev for several helpful comments on earlier versions of the manuscript. SK acknowledges the Russian Science Foundation (RSCF) grant 19-72-20089 for support in the preparation of the $N$-body model partially carried by using the equipment of the shared research facilities of HPC computing resources at Lomonosov Moscow State University (project RFMEFI62117X0011). \newline
This work has made use of data from the European Space Agency (ESA) mission \Gaia (\url{https://www.cosmos.esa.int/gaia}), processed by the \Gaia Data Processing and Analysis Consortium (DPAC, \url{https://www.cosmos.esa.int/web/gaia/dpac/consortium}). Funding for the DPAC has been provided by national institutions, in particular the institutions participating in the \Gaia Multilateral Agreement. 
\newline
Funding for the Sloan Digital Sky Survey IV has been provided by the Alfred P. Sloan Foundation, the U.S. Department of Energy Office of Science, and the Participating Institutions. 
\newline
Guoshoujing Telescope (the Large Sky Area Multi-Object Fiber Spectroscopic Telescope LAMOST) is a National Major Scientific Project built by the Chinese Academy of Sciences. Funding for the project has been provided by the National Development and Reform Commission. LAMOST is operated and managed by the National Astronomical Observatories, Chinese Academy of Sciences.
\newline
{\it Software:} \texttt{IPython} \citep{2007CSE.....9c..21P}, \texttt{Astropy} \citep{2013A&A...558A..33A, 2018AJ....156..123A}, \texttt{NumPy} \citep{2011CSE....13b..22V}, \texttt{SciPy} \citep{2020SciPy-NMeth}, \texttt{AGAMA} \citep{2019MNRAS.482.1525V}, \texttt{Matplotlib} \citep{2007CSE.....9...90H}, \texttt{Pandas} \citep{mckinney-proc-scipy-2010}, TOPCAT~\citep{2005ASPC..347...29T}.
\end{acknowledgements}

\bibliographystyle{aa}
\bibliography{references-1}

\begin{thebibliography}{168}
\expandafter\ifx\csname natexlab\endcsname\relax\def\natexlab#1{#1}\fi

\bibitem[{{Ahumada} {et~al.}(2020){Ahumada}, {Prieto}, {Almeida}, {Anders},
  {Anderson}, {Andrews}, {Anguiano}, {Arcodia}, {Armengaud}, {Aubert}, {Avila},
  {Avila-Reese}, {Badenes}, {Balland }, {Barger}, {Barrera-Ballesteros},
  {Basu}, {Bautista}, {Beaton}, {Beers}, {Benavides}, {Bender}, {Bernardi},
  {Bershady}, {Beutler}, {Bidin}, {Bird}, {Bizyaev}, {Blanc}, {Blanton},
  {Boquien}, {Borissova}, {Bovy}, {Brand t}, {Brinkmann}, {Brownstein},
  {Bundy}, {Bureau}, {Burgasser}, {Burtin}, {Cano-D{\'\i}az}, {Capasso},
  {Cappellari}, {Carrera}, {Chabanier}, {Chaplin}, {Chapman}, {Cherinka},
  {Chiappini}, {Doohyun Choi}, {Chojnowski}, {Chung}, {Clerc}, {Coffey},
  {Comerford}, {Comparat}, {da Costa}, {Cousinou}, {Covey}, {Crane}, {Cunha},
  {Ilha}, {Dai}, {Damsted}, {Darling}, {Davidson}, {Davies}, {Dawson}, {De},
  {de la Macorra}, {De Lee}, {Queiroz}, {Deconto Machado}, {de la Torre},
  {Dell'Agli}, {du Mas des Bourboux}, {Diamond-Stanic}, {Dillon}, {Donor},
  {Drory}, {Duckworth}, {Dwelly}, {Ebelke}, {Eftekharzadeh}, {Davis Eigenbrot},
  {Elsworth}, {Eracleous}, {Erfanianfar}, {Escoffier}, {Fan}, {Farr},
  {Fern{\'a}ndez-Trincado}, {Feuillet}, {Finoguenov}, {Fofie},
  {Fraser-McKelvie}, {Frinchaboy}, {Fromenteau}, {Fu}, {Galbany}, {Garcia},
  {Garc{\'\i}a-Hern{\'a}ndez}, {Oehmichen}, {Ge}, {Maia}, {Geisler}, {Gelfand
  }, {Goddy}, {Gonzalez-Perez}, {Grabowski}, {Green}, {Grier}, {Guo}, {Guy},
  {Harding}, {Hasselquist}, {Hawken}, {Hayes}, {Hearty}, {Hekker}, {Hogg},
  {Holtzman}, {Horta}, {Hou}, {Hsieh}, {Huber}, {Hunt}, {Chitham}, {Imig},
  {Jaber}, {Angel}, {Johnson}, {Jones}, {J{\"o}nsson}, {Jullo}, {Kim},
  {Kinemuchi}, {Kirkpatrick}, {Kite}, {Klaene}, {Kneib}, {Kollmeier}, {Kong},
  {Kounkel}, {Krishnarao}, {Lacerna}, {Lan}, {Lane}, {Law}, {Le Goff}, {Leung},
  {Lewis}, {Li}, {Lian}, {Lin}, {Long}, {Longa-Pe{\~n}a}, {Lundgren}, {Lyke},
  {Ted Mackereth}, {MacLeod}, {Majewski}, {Manchado}, {Maraston}, {Martini},
  {Masseron}, {Masters}, {Mathur}, {McDermid}, {Merloni}, {Merrifield},
  {M{\'e}sz{\'a}ros}, {Miglio}, {Minniti}, {Minsley}, {Miyaji}, {Mohammad},
  {Mosser}, {Mueller}, {Muna}, {Mu{\~n}oz-Guti{\'e}rrez}, {Myers}, {Nadathur},
  {Nair}, {Nandra}, {do Nascimento}, {Nevin}, {Newman}, {Nidever}, {Nitschelm},
  {Noterdaeme}, {O'Connell}, {Olmstead}, {Oravetz}, {Oravetz}, {Osorio},
  {Pace}, {Padilla}, {Palanque-Delabrouille}, {Palicio}, {Pan}, {Pan},
  {Parker}, {Paviot}, {Peirani}, {Ram{\'r}ez}, {Penny}, {Percival},
  {Perez-Fournon}, {P{\'e}rez-R{\`a}fols}, {Petitjean}, {Pieri},
  {Pinsonneault}, {Poovelil}, {Povick}, {Prakash}, {Price-Whelan}, {Raddick},
  {Raichoor}, {Ray}, {Rembold}, {Rezaie}, {Riffel}, {Riffel}, {Rix}, {Robin},
  {Roman-Lopes}, {Rom{\'a}n-Z{\'u}{\~n}iga}, {Rose}, {Ross}, {Rossi}, {Rowland
  s}, {Rubin}, {Salvato}, {S{\'a}nchez}, {S{\'a}nchez-Menguiano},
  {S{\'a}nchez-Gallego}, {Sayres}, {Schaefer}, {Schiavon}, {Schimoia},
  {Schlafly}, {Schlegel}, {Schneider}, {Schultheis}, {Schwope}, {Seo},
  {Serenelli}, {Shafieloo}, {Shamsi}, {Shao}, {Shen}, {Shetrone}, {Shirley},
  {Aguirre}, {Simon}, {Skrutskie}, {Slosar}, {Smethurst}, {Sobeck}, {Sodi},
  {Souto}, {Stark}, {Stassun}, {Steinmetz}, {Stello}, {Stermer},
  {Storchi-Bergmann}, {Streblyanska}, {Stringfellow}, {Stutz}, {Su{\'a}rez},
  {Sun}, {Taghizadeh-Popp}, {Talbot}, {Tayar}, {Thakar}, {Theriault}, {Thomas},
  {Thomas}, {Tinker}, {Tojeiro}, {Toledo}, {Tremonti}, {Troup}, {Tuttle},
  {Unda-Sanzana}, {Valentini}, {Vargas-Gonz{\'a}lez}, {Vargas-Maga{\~n}a},
  {V{\'a}zquez-Mata}, {Vivek}, {Wake}, {Wang}, {Weaver}, {Weijmans}, {Wild},
  {Wilson}, {Wilson}, {Wolthuis}, {Wood-Vasey}, {Yan}, {Yang}, {Y{\`e}che},
  {Zamora}, {Zarrouk}, {Zasowski}, {Zhang}, {Zhao}, {Zhao}, {Zheng}, {Zheng},
  {Zhu}, \& {Zou}}]{2020ApJS..249....3A}
{Ahumada}, R., {Prieto}, C.~A., {Almeida}, A., {et~al.} 2020, \apjs, 249, 3

\bibitem[{{Antoja} {et~al.}(2008){Antoja}, {Figueras}, {Fern{\'a}ndez}, \&
  {Torra}}]{2008A&A...490..135A}
{Antoja}, T., {Figueras}, F., {Fern{\'a}ndez}, D., \& {Torra}, J. 2008, \aap,
  490, 135

\bibitem[{{Antoja} {et~al.}(2011){Antoja}, {Figueras}, {Romero-G{\'o}mez},
  {Pichardo}, {Valenzuela}, \& {Moreno}}]{2011MNRAS.418.1423A}
{Antoja}, T., {Figueras}, F., {Romero-G{\'o}mez}, M., {et~al.} 2011, \mnras,
  418, 1423

\bibitem[{{Antoja} {et~al.}(2014){Antoja}, {Helmi}, {Dehnen}, {Bienaym{\'e}},
  {Bland-Hawthorn}, {Famaey}, {Freeman}, {Gibson}, {Gilmore}, {Grebel},
  {Kordopatis}, {Kunder}, {Minchev}, {Munari}, {Navarro}, {Parker}, {Reid},
  {Seabroke}, {Siebert}, {Steinmetz}, {Watson}, {Wyse}, \&
  {Zwitter}}]{2014A&A...563A..60A}
{Antoja}, T., {Helmi}, A., {Dehnen}, W., {et~al.} 2014, \aap, 563, A60

\bibitem[{{Antoja} {et~al.}(2018){Antoja}, {Helmi}, {Romero-G{\'o}mez}, {Katz},
  {Babusiaux}, {Drimmel}, {Evans}, {Figueras}, {Poggio}, {Reyl{\'e}}, {Robin},
  {Seabroke}, \& {Soubiran}}]{2018Natur.561..360A}
{Antoja}, T., {Helmi}, A., {Romero-G{\'o}mez}, M., {et~al.} 2018, \nat, 561,
  360

\bibitem[{{Antoja} {et~al.}(2017){Antoja}, {Kordopatis}, {Helmi}, {Monari},
  {Famaey}, {Wyse}, {Grebel}, {Steinmetz}, {Bland-Hawthorn}, {Gibson},
  {Bienaym{\'e}}, {Navarro}, {Parker}, {Reid}, {Seabroke}, {Siebert},
  {Siviero}, \& {Zwitter}}]{2017A&A...601A..59A}
{Antoja}, T., {Kordopatis}, G., {Helmi}, A., {et~al.} 2017, \aap, 601, A59

\bibitem[{{Antoja} {et~al.}(2015){Antoja}, {Monari}, {Helmi}, {Bienaym{\'e}},
  {Bland-Hawthorn}, {Famaey}, {Gibson}, {Grebel}, {Kordopatis}, {Munari},
  {Navarro}, {Parker}, {Reid}, {Seabroke}, {Steinmetz}, \&
  {Zwitter}}]{2015ApJ...800L..32A}
{Antoja}, T., {Monari}, G., {Helmi}, A., {et~al.} 2015, \apjl, 800, L32

\bibitem[{{Antoja} {et~al.}(2016){Antoja}, {Roca-F{\`a}brega}, {de Bruijne}, \&
  {Prusti}}]{2016A&A...589A..13A}
{Antoja}, T., {Roca-F{\`a}brega}, S., {de Bruijne}, J., \& {Prusti}, T. 2016,
  \aap, 589, A13

\bibitem[{{Antoja} {et~al.}(2009){Antoja}, {Valenzuela}, {Pichardo}, {Moreno},
  {Figueras}, \& {Fern{\'a}ndez}}]{2009ApJ...700L..78A}
{Antoja}, T., {Valenzuela}, O., {Pichardo}, B., {et~al.} 2009, \apjl, 700, L78

\bibitem[{{Arifyanto} \& {Fuchs}(2006)}]{2006A&A...449..533A}
{Arifyanto}, M.~I. \& {Fuchs}, B. 2006, \aap, 449, 533

\bibitem[{{Asano} {et~al.}(2020){Asano}, {Fujii}, {Baba}, {B{\'e}dorf},
  {Sellentin}, \& {Portegies Zwart}}]{2020MNRAS.499.2416A}
{Asano}, T., {Fujii}, M.~S., {Baba}, J., {et~al.} 2020, \mnras, 499, 2416

\bibitem[{{Astropy Collaboration} {et~al.}(2018){Astropy Collaboration},
  {Price-Whelan}, {Sip{\H{o}}cz}, {G{\"u}nther}, {Lim}, {Crawford}, {Conseil},
  {Shupe}, {Craig}, {Dencheva}, {Ginsburg}, {Vand erPlas}, {Bradley},
  {P{\'e}rez-Su{\'a}rez}, {de Val-Borro}, {Aldcroft}, {Cruz}, {Robitaille},
  {Tollerud}, {Ardelean}, {Babej}, {Bach}, {Bachetti}, {Bakanov}, {Bamford},
  {Barentsen}, {Barmby}, {Baumbach}, {Berry}, {Biscani}, {Boquien}, {Bostroem},
  {Bouma}, {Brammer}, {Bray}, {Breytenbach}, {Buddelmeijer}, {Burke},
  {Calderone}, {Cano Rodr{\'\i}guez}, {Cara}, {Cardoso}, {Cheedella}, {Copin},
  {Corrales}, {Crichton}, {D'Avella}, {Deil}, {Depagne}, {Dietrich}, {Donath},
  {Droettboom}, {Earl}, {Erben}, {Fabbro}, {Ferreira}, {Finethy}, {Fox},
  {Garrison}, {Gibbons}, {Goldstein}, {Gommers}, {Greco}, {Greenfield},
  {Groener}, {Grollier}, {Hagen}, {Hirst}, {Homeier}, {Horton}, {Hosseinzadeh},
  {Hu}, {Hunkeler}, {Ivezi{\'c}}, {Jain}, {Jenness}, {Kanarek}, {Kendrew},
  {Kern}, {Kerzendorf}, {Khvalko}, {King}, {Kirkby}, {Kulkarni}, {Kumar},
  {Lee}, {Lenz}, {Littlefair}, {Ma}, {Macleod}, {Mastropietro}, {McCully},
  {Montagnac}, {Morris}, {Mueller}, {Mumford}, {Muna}, {Murphy}, {Nelson},
  {Nguyen}, {Ninan}, {N{\"o}the}, {Ogaz}, {Oh}, {Parejko}, {Parley}, {Pascual},
  {Patil}, {Patil}, {Plunkett}, {Prochaska}, {Rastogi}, {Reddy Janga},
  {Sabater}, {Sakurikar}, {Seifert}, {Sherbert}, {Sherwood-Taylor}, {Shih},
  {Sick}, {Silbiger}, {Singanamalla}, {Singer}, {Sladen}, {Sooley},
  {Sornarajah}, {Streicher}, {Teuben}, {Thomas}, {Tremblay}, {Turner},
  {Terr{\'o}n}, {van Kerkwijk}, {de la Vega}, {Watkins}, {Weaver}, {Whitmore},
  {Woillez}, {Zabalza}, \& {Astropy Contributors}}]{2018AJ....156..123A}
{Astropy Collaboration}, {Price-Whelan}, A.~M., {Sip{\H{o}}cz}, B.~M., {et~al.}
  2018, \aj, 156, 123

\bibitem[{{Astropy Collaboration} {et~al.}(2013){Astropy Collaboration},
  {Robitaille}, {Tollerud}, {Greenfield}, {Droettboom}, {Bray}, {Aldcroft},
  {Davis}, {Ginsburg}, {Price-Whelan}, {Kerzendorf}, {Conley}, {Crighton},
  {Barbary}, {Muna}, {Ferguson}, {Grollier}, {Parikh}, {Nair}, {Unther},
  {Deil}, {Woillez}, {Conseil}, {Kramer}, {Turner}, {Singer}, {Fox}, {Weaver},
  {Zabalza}, {Edwards}, {Azalee Bostroem}, {Burke}, {Casey}, {Crawford},
  {Dencheva}, {Ely}, {Jenness}, {Labrie}, {Lim}, {Pierfederici}, {Pontzen},
  {Ptak}, {Refsdal}, {Servillat}, \& {Streicher}}]{2013A&A...558A..33A}
{Astropy Collaboration}, {Robitaille}, T.~P., {Tollerud}, E.~J., {et~al.} 2013,
  \aap, 558, A33

\bibitem[{{Athanassoula}(2002)}]{2002ApJ...569L..83A}
{Athanassoula}, E. 2002, \apjl, 569, L83

\bibitem[{{Barros} {et~al.}(2020){Barros}, {P{\'e}rez-Villegas}, {L{\'e}pine},
  {Michtchenko}, \& {Vieira}}]{2020ApJ...888...75B}
{Barros}, D.~A., {P{\'e}rez-Villegas}, A., {L{\'e}pine}, J. R.~D.,
  {Michtchenko}, T.~A., \& {Vieira}, R. S.~S. 2020, \apj, 888, 75

\bibitem[{{Belokurov} {et~al.}(2018){Belokurov}, {Erkal}, {Evans}, {Koposov},
  \& {Deason}}]{2018MNRAS.478..611B}
{Belokurov}, V., {Erkal}, D., {Evans}, N.~W., {Koposov}, S.~E., \& {Deason},
  A.~J. 2018, \mnras, 478, 611

\bibitem[{{Belokurov} {et~al.}(2020){Belokurov}, {Sanders}, {Fattahi}, {Smith},
  {Deason}, {Evans}, \& {Grand}}]{2020MNRAS.494.3880B}
{Belokurov}, V., {Sanders}, J.~L., {Fattahi}, A., {et~al.} 2020, \mnras, 494,
  3880

\bibitem[{{Bensby} \& {Feltzing}(2006)}]{2006MNRAS.367.1181B}
{Bensby}, T. \& {Feltzing}, S. 2006, \mnras, 367, 1181

\bibitem[{{Bensby} {et~al.}(2014){Bensby}, {Feltzing}, \&
  {Oey}}]{2014A&A...562A..71B}
{Bensby}, T., {Feltzing}, S., \& {Oey}, M.~S. 2014, \aap, 562, A71

\bibitem[{{Bensby} {et~al.}(2007){Bensby}, {Oey}, {Feltzing}, \&
  {Gustafsson}}]{2007ApJ...655L..89B}
{Bensby}, T., {Oey}, M.~S., {Feltzing}, S., \& {Gustafsson}, B. 2007, \apjl,
  655, L89

\bibitem[{{Binney}(2020)}]{2020MNRAS.495..895B}
{Binney}, J. 2020, \mnras, 495, 895

\bibitem[{{Bland-Hawthorn} \& {Gerhard}(2016)}]{2016ARA&A..54..529B}
{Bland-Hawthorn}, J. \& {Gerhard}, O. 2016, \araa, 54, 529

\bibitem[{{Bland-Hawthorn} {et~al.}(2019){Bland-Hawthorn}, {Sharma},
  {Tepper-Garcia}, {Binney}, {Freeman}, {Hayden}, {Kos}, {De Silva}, {Ellis},
  {Lewis}, {Asplund}, {Buder}, {Casey}, {D'Orazi}, {Duong}, {Khanna}, {Lin},
  {Lind}, {Martell}, {Ness}, {Simpson}, {Zucker}, {Zwitter}, {Kafle},
  {Quillen}, {Ting}, \& {Wyse}}]{2019MNRAS.486.1167B}
{Bland-Hawthorn}, J., {Sharma}, S., {Tepper-Garcia}, T., {et~al.} 2019, \mnras,
  486, 1167

\bibitem[{{Bland-Hawthorn} \& {Tepper-Garc{\'\i}a}(2021)}]{2021MNRAS.504.3168B}
{Bland-Hawthorn}, J. \& {Tepper-Garc{\'\i}a}, T. 2021, \mnras, 504, 3168

\bibitem[{{Bovy} {et~al.}(2019){Bovy}, {Leung}, {Hunt}, {Mackereth},
  {Garc{\'\i}a-Hern{\'a}ndez}, \& {Roman-Lopes}}]{2019MNRAS.490.4740B}
{Bovy}, J., {Leung}, H.~W., {Hunt}, J. A.~S., {et~al.} 2019, \mnras, 490, 4740

\bibitem[{{Buder} {et~al.}(2021){Buder}, {Sharma}, {Kos}, {Amarsi},
  {Nordlander}, {Lind}, {Martell}, {Asplund}, {Bland-Hawthorn}, {Casey}, {de
  Silva}, {D'Orazi}, {Freeman}, {Hayden}, {Lewis}, {Lin}, {Schlesinger},
  {Simpson}, {Stello}, {Zucker}, {Zwitter}, {Beeson}, {Buck}, {Casagrande},
  {Clark}, {{\v{C}}otar}, {da Costa}, {de Grijs}, {Feuillet}, {Horner},
  {Kafle}, {Khanna}, {Kobayashi}, {Liu}, {Montet}, {Nandakumar}, {Nataf},
  {Ness}, {Spina}, {Tepper-Garc{\'\i}a}, {Ting}, {Traven},
  {Vogrin{\v{c}}i{\v{c}}}, {Wittenmyer}, {Wyse}, {{\v{Z}}erjal}, \& {GALAH
  Collaboration}}]{2021MNRAS.506..150B}
{Buder}, S., {Sharma}, S., {Kos}, J., {et~al.} 2021, \mnras, 506, 150

\bibitem[{{Carrillo} {et~al.}(2019){Carrillo}, {Minchev}, {Steinmetz},
  {Monari}, {Laporte}, {Anders}, {Queiroz}, {Chiappini}, {Khalatyan}, {Martig},
  {McMillan}, {Santiago}, \& {Youakim}}]{2019MNRAS.490..797C}
{Carrillo}, I., {Minchev}, I., {Steinmetz}, M., {et~al.} 2019, \mnras, 490, 797

\bibitem[{{Ceverino} \& {Klypin}(2007)}]{2007MNRAS.379.1155C}
{Ceverino}, D. \& {Klypin}, A. 2007, \mnras, 379, 1155

\bibitem[{{Chiappini} {et~al.}(1997){Chiappini}, {Matteucci}, \&
  {Gratton}}]{1997ApJ...477..765C}
{Chiappini}, C., {Matteucci}, F., \& {Gratton}, R. 1997, \apj, 477, 765

\bibitem[{{Chiba} {et~al.}(2021){Chiba}, {Friske}, \&
  {Sch{\"o}nrich}}]{2021MNRAS.500.4710C}
{Chiba}, R., {Friske}, J. K.~S., \& {Sch{\"o}nrich}, R. 2021, \mnras, 500, 4710

\bibitem[{{Chiba} \& {Sch{\"o}nrich}(2021)}]{2021MNRAS.505.2412C}
{Chiba}, R. \& {Sch{\"o}nrich}, R. 2021, \mnras, 505, 2412

\bibitem[{{Clarke} {et~al.}(2019){Clarke}, {Wegg}, {Gerhard}, {Smith}, {Lucas},
  \& {Wylie}}]{2019MNRAS.489.3519C}
{Clarke}, J.~P., {Wegg}, C., {Gerhard}, O., {et~al.} 2019, \mnras, 489, 3519

\bibitem[{{Cohen} {et~al.}(1986){Cohen}, {Dame}, \&
  {Thaddeus}}]{1986ApJS...60..695C}
{Cohen}, R.~S., {Dame}, T.~M., \& {Thaddeus}, P. 1986, \apjs, 60, 695

\bibitem[{{Dame} {et~al.}(1986){Dame}, {Elmegreen}, {Cohen}, \&
  {Thaddeus}}]{1986ApJ...305..892D}
{Dame}, T.~M., {Elmegreen}, B.~G., {Cohen}, R.~S., \& {Thaddeus}, P. 1986,
  \apj, 305, 892

\bibitem[{{De Silva} {et~al.}(2007){De Silva}, {Freeman}, {Bland-Hawthorn},
  {Asplund}, \& {Bessell}}]{2007AJ....133..694D}
{De Silva}, G.~M., {Freeman}, K.~C., {Bland-Hawthorn}, J., {Asplund}, M., \&
  {Bessell}, M.~S. 2007, \aj, 133, 694

\bibitem[{{Debattista} {et~al.}(2017){Debattista}, {Ness}, {Gonzalez},
  {Freeman}, {Zoccali}, \& {Minniti}}]{2017MNRAS.469.1587D}
{Debattista}, V.~P., {Ness}, M., {Gonzalez}, O.~A., {et~al.} 2017, \mnras, 469,
  1587

\bibitem[{{Dehnen}(1998)}]{1998AJ....115.2384D}
{Dehnen}, W. 1998, \aj, 115, 2384

\bibitem[{{Dehnen}(2000)}]{2000AJ....119..800D}
{Dehnen}, W. 2000, \aj, 119, 800

\bibitem[{{Di Matteo} {et~al.}(2019{\natexlab{a}}){Di Matteo}, {Fragkoudi},
  {Khoperskov}, {Ciambur}, {Haywood}, {Combes}, \&
  {G{\'o}mez}}]{2019A&A...628A..11D}
{Di Matteo}, P., {Fragkoudi}, F., {Khoperskov}, S., {et~al.}
  2019{\natexlab{a}}, \aap, 628, A11

\bibitem[{{Di Matteo} {et~al.}(2019{\natexlab{b}}){Di Matteo}, {Haywood},
  {Lehnert}, {Katz}, {Khoperskov}, {Snaith}, {G{\'o}mez}, \&
  {Robichon}}]{2019A&A...632A...4D}
{Di Matteo}, P., {Haywood}, M., {Lehnert}, M.~D., {et~al.} 2019{\natexlab{b}},
  \aap, 632, A4

\bibitem[{{Dobbs} {et~al.}(2010){Dobbs}, {Theis}, {Pringle}, \&
  {Bate}}]{2010MNRAS.403..625D}
{Dobbs}, C.~L., {Theis}, C., {Pringle}, J.~E., \& {Bate}, M.~R. 2010, \mnras,
  403, 625

\bibitem[{{D'Onghia} \& {L. Aguerri}(2020)}]{2020ApJ...890..117D}
{D'Onghia}, E. \& {L. Aguerri}, J.~A. 2020, \apj, 890, 117

\bibitem[{{Eggen}(1965)}]{1965gast.book..111E}
{Eggen}, O.~J. 1965, {Moving Groups of Stars}, 111

\bibitem[{{Eilers} {et~al.}(2019){Eilers}, {Hogg}, {Rix}, \&
  {Ness}}]{2019ApJ...871..120E}
{Eilers}, A.-C., {Hogg}, D.~W., {Rix}, H.-W., \& {Ness}, M.~K. 2019, \apj, 871,
  120

\bibitem[{{Famaey} {et~al.}(2008){Famaey}, {Siebert}, \&
  {Jorissen}}]{2008A&A...483..453F}
{Famaey}, B., {Siebert}, A., \& {Jorissen}, A. 2008, \aap, 483, 453

\bibitem[{{Faure} {et~al.}(2014){Faure}, {Siebert}, \&
  {Famaey}}]{2014MNRAS.440.2564F}
{Faure}, C., {Siebert}, A., \& {Famaey}, B. 2014, \mnras, 440, 2564

\bibitem[{{Fragkoudi} {et~al.}(2020){Fragkoudi}, {Grand}, {Pakmor},
  {Bl{\'a}zquez-Calero}, {Gargiulo}, {Gomez}, {Marinacci}, {Monachesi}, {Ness},
  {Perez}, {Tissera}, \& {White}}]{2020MNRAS.494.5936F}
{Fragkoudi}, F., {Grand}, R.~J.~J., {Pakmor}, R., {et~al.} 2020, \mnras, 494,
  5936

\bibitem[{{Fragkoudi} {et~al.}(2019){Fragkoudi}, {Katz}, {Trick}, {White}, {Di
  Matteo}, {Sormani}, {Khoperskov}, {Haywood}, {Hall{\'e}}, \&
  {G{\'o}mez}}]{2019MNRAS.488.3324F}
{Fragkoudi}, F., {Katz}, D., {Trick}, W., {et~al.} 2019, \mnras, 488, 3324

\bibitem[{{Friske} \& {Sch{\"o}nrich}(2019)}]{2019MNRAS.490.5414F}
{Friske}, J. K.~S. \& {Sch{\"o}nrich}, R. 2019, \mnras, 490, 5414

\bibitem[{{Fujii} {et~al.}(2019){Fujii}, {B{\'e}dorf}, {Baba}, \& {Portegies
  Zwart}}]{2019MNRAS.482.1983F}
{Fujii}, M.~S., {B{\'e}dorf}, J., {Baba}, J., \& {Portegies Zwart}, S. 2019,
  \mnras, 482, 1983

\bibitem[{{Fukushige} {et~al.}(2005){Fukushige}, {Makino}, \&
  {Kawai}}]{2005PASJ...57.1009F}
{Fukushige}, T., {Makino}, J., \& {Kawai}, A. 2005, \pasj, 57, 1009

\bibitem[{{Fux}(2001)}]{2001A&A...373..511F}
{Fux}, R. 2001, \aap, 373, 511

\bibitem[{{Gaia Collaboration} {et~al.}(2021{\natexlab{a}}){Gaia
  Collaboration}, {Antoja}, {McMillan}, {Kordopatis}, {Ramos}, {Helmi},
  {Balbinot}, {Cantat-Gaudin}, {Chemin}, {Figueras}, {Jordi}, {Khanna},
  {Romero-G{\'o}mez}, {Seabroke}, {Brown}, {Vallenari}, {Prusti}, {de Bruijne},
  {Babusiaux}, {Biermann}, {Creevey}, {Evans}, {Eyer}, {Hutton}, {Jansen},
  {Klioner}, {Lammers}, {Lindegren}, {Luri}, {Mignard}, {Panem}, {Pourbaix},
  {Randich}, {Sartoretti}, {Soubiran}, {Walton}, {Arenou}, {Bailer-Jones},
  {Bastian}, {Cropper}, {Drimmel}, {Katz}, {Lattanzi}, {van Leeuwen}, {Bakker},
  {Casta{\~n}eda}, {De Angeli}, {Ducourant}, {Fabricius}, {Fouesneau},
  {Fr{\'e}mat}, {Guerra}, {Guerrier}, {Guiraud}, {Jean-Antoine Piccolo},
  {Masana}, {Messineo}, {Mowlavi}, {Nicolas}, {Nienartowicz}, {Pailler},
  {Panuzzo}, {Riclet}, {Roux}, {Sordo}, {Tanga}, {Th{\'e}venin},
  {Gracia-Abril}, {Portell}, {Teyssier}, {Altmann}, {Andrae}, {Bellas-Velidis},
  {Benson}, {Berthier}, {Blomme}, {Brugaletta}, {Burgess}, {Busso}, {Carry},
  {Cellino}, {Cheek}, {Clementini}, {Damerdji}, {Davidson}, {Delchambre},
  {Dell'Oro}, {Fern{\'a}ndez-Hern{\'a}ndez}, {Galluccio}, {Garc{\'\i}a-Lario},
  {Garcia-Reinaldos}, {Gonz{\'a}lez-N{\'u}{\~n}ez}, {Gosset}, {Haigron},
  {Halbwachs}, {Hambly}, {Harrison}, {Hatzidimitriou}, {Heiter},
  {Hern{\'a}ndez}, {Hestroffer}, {Hodgkin}, {Holl}, {Jan{\ss}en}, {Jevardat de
  Fombelle}, {Jordan}, {Krone-Martins}, {Lanzafame}, {L{\"o}ffler}, {Lorca},
  {Manteiga}, {Marchal}, {Marrese}, {Moitinho}, {Mora}, {Muinonen}, {Osborne},
  {Pancino}, {Pauwels}, {Recio-Blanco}, {Richards}, {Riello}, {Rimoldini},
  {Robin}, {Roegiers}, {Rybizki}, {Sarro}, {Siopis}, {Smith}, {Sozzetti},
  {Ulla}, {Utrilla}, {van Leeuwen}, {van Reeven}, {Abbas}, {Abreu Aramburu},
  {Accart}, {Aerts}, {Aguado}, {Ajaj}, {Altavilla}, {{\'A}lvarez}, {{\'A}lvarez
  Cid-Fuentes}, {Alves}, {Anderson}, {Varela}, {Audard}, {Baines}, {Baker},
  {Balaguer-N{\'u}{\~n}ez}, {Balog}, {Barache}, {Barbato}, {Barros}, {Barstow},
  {Bartolom{\'e}}, {Bassilana}, {Bauchet}, {Baudesson-Stella}, {Becciani},
  {Bellazzini}, {Bernet}, {Bertone}, {Bianchi}, {Blanco-Cuaresma}, {Boch},
  {Bombrun}, {Bossini}, {Bouquillon}, {Bragaglia}, {Bramante}, {Breedt},
  {Bressan}, {Brouillet}, {Bucciarelli}, {Burlacu}, {Busonero}, {Butkevich},
  {Buzzi}, {Caffau}, {Cancelliere}, {C{\'a}novas}, {Carballo}, {Carlucci},
  {Carnerero}, {Carrasco}, {Casamiquela}, {Castellani}, {Castro-Ginard},
  {Castro Sampol}, {Chaoul}, {Charlot}, {Chiavassa}, {Cioni}, {Comoretto},
  {Cooper}, {Cornez}, {Cowell}, {Crifo}, {Crosta}, {Crowley}, {Dafonte},
  {Dapergolas}, {David}, {David}, {de Laverny}, {De Luise}, {De March}, {De
  Ridder}, {de Souza}, {de Teodoro}, {de Torres}, {del Peloso}, {del Pozo},
  {Delgado}, {Delgado}, {Delisle}, {Di Matteo}, {Diakite}, {Diener},
  {Distefano}, {Dolding}, {Eappachen}, {Enke}, {Esquej}, {Fabre}, {Fabrizio},
  {Faigler}, {Fedorets}, {Fernique}, {Fienga}, {Fouron}, {Fragkoudi}, {Fraile},
  {Franke}, {Gai}, {Garabato}, {Garcia-Gutierrez}, {Garc{\'\i}a-Torres},
  {Garofalo}, {Gavras}, {Gerlach}, {Geyer}, {Giacobbe}, {Gilmore}, {Girona},
  {Giuffrida}, {Gomez}, {Gonzalez-Santamaria}, {Gonz{\'a}lez-Vidal}, {Granvik},
  {Guti{\'e}rrez-S{\'a}nchez}, {Guy}, {Hauser}, {Haywood}, {Hidalgo}, {Hilger},
  {H{\l}adczuk}, {Hobbs}, {Holland}, {Huckle}, {Jasniewicz}, {Jonker},
  {Juaristi Campillo}, {Julbe}, {Karbevska}, {Kervella}, {Kochoska},
  {Kontizas}, {Korn}, {Kostrzewa-Rutkowska}, {Kruszy{\'n}ska}, {Lambert},
  {Lanza}, {Lasne}, {Le Campion}, {Le Fustec}, {Lebreton}, {Lebzelter},
  {Leccia}, {Leclerc}, {Lecoeur-Taibi}, {Liao}, {Licata}, {Lindstr{\o}m},
  {Lister}, {Livanou}, {Lobel}, {Madrero Pardo}, {Managau}, {Mann}, {Marchant},
  {Marconi}, {Marcos Santos}, {Marinoni}, {Marocco}, {Marshall}, {Martin Polo},
  {Mart{\'\i}n-Fleitas}, {Masip}, {Massari}, {Mastrobuono-Battisti}, {Mazeh},
  {Messina}, {Michalik}, {Millar}, {Mints}, {Molina}, {Molinaro}, {Moln{\'a}r},
  {Montegriffo}, {Mor}, {Morbidelli}, {Morel}, {Morris}, {Mulone}, {Munoz},
  {Muraveva}, {Murphy}, {Musella}, {Noval}, {Ord{\'e}novic}, {Orr{\`u}},
  {Osinde}, {Pagani}, {Pagano}, {Palaversa}, {Palicio}, {Panahi}, {Pawlak},
  {Pe{\~n}alosa Esteller}, {Penttil{\"a}}, {Piersimoni}, {Pineau}, {Plachy},
  {Plum}, {Poggio}, {Poretti}, {Poujoulet}, {Pr{\v{s}}a}, {Pulone}, {Racero},
  {Ragaini}, {Rainer}, {Raiteri}, {Rambaux}, {Ramos-Lerate}, {Re Fiorentin},
  {Regibo}, {Reyl{\'e}}, {Ripepi}, {Riva}, {Rixon}, {Robichon}, {Robin},
  {Roelens}, {Rohrbasser}, {Rowell}, {Royer}, {Rybicki}, {Sadowski},
  {Sagrist{\`a} Sell{\'e}s}, {Sahlmann}, {Salgado}, {Salguero}, {Samaras},
  {Sanchez Gimenez}, {Sanna}, {Santove{\~n}a}, {Sarasso}, {Schultheis},
  {Sciacca}, {Segol}, {Segovia}, {S{\'e}gransan}, {Semeux}, {Siddiqui},
  {Siebert}, {Siltala}, {Slezak}, {Smart}, {Solano}, {Solitro}, {Souami},
  {Souchay}, {Spagna}, {Spoto}, {Steele}, {Steidelm{\"u}ller}, {Stephenson},
  {S{\"u}veges}, {Szabados}, {Szegedi-Elek}, {Taris}, {Tauran}, {Taylor},
  {Teixeira}, {Thuillot}, {Tonello}, {Torra}, {Torra}, {Turon}, {Unger},
  {Vaillant}, {van Dillen}, {Vanel}, {Vecchiato}, {Viala}, {Vicente},
  {Voutsinas}, {Weiler}, {Wevers}, {Wyrzykowski}, {Yoldas}, {Yvard}, {Zhao},
  {Zorec}, {Zucker}, {Zurbach}, \& {Zwitter}}]{2021A&A...649A...8G}
{Gaia Collaboration}, {Antoja}, T., {McMillan}, P.~J., {et~al.}
  2021{\natexlab{a}}, \aap, 649, A8

\bibitem[{{Gaia Collaboration} {et~al.}(2018{\natexlab{a}}){Gaia
  Collaboration}, {Brown}, {Vallenari}, {Prusti}, {de Bruijne}, {Babusiaux},
  {Bailer-Jones}, {Biermann}, {Evans}, {Eyer}, {Jansen}, {Jordi}, {Klioner},
  {Lammers}, {Lindegren}, {Luri}, {Mignard}, {Panem}, {Pourbaix}, {Randich},
  {Sartoretti}, {Siddiqui}, {Soubiran}, {van Leeuwen}, {Walton}, {Arenou},
  {Bastian}, {Cropper}, {Drimmel}, {Katz}, {Lattanzi}, {Bakker}, {Cacciari},
  {Casta{\~n}eda}, {Chaoul}, {Cheek}, {De Angeli}, {Fabricius}, {Guerra},
  {Holl}, {Masana}, {Messineo}, {Mowlavi}, {Nienartowicz}, {Panuzzo},
  {Portell}, {Riello}, {Seabroke}, {Tanga}, {Th{\'e}venin}, {Gracia-Abril},
  {Comoretto}, {Garcia-Reinaldos}, {Teyssier}, {Altmann}, {Andrae}, {Audard},
  {Bellas-Velidis}, {Benson}, {Berthier}, {Blomme}, {Burgess}, {Busso},
  {Carry}, {Cellino}, {Clementini}, {Clotet}, {Creevey}, {Davidson}, {De
  Ridder}, {Delchambre}, {Dell'Oro}, {Ducourant},
  {Fern{\'a}ndez-Hern{\'a}ndez}, {Fouesneau}, {Fr{\'e}mat}, {Galluccio},
  {Garc{\'\i}a-Torres}, {Gonz{\'a}lez-N{\'u}{\~n}ez}, {Gonz{\'a}lez-Vidal},
  {Gosset}, {Guy}, {Halbwachs}, {Hambly}, {Harrison}, {Hern{\'a}ndez},
  {Hestroffer}, {Hodgkin}, {Hutton}, {Jasniewicz}, {Jean-Antoine-Piccolo},
  {Jordan}, {Korn}, {Krone-Martins}, {Lanzafame}, {Lebzelter}, {L{\"o}ffler},
  {Manteiga}, {Marrese}, {Mart{\'\i}n-Fleitas}, {Moitinho}, {Mora}, {Muinonen},
  {Osinde}, {Pancino}, {Pauwels}, {Petit}, {Recio-Blanco}, {Richards},
  {Rimoldini}, {Robin}, {Sarro}, {Siopis}, {Smith}, {Sozzetti}, {S{\"u}veges},
  {Torra}, {van Reeven}, {Abbas}, {Abreu Aramburu}, {Accart}, {Aerts},
  {Altavilla}, {{\'A}lvarez}, {Alvarez}, {Alves}, {Anderson}, {Andrei},
  {Anglada Varela}, {Antiche}, {Antoja}, {Arcay}, {Astraatmadja}, {Bach},
  {Baker}, {Balaguer-N{\'u}{\~n}ez}, {Balm}, {Barache}, {Barata}, {Barbato},
  {Barblan}, {Barklem}, {Barrado}, {Barros}, {Barstow}, {Bartholom{\'e}
  Mu{\~n}oz}, {Bassilana}, {Becciani}, {Bellazzini}, {Berihuete}, {Bertone},
  {Bianchi}, {Bienaym{\'e}}, {Blanco-Cuaresma}, {Boch}, {Boeche}, {Bombrun},
  {Borrachero}, {Bossini}, {Bouquillon}, {Bourda}, {Bragaglia}, {Bramante},
  {Breddels}, {Bressan}, {Brouillet}, {Br{\"u}semeister}, {Brugaletta},
  {Bucciarelli}, {Burlacu}, {Busonero}, {Butkevich}, {Buzzi}, {Caffau},
  {Cancelliere}, {Cannizzaro}, {Cantat-Gaudin}, {Carballo}, {Carlucci},
  {Carrasco}, {Casamiquela}, {Castellani}, {Castro-Ginard}, {Charlot},
  {Chemin}, {Chiavassa}, {Cocozza}, {Costigan}, {Cowell}, {Crifo}, {Crosta},
  {Crowley}, {Cuypers}, {Dafonte}, {Damerdji}, {Dapergolas}, {David}, {David},
  {de Laverny}, {De Luise}, {De March}, {de Martino}, {de Souza}, {de Torres},
  {Debosscher}, {del Pozo}, {Delbo}, {Delgado}, {Delgado}, {Di Matteo},
  {Diakite}, {Diener}, {Distefano}, {Dolding}, {Drazinos}, {Dur{\'a}n},
  {Edvardsson}, {Enke}, {Eriksson}, {Esquej}, {Eynard Bontemps}, {Fabre},
  {Fabrizio}, {Faigler}, {Falc{\~a}o}, {Farr{\`a}s Casas}, {Federici},
  {Fedorets}, {Fernique}, {Figueras}, {Filippi}, {Findeisen}, {Fonti},
  {Fraile}, {Fraser}, {Fr{\'e}zouls}, {Gai}, {Galleti}, {Garabato},
  {Garc{\'\i}a-Sedano}, {Garofalo}, {Garralda}, {Gavel}, {Gavras}, {Gerssen},
  {Geyer}, {Giacobbe}, {Gilmore}, {Girona}, {Giuffrida}, {Glass}, {Gomes},
  {Granvik}, {Gueguen}, {Guerrier}, {Guiraud}, {Guti{\'e}rrez-S{\'a}nchez},
  {Haigron}, {Hatzidimitriou}, {Hauser}, {Haywood}, {Heiter}, {Helmi}, {Heu},
  {Hilger}, {Hobbs}, {Hofmann}, {Holland}, {Huckle}, {Hypki}, {Icardi},
  {Jan{\ss}en}, {Jevardat de Fombelle}, {Jonker}, {Juh{\'a}sz}, {Julbe},
  {Karampelas}, {Kewley}, {Klar}, {Kochoska}, {Kohley}, {Kolenberg},
  {Kontizas}, {Kontizas}, {Koposov}, {Kordopatis}, {Kostrzewa-Rutkowska},
  {Koubsky}, {Lambert}, {Lanza}, {Lasne}, {Lavigne}, {Le Fustec}, {Le
  Poncin-Lafitte}, {Lebreton}, {Leccia}, {Leclerc}, {Lecoeur-Taibi},
  {Lenhardt}, {Leroux}, {Liao}, {Licata}, {Lindstr{\o}m}, {Lister}, {Livanou},
  {Lobel}, {L{\'o}pez}, {Managau}, {Mann}, {Mantelet}, {Marchal}, {Marchant},
  {Marconi}, {Marinoni}, {Marschalk{\'o}}, {Marshall}, {Martino}, {Marton},
  {Mary}, {Massari}, {Matijevi{\v{c}}}, {Mazeh}, {McMillan}, {Messina},
  {Michalik}, {Millar}, {Molina}, {Molinaro}, {Moln{\'a}r}, {Montegriffo},
  {Mor}, {Morbidelli}, {Morel}, {Morris}, {Mulone}, {Muraveva}, {Musella},
  {Nelemans}, {Nicastro}, {Noval}, {O'Mullane}, {Ord{\'e}novic},
  {Ord{\'o}{\~n}ez-Blanco}, {Osborne}, {Pagani}, {Pagano}, {Pailler},
  {Palacin}, {Palaversa}, {Panahi}, {Pawlak}, {Piersimoni}, {Pineau}, {Plachy},
  {Plum}, {Poggio}, {Poujoulet}, {Pr{\v{s}}a}, {Pulone}, {Racero}, {Ragaini},
  {Rambaux}, {Ramos-Lerate}, {Regibo}, {Reyl{\'e}}, {Riclet}, {Ripepi}, {Riva},
  {Rivard}, {Rixon}, {Roegiers}, {Roelens}, {Romero-G{\'o}mez}, {Rowell},
  {Royer}, {Ruiz-Dern}, {Sadowski}, {Sagrist{\`a} Sell{\'e}s}, {Sahlmann},
  {Salgado}, {Salguero}, {Sanna}, {Santana-Ros}, {Sarasso}, {Savietto},
  {Schultheis}, {Sciacca}, {Segol}, {Segovia}, {S{\'e}gransan}, {Shih},
  {Siltala}, {Silva}, {Smart}, {Smith}, {Solano}, {Solitro}, {Sordo}, {Soria
  Nieto}, {Souchay}, {Spagna}, {Spoto}, {Stampa}, {Steele},
  {Steidelm{\"u}ller}, {Stephenson}, {Stoev}, {Suess}, {Surdej}, {Szabados},
  {Szegedi-Elek}, {Tapiador}, {Taris}, {Tauran}, {Taylor}, {Teixeira},
  {Terrett}, {Teyssand ier}, {Thuillot}, {Titarenko}, {Torra Clotet}, {Turon},
  {Ulla}, {Utrilla}, {Uzzi}, {Vaillant}, {Valentini}, {Valette}, {van Elteren},
  {Van Hemelryck}, {van Leeuwen}, {Vaschetto}, {Vecchiato}, {Veljanoski},
  {Viala}, {Vicente}, {Vogt}, {von Essen}, {Voss}, {Votruba}, {Voutsinas},
  {Walmsley}, {Weiler}, {Wertz}, {Wevers}, {Wyrzykowski}, {Yoldas},
  {{\v{Z}}erjal}, {Ziaeepour}, {Zorec}, {Zschocke}, {Zucker}, {Zurbach}, \&
  {Zwitter}}]{2018A&A...616A...1G}
{Gaia Collaboration}, {Brown}, A.~G.~A., {Vallenari}, A., {et~al.}
  2018{\natexlab{a}}, \aap, 616, A1

\bibitem[{{Gaia Collaboration} {et~al.}(2021{\natexlab{b}}){Gaia
  Collaboration}, {Brown}, {Vallenari}, {Prusti}, {de Bruijne}, {Babusiaux},
  {Biermann}, {Creevey}, {Evans}, {Eyer}, {Hutton}, {Jansen}, {Jordi},
  {Klioner}, {Lammers}, {Lindegren}, {Luri}, {Mignard}, {Panem}, {Pourbaix},
  {Randich}, {Sartoretti}, {Soubiran}, {Walton}, {Arenou}, {Bailer-Jones},
  {Bastian}, {Cropper}, {Drimmel}, {Katz}, {Lattanzi}, {van Leeuwen}, {Bakker},
  {Cacciari}, {Casta{\~n}eda}, {De Angeli}, {Ducourant}, {Fabricius},
  {Fouesneau}, {Fr{\'e}mat}, {Guerra}, {Guerrier}, {Guiraud}, {Jean-Antoine
  Piccolo}, {Masana}, {Messineo}, {Mowlavi}, {Nicolas}, {Nienartowicz},
  {Pailler}, {Panuzzo}, {Riclet}, {Roux}, {Seabroke}, {Sordo}, {Tanga},
  {Th{\'e}venin}, {Gracia-Abril}, {Portell}, {Teyssier}, {Altmann}, {Andrae},
  {Bellas-Velidis}, {Benson}, {Berthier}, {Blomme}, {Brugaletta}, {Burgess},
  {Busso}, {Carry}, {Cellino}, {Cheek}, {Clementini}, {Damerdji}, {Davidson},
  {Delchambre}, {Dell'Oro}, {Fern{\'a}ndez-Hern{\'a}ndez}, {Galluccio},
  {Garc{\'\i}a-Lario}, {Garcia-Reinaldos}, {Gonz{\'a}lez-N{\'u}{\~n}ez},
  {Gosset}, {Haigron}, {Halbwachs}, {Hambly}, {Harrison}, {Hatzidimitriou},
  {Heiter}, {Hern{\'a}ndez}, {Hestroffer}, {Hodgkin}, {Holl}, {Jan{\ss}en},
  {Jevardat de Fombelle}, {Jordan}, {Krone-Martins}, {Lanzafame},
  {L{\"o}ffler}, {Lorca}, {Manteiga}, {Marchal}, {Marrese}, {Moitinho}, {Mora},
  {Muinonen}, {Osborne}, {Pancino}, {Pauwels}, {Petit}, {Recio-Blanco},
  {Richards}, {Riello}, {Rimoldini}, {Robin}, {Roegiers}, {Rybizki}, {Sarro},
  {Siopis}, {Smith}, {Sozzetti}, {Ulla}, {Utrilla}, {van Leeuwen}, {van
  Reeven}, {Abbas}, {Abreu Aramburu}, {Accart}, {Aerts}, {Aguado}, {Ajaj},
  {Altavilla}, {{\'A}lvarez}, {{\'A}lvarez Cid-Fuentes}, {Alves}, {Anderson},
  {Anglada Varela}, {Antoja}, {Audard}, {Baines}, {Baker},
  {Balaguer-N{\'u}{\~n}ez}, {Balbinot}, {Balog}, {Barache}, {Barbato},
  {Barros}, {Barstow}, {Bartolom{\'e}}, {Bassilana}, {Bauchet},
  {Baudesson-Stella}, {Becciani}, {Bellazzini}, {Bernet}, {Bertone}, {Bianchi},
  {Blanco-Cuaresma}, {Boch}, {Bombrun}, {Bossini}, {Bouquillon}, {Bragaglia},
  {Bramante}, {Breedt}, {Bressan}, {Brouillet}, {Bucciarelli}, {Burlacu},
  {Busonero}, {Butkevich}, {Buzzi}, {Caffau}, {Cancelliere}, {C{\'a}novas},
  {Cantat-Gaudin}, {Carballo}, {Carlucci}, {Carnerero}, {Carrasco},
  {Casamiquela}, {Castellani}, {Castro-Ginard}, {Castro Sampol}, {Chaoul},
  {Charlot}, {Chemin}, {Chiavassa}, {Cioni}, {Comoretto}, {Cooper}, {Cornez},
  {Cowell}, {Crifo}, {Crosta}, {Crowley}, {Dafonte}, {Dapergolas}, {David},
  {David}, {de Laverny}, {De Luise}, {De March}, {De Ridder}, {de Souza}, {de
  Teodoro}, {de Torres}, {del Peloso}, {del Pozo}, {Delbo}, {Delgado},
  {Delgado}, {Delisle}, {Di Matteo}, {Diakite}, {Diener}, {Distefano},
  {Dolding}, {Eappachen}, {Edvardsson}, {Enke}, {Esquej}, {Fabre}, {Fabrizio},
  {Faigler}, {Fedorets}, {Fernique}, {Fienga}, {Figueras}, {Fouron},
  {Fragkoudi}, {Fraile}, {Franke}, {Gai}, {Garabato}, {Garcia-Gutierrez},
  {Garc{\'\i}a-Torres}, {Garofalo}, {Gavras}, {Gerlach}, {Geyer}, {Giacobbe},
  {Gilmore}, {Girona}, {Giuffrida}, {Gomel}, {Gomez}, {Gonzalez-Santamaria},
  {Gonz{\'a}lez-Vidal}, {Granvik}, {Guti{\'e}rrez-S{\'a}nchez}, {Guy},
  {Hauser}, {Haywood}, {Helmi}, {Hidalgo}, {Hilger}, {H{\l}adczuk}, {Hobbs},
  {Holland}, {Huckle}, {Jasniewicz}, {Jonker}, {Juaristi Campillo}, {Julbe},
  {Karbevska}, {Kervella}, {Khanna}, {Kochoska}, {Kontizas}, {Kordopatis},
  {Korn}, {Kostrzewa-Rutkowska}, {Kruszy{\'n}ska}, {Lambert}, {Lanza}, {Lasne},
  {Le Campion}, {Le Fustec}, {Lebreton}, {Lebzelter}, {Leccia}, {Leclerc},
  {Lecoeur-Taibi}, {Liao}, {Licata}, {Lindstr{\o}m}, {Lister}, {Livanou},
  {Lobel}, {Madrero Pardo}, {Managau}, {Mann}, {Marchant}, {Marconi}, {Marcos
  Santos}, {Marinoni}, {Marocco}, {Marshall}, {Martin Polo},
  {Mart{\'\i}n-Fleitas}, {Masip}, {Massari}, {Mastrobuono-Battisti}, {Mazeh},
  {McMillan}, {Messina}, {Michalik}, {Millar}, {Mints}, {Molina}, {Molinaro},
  {Moln{\'a}r}, {Montegriffo}, {Mor}, {Morbidelli}, {Morel}, {Morris},
  {Mulone}, {Munoz}, {Muraveva}, {Murphy}, {Musella}, {Noval}, {Ord{\'e}novic},
  {Orr{\`u}}, {Osinde}, {Pagani}, {Pagano}, {Palaversa}, {Palicio}, {Panahi},
  {Pawlak}, {Pe{\~n}alosa Esteller}, {Penttil{\"a}}, {Piersimoni}, {Pineau},
  {Plachy}, {Plum}, {Poggio}, {Poretti}, {Poujoulet}, {Pr{\v{s}}a}, {Pulone},
  {Racero}, {Ragaini}, {Rainer}, {Raiteri}, {Rambaux}, {Ramos}, {Ramos-Lerate},
  {Re Fiorentin}, {Regibo}, {Reyl{\'e}}, {Ripepi}, {Riva}, {Rixon}, {Robichon},
  {Robin}, {Roelens}, {Rohrbasser}, {Romero-G{\'o}mez}, {Rowell}, {Royer},
  {Rybicki}, {Sadowski}, {Sagrist{\`a} Sell{\'e}s}, {Sahlmann}, {Salgado},
  {Salguero}, {Samaras}, {Sanchez Gimenez}, {Sanna}, {Santove{\~n}a},
  {Sarasso}, {Schultheis}, {Sciacca}, {Segol}, {Segovia}, {S{\'e}gransan},
  {Semeux}, {Shahaf}, {Siddiqui}, {Siebert}, {Siltala}, {Slezak}, {Smart},
  {Solano}, {Solitro}, {Souami}, {Souchay}, {Spagna}, {Spoto}, {Steele},
  {Steidelm{\"u}ller}, {Stephenson}, {S{\"u}veges}, {Szabados}, {Szegedi-Elek},
  {Taris}, {Tauran}, {Taylor}, {Teixeira}, {Thuillot}, {Tonello}, {Torra},
  {Torra}, {Turon}, {Unger}, {Vaillant}, {van Dillen}, {Vanel}, {Vecchiato},
  {Viala}, {Vicente}, {Voutsinas}, {Weiler}, {Wevers}, {Wyrzykowski}, {Yoldas},
  {Yvard}, {Zhao}, {Zorec}, {Zucker}, {Zurbach}, \&
  {Zwitter}}]{2021A&A...649A...1G}
{Gaia Collaboration}, {Brown}, A.~G.~A., {Vallenari}, A., {et~al.}
  2021{\natexlab{b}}, \aap, 649, A1

\bibitem[{{Gaia Collaboration} {et~al.}(2018{\natexlab{b}}){Gaia
  Collaboration}, {Katz}, {Antoja}, {Romero-G{\'o}mez}, {Drimmel}, {Reyl{\'e}},
  {Seabroke}, {Soubiran}, {Babusiaux}, {Di Matteo}, {Figueras}, {Poggio},
  {Robin}, {Evans}, {Brown}, {Vallenari}, {Prusti}, {de Bruijne},
  {Bailer-Jones}, {Biermann}, {Eyer}, {Jansen}, {Jordi}, {Klioner}, {Lammers},
  {Lindegren}, {Luri}, {Mignard}, {Panem}, {Pourbaix}, {Randich}, {Sartoretti},
  {Siddiqui}, {van Leeuwen}, {Walton}, {Arenou}, {Bastian}, {Cropper},
  {Lattanzi}, {Bakker}, {Cacciari}, {Casta n}, {Chaoul}, {Cheek}, {De Angeli},
  {Fabricius}, {Guerra}, {Holl}, {Masana}, {Messineo}, {Mowlavi},
  {Nienartowicz}, {Panuzzo}, {Portell}, {Riello}, {Tanga}, {Th{\'e}venin},
  {Gracia-Abril}, {Comoretto}, {Garcia-Reinaldos}, {Teyssier}, {Altmann},
  {Andrae}, {Audard}, {Bellas-Velidis}, {Benson}, {Berthier}, {Blomme},
  {Burgess}, {Busso}, {Carry}, {Cellino}, {Clementini}, {Clotet}, {Creevey},
  {Davidson}, {De Ridder}, {Delchambre}, {Dell'Oro}, {Ducourant},
  {Fern{\'a}ndez-Hern{\'a}ndez}, {Fouesneau}, {Fr{\'e}mat}, {Galluccio},
  {Garc{\'\i}a-Torres}, {Gonz{\'a}lez-N{\'u}{\~n}ez}, {Gonz{\'a}lez-Vidal},
  {Gosset}, {Guy}, {Halbwachs}, {Hambly}, {Harrison}, {Hern{\'a}ndez},
  {Hestroffer}, {Hodgkin}, {Hutton}, {Jasniewicz}, {Jean-Antoine-Piccolo},
  {Jordan}, {Korn}, {Krone-Martins}, {Lanzafame}, {Lebzelter}, {L{\"o}ffler},
  {Manteiga}, {Marrese}, {Mart{\'\i}n-Fleitas}, {Moitinho}, {Mora}, {Muinonen},
  {Osinde}, {Pancino}, {Pauwels}, {Petit}, {Recio-Blanco}, {Richards},
  {Rimoldini}, {Sarro}, {Siopis}, {Smith}, {Sozzetti}, {S{\"u}veges}, {Torra},
  {van Reeven}, {Abbas}, {Abreu Aramburu}, {Accart}, {Aerts}, {Altavilla},
  {{\'A}lvarez}, {Alvarez}, {Alves}, {Anderson}, {Andrei}, {Anglada Varela},
  {Antiche}, {Arcay}, {Astraatmadja}, {Bach}, {Baker},
  {Balaguer-N{\'u}{\~n}ez}, {Balm}, {Barache}, {Barata}, {Barbato}, {Barblan},
  {Barklem}, {Barrado}, {Barros}, {Barstow}, {Bartholom{\'e} Mu{\~n}oz},
  {Bassilana}, {Becciani}, {Bellazzini}, {Berihuete}, {Bertone}, {Bianchi},
  {Bienaym{\'e}}, {Blanco-Cuaresma}, {Boch}, {Boeche}, {Bombrun}, {Borrachero},
  {Bossini}, {Bouquillon}, {Bourda}, {Bragaglia}, {Bramante}, {Breddels},
  {Bressan}, {Brouillet}, {Br{\"u}semeister}, {Brugaletta}, {Bucciarelli},
  {Burlacu}, {Busonero}, {Butkevich}, {Buzzi}, {Caffau}, {Cancelliere},
  {Cannizzaro}, {Cantat-Gaudin}, {Carballo}, {Carlucci}, {Carrasco},
  {Casamiquela}, {Castellani}, {Castro-Ginard}, {Charlot}, {Chemin},
  {Chiavassa}, {Cocozza}, {Costigan}, {Cowell}, {Crifo}, {Crosta}, {Crowley},
  {Cuypers}, {Dafonte}, {Damerdji}, {Dapergolas}, {David}, {David}, {de
  Laverny}, {De Luise}, {De March}, {de Souza}, {de Torres}, {Debosscher}, {del
  Pozo}, {Delbo}, {Delgado}, {Delgado}, {Diakite}, {Diener}, {Distefano},
  {Dolding}, {Drazinos}, {Dur{\'a}n}, {Edvardsson}, {Enke}, {Eriksson},
  {Esquej}, {Eynard Bontemps}, {Fabre}, {Fabrizio}, {Faigler}, {Falc a},
  {Farr{\`a}s Casas}, {Federici}, {Fedorets}, {Fernique}, {Filippi},
  {Findeisen}, {Fonti}, {Fraile}, {Fraser}, {Fr{\'e}zouls}, {Gai}, {Galleti},
  {Garabato}, {Garc{\'\i}a-Sedano}, {Garofalo}, {Garralda}, {Gavel}, {Gavras},
  {Gerssen}, {Geyer}, {Giacobbe}, {Gilmore}, {Girona}, {Giuffrida}, {Glass},
  {Gomes}, {Granvik}, {Gueguen}, {Guerrier}, {Guiraud}, {Guti{\'e}}, {Haigron},
  {Hatzidimitriou}, {Hauser}, {Haywood}, {Heiter}, {Helmi}, {Heu}, {Hilger},
  {Hobbs}, {Hofmann}, {Holland }, {Huckle}, {Hypki}, {Icardi}, {Jan{\ss}en},
  {Jevardat de Fombelle}, {Jonker}, {Juh{\'a}sz}, {Julbe}, {Karampelas},
  {Kewley}, {Klar}, {Kochoska}, {Kohley}, {Kolenberg}, {Kontizas}, {Kontizas},
  {Koposov}, {Kordopatis}, {Kostrzewa-Rutkowska}, {Koubsky}, {Lambert},
  {Lanza}, {Lasne}, {Lavigne}, {Le Fustec}, {Le Poncin-Lafitte}, {Lebreton},
  {Leccia}, {Leclerc}, {Lecoeur-Taibi}, {Lenhardt}, {Leroux}, {Liao}, {Licata},
  {Lindstr{\o}m}, {Lister}, {Livanou}, {Lobel}, {L{\'o}pez}, {Managau}, {Mann},
  {Mantelet}, {Marchal}, {Marchant}, {Marconi}, {Marinoni}, {Marschalk{\'o}},
  {Marshall}, {Martino}, {Marton}, {Mary}, {Massari}, {Matijevi{\v{c}}},
  {Mazeh}, {McMillan}, {Messina}, {Michalik}, {Millar}, {Molina}, {Molinaro},
  {Moln{\'a}r}, {Montegriffo}, {Mor}, {Morbidelli}, {Morel}, {Morris},
  {Mulone}, {Muraveva}, {Musella}, {Nelemans}, {Nicastro}, {Noval},
  {O'Mullane}, {Ord{\'e}novic}, {Ord{\'o}{\~n}ez-Blanco}, {Osborne}, {Pagani},
  {Pagano}, {Pailler}, {Palacin}, {Palaversa}, {Panahi}, {Pawlak},
  {Piersimoni}, {Pineau}, {Plachy}, {Plum}, {Poujoulet}, {Pr{\v{s}}a},
  {Pulone}, {Racero}, {Ragaini}, {Rambaux}, {Ramos-Lerate}, {Regibo}, {Riclet},
  {Ripepi}, {Riva}, {Rivard}, {Rixon}, {Roegiers}, {Roelens}, {Rowell},
  {Royer}, {Ruiz-Dern}, {Sadowski}, {Sagrist{\`a} Sell{\'e}s}, {Sahlmann},
  {Salgado}, {Salguero}, {Sanna}, {Santana-Ros}, {Sarasso}, {Savietto},
  {Schultheis}, {Sciacca}, {Segol}, {Segovia}, {S{\'e}gransan}, {Shih},
  {Siltala}, {Silva}, {Smart}, {Smith}, {Solano}, {Solitro}, {Sordo}, {Soria
  Nieto}, {Souchay}, {Spagna}, {Spoto}, {Stampa}, {Steele},
  {Steidelm{\"u}ller}, {Stephenson}, {Stoev}, {Suess}, {Surdej}, {Szabados},
  {Szegedi-Elek}, {Tapiador}, {Taris}, {Tauran}, {Taylor}, {Teixeira},
  {Terrett}, {Teyssand ier}, {Thuillot}, {Titarenko}, {Torra Clotet}, {Turon},
  {Ulla}, {Utrilla}, {Uzzi}, {Vaillant}, {Valentini}, {Valette}, {van Elteren},
  {Van Hemelryck}, {van Leeuwen}, {Vaschetto}, {Vecchiato}, {Veljanoski},
  {Viala}, {Vicente}, {Vogt}, {von Essen}, {Voss}, {Votruba}, {Voutsinas},
  {Walmsley}, {Weiler}, {Wertz}, {Wevers}, {Wyrzykowski}, {Yoldas},
  {{\v{Z}}erjal}, {Ziaeepour}, {Zorec}, {Zschocke}, {Zucker}, {Zurbach}, \&
  {Zwitter}}]{2018A&A...616A..11G}
{Gaia Collaboration}, {Katz}, D., {Antoja}, T., {et~al.} 2018{\natexlab{b}},
  \aap, 616, A11

\bibitem[{{Georgelin} \& {Georgelin}(1976)}]{1976A&A....49...57G}
{Georgelin}, Y.~M. \& {Georgelin}, Y.~P. 1976, \aap, 49, 57

\bibitem[{{Gerin} {et~al.}(1990){Gerin}, {Combes}, \&
  {Athanassoula}}]{1990A&A...230...37G}
{Gerin}, M., {Combes}, F., \& {Athanassoula}, E. 1990, \aap, 230, 37

\bibitem[{{G{\'o}mez} {et~al.}(2013){G{\'o}mez}, {Minchev}, {O'Shea}, {Beers},
  {Bullock}, \& {Purcell}}]{2013MNRAS.429..159G}
{G{\'o}mez}, F.~A., {Minchev}, I., {O'Shea}, B.~W., {et~al.} 2013, \mnras, 429,
  159

\bibitem[{{G{\'o}mez} {et~al.}(2012){G{\'o}mez}, {Minchev}, {Villalobos},
  {O'Shea}, \& {Williams}}]{2012MNRAS.419.2163G}
{G{\'o}mez}, F.~A., {Minchev}, I., {Villalobos}, {\'A}., {O'Shea}, B.~W., \&
  {Williams}, M. E.~K. 2012, \mnras, 419, 2163

\bibitem[{{Grabelsky} {et~al.}(1988){Grabelsky}, {Cohen}, {Bronfman}, \&
  {Thaddeus}}]{1988ApJ...331..181G}
{Grabelsky}, D.~A., {Cohen}, R.~S., {Bronfman}, L., \& {Thaddeus}, P. 1988,
  \apj, 331, 181

\bibitem[{{Grand} {et~al.}(2015){Grand}, {Bovy}, {Kawata}, {Hunt}, {Famaey},
  {Siebert}, {Monari}, \& {Cropper}}]{2015MNRAS.453.1867G}
{Grand}, R. J.~J., {Bovy}, J., {Kawata}, D., {et~al.} 2015, \mnras, 453, 1867

\bibitem[{{Grand} {et~al.}(2016){Grand}, {Springel}, {Kawata}, {Minchev},
  {S{\'a}nchez-Bl{\'a}zquez}, {G{\'o}mez}, {Marinacci}, {Pakmor}, \&
  {Campbell}}]{2016MNRAS.460L..94G}
{Grand}, R. J.~J., {Springel}, V., {Kawata}, D., {et~al.} 2016, \mnras, 460,
  L94

\bibitem[{{Gravity Collaboration} {et~al.}(2018){Gravity Collaboration},
  {Abuter}, {Amorim}, {Anugu}, {Baub{\"o}ck}, {Benisty}, {Berger}, {Blind},
  {Bonnet}, {Brandner}, {Buron}, {Collin}, {Chapron}, {Cl{\'e}net}, {Coud{\'e}
  Du Foresto}, {de Zeeuw}, {Deen}, {Delplancke-Str{\"o}bele}, {Dembet},
  {Dexter}, {Duvert}, {Eckart}, {Eisenhauer}, {Finger}, {F{\"o}rster
  Schreiber}, {F{\'e}dou}, {Garcia}, {Garcia Lopez}, {Gao}, {Gendron},
  {Genzel}, {Gillessen}, {Gordo}, {Habibi}, {Haubois}, {Haug}, {Hau{\ss}mann},
  {Henning}, {Hippler}, {Horrobin}, {Hubert}, {Hubin}, {Jimenez Rosales},
  {Jochum}, {Jocou}, {Kaufer}, {Kellner}, {Kendrew}, {Kervella}, {Kok},
  {Kulas}, {Lacour}, {Lapeyr{\`e}re}, {Lazareff}, {Le Bouquin}, {L{\'e}na},
  {Lippa}, {Lenzen}, {M{\'e}rand}, {M{\"u}ler}, {Neumann}, {Ott}, {Palanca},
  {Paumard}, {Pasquini}, {Perraut}, {Perrin}, {Pfuhl}, {Plewa}, {Rabien},
  {Ram{\'\i}rez}, {Ramos}, {Rau}, {Rodr{\'\i}guez-Coira}, {Rohloff}, {Rousset},
  {Sanchez-Bermudez}, {Scheithauer}, {Sch{\"o}ller}, {Schuler}, {Spyromilio},
  {Straub}, {Straubmeier}, {Sturm}, {Tacconi}, {Tristram}, {Vincent}, {von
  Fellenberg}, {Wank}, {Waisberg}, {Widmann}, {Wieprecht}, {Wiest},
  {Wiezorrek}, {Woillez}, {Yazici}, {Ziegler}, \& {Zins}}]{2018A&A...615L..15G}
{Gravity Collaboration}, {Abuter}, R., {Amorim}, A., {et~al.} 2018, Astronomy
  and Astrophysics, 615, L15

\bibitem[{{Halle} {et~al.}(2015){Halle}, {Di Matteo}, {Haywood}, \&
  {Combes}}]{2015A&A...578A..58H}
{Halle}, A., {Di Matteo}, P., {Haywood}, M., \& {Combes}, F. 2015, \aap, 578,
  A58

\bibitem[{{Hayden} {et~al.}(2020){Hayden}, {Bland-Hawthorn}, {Sharma},
  {Freeman}, {Kos}, {Buder}, {Anguiano}, {Asplund}, {Chen}, {De Silva},
  {Khanna}, {Lin}, {Horner}, {Martell}, {Ting}, {Wyse}, {Zucker}, \&
  {Zwitter}}]{2020MNRAS.493.2952H}
{Hayden}, M.~R., {Bland-Hawthorn}, J., {Sharma}, S., {et~al.} 2020, \mnras,
  493, 2952

\bibitem[{{Haywood} {et~al.}(2013){Haywood}, {Di Matteo}, {Lehnert}, {Katz}, \&
  {G{\'o}mez}}]{2013A&A...560A.109H}
{Haywood}, M., {Di Matteo}, P., {Lehnert}, M.~D., {Katz}, D., \& {G{\'o}mez},
  A. 2013, \aap, 560, A109

\bibitem[{{Haywood} {et~al.}(2019){Haywood}, {Snaith}, {Lehnert}, {Di Matteo},
  \& {Khoperskov}}]{2019A&A...625A.105H}
{Haywood}, M., {Snaith}, O., {Lehnert}, M.~D., {Di Matteo}, P., \&
  {Khoperskov}, S. 2019, \aap, 625, A105

\bibitem[{{Helmi} {et~al.}(2006){Helmi}, {Navarro}, {Nordstr{\"o}m},
  {Holmberg}, {Abadi}, \& {Steinmetz}}]{2006MNRAS.365.1309H}
{Helmi}, A., {Navarro}, J.~F., {Nordstr{\"o}m}, B., {et~al.} 2006, \mnras, 365,
  1309

\bibitem[{{Helmi} {et~al.}(2014){Helmi}, {Williams}, {Freeman}, {Bland
  -Hawthorn}, \& {De Silva}}]{2014ApJ...791..135H}
{Helmi}, A., {Williams}, M., {Freeman}, K.~C., {Bland -Hawthorn}, J., \& {De
  Silva}, G. 2014, \apj, 791, 135

\bibitem[{{Hottier} {et~al.}(2020){Hottier}, {Babusiaux}, \&
  {Arenou}}]{2020A&A...641A..79H}
{Hottier}, C., {Babusiaux}, C., \& {Arenou}, F. 2020, \aap, 641, A79

\bibitem[{{Hou} \& {Han}(2014)}]{2014A&A...569A.125H}
{Hou}, L.~G. \& {Han}, J.~L. 2014, \aap, 569, A125

\bibitem[{{Hou} {et~al.}(2009){Hou}, {Han}, \& {Shi}}]{2009A&A...499..473H}
{Hou}, L.~G., {Han}, J.~L., \& {Shi}, W.~B. 2009, \aap, 499, 473

\bibitem[{{Hunt} {et~al.}(2019){Hunt}, {Bub}, {Bovy}, {Mackereth}, {Trick}, \&
  {Kawata}}]{2019MNRAS.490.1026H}
{Hunt}, J. A.~S., {Bub}, M.~W., {Bovy}, J., {et~al.} 2019, \mnras, 490, 1026

\bibitem[{{Hunt} {et~al.}(2018){Hunt}, {Hong}, {Bovy}, {Kawata}, \&
  {Grand}}]{2018MNRAS.481.3794H}
{Hunt}, J. A.~S., {Hong}, J., {Bovy}, J., {Kawata}, D., \& {Grand}, R. J.~J.
  2018, \mnras, 481, 3794

\bibitem[{{Hunt} {et~al.}(2020){Hunt}, {Johnston}, {Pettitt}, {Cunningham},
  {Kawata}, \& {Hogg}}]{2020MNRAS.497..818H}
{Hunt}, J. A.~S., {Johnston}, K.~V., {Pettitt}, A.~R., {et~al.} 2020, \mnras,
  497, 818

\bibitem[{{Hunter}(2007)}]{2007CSE.....9...90H}
{Hunter}, J.~D. 2007, Computing in Science and Engineering, 9, 90

\bibitem[{{Jean-Baptiste} {et~al.}(2017){Jean-Baptiste}, {Di Matteo},
  {Haywood}, {G{\'o}mez}, {Montuori}, {Combes}, \&
  {Semelin}}]{2017A&A...604A.106J}
{Jean-Baptiste}, I., {Di Matteo}, P., {Haywood}, M., {et~al.} 2017, \aap, 604,
  A106

\bibitem[{{Kamdar} {et~al.}(2019){Kamdar}, {Conroy}, {Ting}, {Bonaca},
  {Johnson}, \& {Cargile}}]{2019ApJ...884..173K}
{Kamdar}, H., {Conroy}, C., {Ting}, Y.-S., {et~al.} 2019, \apj, 884, 173

\bibitem[{{Katz} {et~al.}(2021){Katz}, {Gomez}, {Haywood}, {Snaith}, \& {Di
  Matteo}}]{2021arXiv210202082K}
{Katz}, D., {Gomez}, A., {Haywood}, M., {Snaith}, O., \& {Di Matteo}, P. 2021,
  arXiv e-prints, arXiv:2102.02082

\bibitem[{{Kawata} {et~al.}(2018){Kawata}, {Baba}, {Ciuc{\v{a}}}, {Cropper},
  {Grand}, {Hunt}, \& {Seabroke}}]{2018MNRAS.479L.108K}
{Kawata}, D., {Baba}, J., {Ciuc{\v{a}}}, I., {et~al.} 2018, \mnras, 479, L108

\bibitem[{{Khanna} {et~al.}(2019){Khanna}, {Sharma}, {Tepper-Garcia}, {Bland
  -Hawthorn}, {Hayden}, {Asplund}, {Buder}, {Chen}, {De Silva}, {Freeman},
  {Kos}, {Lewis}, {Lin}, {Martell}, {Simpson}, {Nordlander}, {Stello}, {Ting},
  {Zucker}, \& {Zwitter}}]{2019MNRAS.489.4962K}
{Khanna}, S., {Sharma}, S., {Tepper-Garcia}, T., {et~al.} 2019, \mnras, 489,
  4962

\bibitem[{{Khoperskov} {et~al.}(2019){Khoperskov}, {Di Matteo}, {Gerhard},
  {Katz}, {Haywood}, {Combes}, {Berczik}, \& {Gomez}}]{2019A&A...622L...6K}
{Khoperskov}, S., {Di Matteo}, P., {Gerhard}, O., {et~al.} 2019, \aap, 622, L6

\bibitem[{{Khoperskov} {et~al.}(2018{\natexlab{a}}){Khoperskov}, {Di Matteo},
  {Haywood}, \& {Combes}}]{2018A&A...611L...2K}
{Khoperskov}, S., {Di Matteo}, P., {Haywood}, M., \& {Combes}, F.
  2018{\natexlab{a}}, \aap, 611, L2

\bibitem[{{Khoperskov} {et~al.}(2020{\natexlab{a}}){Khoperskov}, {Di Matteo},
  {Haywood}, {G{\'o}mez}, \& {Snaith}}]{2020A&A...638A.144K}
{Khoperskov}, S., {Di Matteo}, P., {Haywood}, M., {G{\'o}mez}, A., \& {Snaith},
  O.~N. 2020{\natexlab{a}}, \aap, 638, A144

\bibitem[{{Khoperskov} {et~al.}(2020{\natexlab{b}}){Khoperskov}, {Gerhard}, {Di
  Matteo}, {Haywood}, {Katz}, {Khrapov}, {Khoperskov}, \&
  {Arnaboldi}}]{2020A&A...634L...8K}
{Khoperskov}, S., {Gerhard}, O., {Di Matteo}, P., {et~al.} 2020{\natexlab{b}},
  \aap, 634, L8

\bibitem[{{Khoperskov} {et~al.}(2018{\natexlab{b}}){Khoperskov},
  {Mastrobuono-Battisti}, {Di Matteo}, \& {Haywood}}]{2018A&A...620A.154K}
{Khoperskov}, S., {Mastrobuono-Battisti}, A., {Di Matteo}, P., \& {Haywood}, M.
  2018{\natexlab{b}}, \aap, 620, A154

\bibitem[{{Kushniruk} \& {Bensby}(2019)}]{2019A&A...631A..47K}
{Kushniruk}, I. \& {Bensby}, T. 2019, \aap, 631, A47

\bibitem[{{Kushniruk} {et~al.}(2020){Kushniruk}, {Bensby}, {Feltzing},
  {Sahlholdt}, {Feuillet}, \& {Casagrande}}]{2020A&A...638A.154K}
{Kushniruk}, I., {Bensby}, T., {Feltzing}, S., {et~al.} 2020, \aap, 638, A154

\bibitem[{{Kushniruk} {et~al.}(2017){Kushniruk}, {Schirmer}, \&
  {Bensby}}]{2017A&A...608A..73K}
{Kushniruk}, I., {Schirmer}, T., \& {Bensby}, T. 2017, \aap, 608, A73

\bibitem[{{Lallement} {et~al.}(2019){Lallement}, {Babusiaux}, {Vergely},
  {Katz}, {Arenou}, {Valette}, {Hottier}, \& {Capitanio}}]{2019A&A...625A.135L}
{Lallement}, R., {Babusiaux}, C., {Vergely}, J.~L., {et~al.} 2019, \aap, 625,
  A135

\bibitem[{{Laporte} {et~al.}(2019){Laporte}, {Minchev}, {Johnston}, \&
  {G{\'o}mez}}]{2019MNRAS.485.3134L}
{Laporte}, C. F.~P., {Minchev}, I., {Johnston}, K.~V., \& {G{\'o}mez}, F.~A.
  2019, \mnras, 485, 3134

\bibitem[{{Lee} {et~al.}(2011){Lee}, {Beers}, {An}, {Ivezi{\'c}}, {Just},
  {Rockosi}, {Morrison}, {Johnson}, {Sch{\"o}nrich}, {Bird}, {Yanny},
  {Harding}, \& {Rocha-Pinto}}]{2011ApJ...738..187L}
{Lee}, Y.~S., {Beers}, T.~C., {An}, D., {et~al.} 2011, \apj, 738, 187

\bibitem[{{Leung} \& {Bovy}(2019)}]{2019MNRAS.483.3255L}
{Leung}, H.~W. \& {Bovy}, J. 2019, \mnras, 483, 3255

\bibitem[{{Liang} {et~al.}(2019){Liang}, {Zhao}, {Chen}, {Zuo}, {Zhang}, {Zhu},
  \& {Zhao}}]{2019ApJ...887..193L}
{Liang}, X., {Zhao}, J., {Chen}, Y., {et~al.} 2019, \apj, 887, 193

\bibitem[{{Lindegren} {et~al.}(2021{\natexlab{a}}){Lindegren}, {Bastian},
  {Biermann}, {Bombrun}, {de Torres}, {Gerlach}, {Geyer}, {Hern{\'a}ndez},
  {Hilger}, {Hobbs}, {Klioner}, {Lammers}, {McMillan}, {Ramos-Lerate},
  {Steidelm{\"u}ller}, {Stephenson}, \& {van Leeuwen}}]{2021A&A...649A...4L}
{Lindegren}, L., {Bastian}, U., {Biermann}, M., {et~al.} 2021{\natexlab{a}},
  \aap, 649, A4

\bibitem[{{Lindegren} {et~al.}(2021{\natexlab{b}}){Lindegren}, {Klioner},
  {Hern{\'a}ndez}, {Bombrun}, {Ramos-Lerate}, {Steidelm{\"u}ller}, {Bastian},
  {Biermann}, {de Torres}, {Gerlach}, {Geyer}, {Hilger}, {Hobbs}, {Lammers},
  {McMillan}, {Stephenson}, {Casta{\~n}eda}, {Davidson}, {Fabricius},
  {Gracia-Abril}, {Portell}, {Rowell}, {Teyssier}, {Torra}, {Bartolom{\'e}},
  {Clotet}, {Garralda}, {Gonz{\'a}lez-Vidal}, {Torra}, {Abbas}, {Altmann},
  {Anglada Varela}, {Balaguer-N{\'u}{\~n}ez}, {Balog}, {Barache}, {Becciani},
  {Bernet}, {Bertone}, {Bianchi}, {Bouquillon}, {Brown}, {Bucciarelli},
  {Busonero}, {Butkevich}, {Buzzi}, {Cancelliere}, {Carlucci}, {Charlot},
  {Cioni}, {Crosta}, {Crowley}, {del Peloso}, {del Pozo}, {Drimmel}, {Esquej},
  {Fienga}, {Fraile}, {Gai}, {Garcia-Reinaldos}, {Guerra}, {Hambly}, {Hauser},
  {Jan{\ss}en}, {Jordan}, {Kostrzewa-Rutkowska}, {Lattanzi}, {Liao}, {Licata},
  {Lister}, {L{\"o}ffler}, {Marchant}, {Masip}, {Mignard}, {Mints}, {Molina},
  {Mora}, {Morbidelli}, {Murphy}, {Pagani}, {Panuzzo}, {Pe{\~n}alosa Esteller},
  {Poggio}, {Re Fiorentin}, {Riva}, {Sagrist{\`a} Sell{\'e}s}, {Sanchez
  Gimenez}, {Sarasso}, {Sciacca}, {Siddiqui}, {Smart}, {Souami}, {Spagna},
  {Steele}, {Taris}, {Utrilla}, {van Reeven}, \&
  {Vecchiato}}]{2021A&A...649A...2L}
{Lindegren}, L., {Klioner}, S.~A., {Hern{\'a}ndez}, J., {et~al.}
  2021{\natexlab{b}}, \aap, 649, A2

\bibitem[{{Luo} {et~al.}(2015){Luo}, {Zhao}, {Zhao}, {Deng}, {Liu}, {Jing},
  {Wang}, {Zhang}, {Shi}, {Cui}, {Chu}, {Li}, {Bai}, {Wu}, {Cai}, {Cao}, {Cao},
  {Carlin}, {Chen}, {Chen}, {Chen}, {Chen}, {Chen}, {Chen}, {Chen},
  {Christlieb}, {Chu}, {Cui}, {Dong}, {Du}, {Fan}, {Feng}, {Fu}, {Gao}, {Gong},
  {Gu}, {Guo}, {Han}, {He}, {Hou}, {Hou}, {Hou}, {Hu}, {Hu}, {Hu}, {Huo},
  {Jia}, {Jiang}, {Jiang}, {Jiang}, {Jin}, {Kong}, {Kong}, {Lei}, {Li}, {Li},
  {Li}, {Li}, {Li}, {Li}, {Li}, {Li}, {Li}, {Li}, {Li}, {Li}, {Liang}, {Lin},
  {Liu}, {Liu}, {Liu}, {Liu}, {Lu}, {Luo}, {Mao}, {Newberg}, {Ni}, {Qi}, {Qi},
  {Shen}, {Shi}, {Song}, {Song}, {Su}, {Su}, {Tang}, {Tao}, {Tian}, {Wang},
  {Wang}, {Wang}, {Wang}, {Wang}, {Wang}, {Wang}, {Wang}, {Wang}, {Wang},
  {Wang}, {Wang}, {Wang}, {Wang}, {Wang}, {Wang}, {Wang}, {Wang}, {Wang},
  {Wang}, {Wei}, {Wei}, {Wu}, {Wu}, {Wu}, {Wu}, {Xing}, {Xu}, {Xu}, {Xu},
  {Yan}, {Yang}, {Yang}, {Yang}, {Yang}, {Yao}, {Yu}, {Yuan}, {Yuan}, {Yuan},
  {Yuan}, {Zhai}, {Zhang}, {Zhang}, {Zhang}, {Zhang}, {Zhang}, {Zhang},
  {Zhang}, {Zhang}, {Zhao}, {Zhou}, {Zhou}, {Zhu}, {Zhu}, {Zou}, \&
  {Zuo}}]{2015RAA....15.1095L}
{Luo}, A.~L., {Zhao}, Y.-H., {Zhao}, G., {et~al.} 2015, Research in Astronomy
  and Astrophysics, 15, 1095

\bibitem[{{Mackereth} {et~al.}(2019){Mackereth}, {Schiavon}, {Pfeffer},
  {Hayes}, {Bovy}, {Anguiano}, {Allende Prieto}, {Hasselquist}, {Holtzman},
  {Johnson}, {Majewski}, {O'Connell}, {Shetrone}, {Tissera}, \&
  {Fern{\'a}ndez-Trincado}}]{2019MNRAS.482.3426M}
{Mackereth}, J.~T., {Schiavon}, R.~P., {Pfeffer}, J., {et~al.} 2019, \mnras,
  482, 3426

\bibitem[{{Majewski} {et~al.}(2017){Majewski}, {Schiavon}, {Frinchaboy},
  {Allende Prieto}, {Barkhouser}, {Bizyaev}, {Blank}, {Brunner}, {Burton},
  {Carrera}, {Chojnowski}, {Cunha}, {Epstein}, {Fitzgerald}, {Garc{\'\i}a
  P{\'e}rez}, {Hearty}, {Henderson}, {Holtzman}, {Johnson}, {Lam}, {Lawler},
  {Maseman}, {M{\'e}sz{\'a}ros}, {Nelson}, {Nguyen}, {Nidever}, {Pinsonneault},
  {Shetrone}, {Smee}, {Smith}, {Stolberg}, {Skrutskie}, {Walker}, {Wilson},
  {Zasowski}, {Anders}, {Basu}, {Beland}, {Blanton}, {Bovy}, {Brownstein},
  {Carlberg}, {Chaplin}, {Chiappini}, {Eisenstein}, {Elsworth}, {Feuillet},
  {Fleming}, {Galbraith-Frew}, {Garc{\'\i}a}, {Garc{\'\i}a-Hern{\'a}ndez},
  {Gillespie}, {Girardi}, {Gunn}, {Hasselquist}, {Hayden}, {Hekker}, {Ivans},
  {Kinemuchi}, {Klaene}, {Mahadevan}, {Mathur}, {Mosser}, {Muna}, {Munn},
  {Nichol}, {O'Connell}, {Parejko}, {Robin}, {Rocha-Pinto}, {Schultheis},
  {Serenelli}, {Shane}, {Silva Aguirre}, {Sobeck}, {Thompson}, {Troup},
  {Weinberg}, \& {Zamora}}]{2017AJ....154...94M}
{Majewski}, S.~R., {Schiavon}, R.~P., {Frinchaboy}, P.~M., {et~al.} 2017, \aj,
  154, 94

\bibitem[{{Marshall} {et~al.}(2006){Marshall}, {Robin}, {Reyl{\'e}},
  {Schultheis}, \& {Picaud}}]{2006A&A...453..635M}
{Marshall}, D.~J., {Robin}, A.~C., {Reyl{\'e}}, C., {Schultheis}, M., \&
  {Picaud}, S. 2006, \aap, 453, 635

\bibitem[{{Martinez-Medina} {et~al.}(2019){Martinez-Medina}, {Pichardo},
  {Peimbert}, \& {Valenzuela}}]{2019MNRAS.485L.104M}
{Martinez-Medina}, L., {Pichardo}, B., {Peimbert}, A., \& {Valenzuela}, O.
  2019, \mnras, 485, L104

\bibitem[{{Mayer} \& {Wadsley}(2004)}]{2004MNRAS.347..277M}
{Mayer}, L. \& {Wadsley}, J. 2004, \mnras, 347, 277

\bibitem[{{Michtchenko} {et~al.}(2018){Michtchenko}, {L{\'e}pine}, {Barros}, \&
  {Vieira}}]{2018A&A...615A..10M}
{Michtchenko}, T.~A., {L{\'e}pine}, J.~R.~D., {Barros}, D.~A., \& {Vieira},
  R.~S.~S. 2018, \aap, 615, A10

\bibitem[{{Minchev} {et~al.}(2009){Minchev}, {Quillen}, {Williams}, {Freeman},
  {Nordhaus}, {Siebert}, \& {Bienaym{\'e}}}]{2009MNRAS.396L..56M}
{Minchev}, I., {Quillen}, A.~C., {Williams}, M., {et~al.} 2009, \mnras, 396,
  L56

\bibitem[{{Miyamoto} \& {Nagai}(1975)}]{1975PASJ...27..533M}
{Miyamoto}, M. \& {Nagai}, R. 1975, \pasj, 27, 533

\bibitem[{{Monari} {et~al.}(2018){Monari}, {Famaey}, {Minchev}, {Antoja},
  {Bienaym{\'e}}, {Gibson}, {Grebel}, {Kordopatis}, {McMillan}, {Navarro},
  {Parker}, {Quillen}, {Reid}, {Seabroke}, {Siebert}, {Steinmetz}, {Wyse}, \&
  {Zwitter}}]{2018RNAAS...2...32M}
{Monari}, G., {Famaey}, B., {Minchev}, I., {et~al.} 2018, Research Notes of the
  American Astronomical Society, 2, 32

\bibitem[{{Monari} {et~al.}(2016{\natexlab{a}}){Monari}, {Famaey}, \&
  {Siebert}}]{2016MNRAS.457.2569M}
{Monari}, G., {Famaey}, B., \& {Siebert}, A. 2016{\natexlab{a}}, \mnras, 457,
  2569

\bibitem[{{Monari} {et~al.}(2019{\natexlab{a}}){Monari}, {Famaey}, {Siebert},
  {Bienaym{\'e}}, {Ibata}, {Wegg}, \& {Gerhard}}]{2019A&A...632A.107M}
{Monari}, G., {Famaey}, B., {Siebert}, A., {et~al.} 2019{\natexlab{a}}, \aap,
  632, A107

\bibitem[{{Monari} {et~al.}(2016{\natexlab{b}}){Monari}, {Famaey}, {Siebert},
  {Grand }, {Kawata}, \& {Boily}}]{2016MNRAS.461.3835M}
{Monari}, G., {Famaey}, B., {Siebert}, A., {et~al.} 2016{\natexlab{b}}, \mnras,
  461, 3835

\bibitem[{{Monari} {et~al.}(2019{\natexlab{b}}){Monari}, {Famaey}, {Siebert},
  {Wegg}, \& {Gerhard}}]{2019A&A...626A..41M}
{Monari}, G., {Famaey}, B., {Siebert}, A., {Wegg}, C., \& {Gerhard}, O.
  2019{\natexlab{b}}, \aap, 626, A41

\bibitem[{{Monari} {et~al.}(2017){Monari}, {Kawata}, {Hunt}, \&
  {Famaey}}]{2017MNRAS.466L.113M}
{Monari}, G., {Kawata}, D., {Hunt}, J. A.~S., \& {Famaey}, B. 2017, \mnras,
  466, L113

\bibitem[{{Mr{\'o}z} {et~al.}(2019){Mr{\'o}z}, {Udalski}, {Skowron}, {Skowron},
  {Soszy{\'n}ski}, {Pietrukowicz}, {Szyma{\'n}ski}, {Poleski}, {Koz{\l}owski},
  \& {Ulaczyk}}]{2019ApJ...870L..10M}
{Mr{\'o}z}, P., {Udalski}, A., {Skowron}, D.~M., {et~al.} 2019, \apjl, 870, L10

\bibitem[{{Navarro} {et~al.}(2004){Navarro}, {Helmi}, \&
  {Freeman}}]{2004ApJ...601L..43N}
{Navarro}, J.~F., {Helmi}, A., \& {Freeman}, K.~C. 2004, \apjl, 601, L43

\bibitem[{{Nissen} \& {Schuster}(2010)}]{2010A&A...511L..10N}
{Nissen}, P.~E. \& {Schuster}, W.~J. 2010, \aap, 511, L10

\bibitem[{{Nordstr{\"o}m} {et~al.}(2004){Nordstr{\"o}m}, {Mayor}, {Andersen},
  {Holmberg}, {Pont}, {J{\o}rgensen}, {Olsen}, {Udry}, \&
  {Mowlavi}}]{2004A&A...418..989N}
{Nordstr{\"o}m}, B., {Mayor}, M., {Andersen}, J., {et~al.} 2004, \aap, 418, 989

\bibitem[{{Perez} \& {Granger}(2007)}]{2007CSE.....9c..21P}
{Perez}, F. \& {Granger}, B.~E. 2007, Computing in Science and Engineering, 9,
  21

\bibitem[{{P{\'e}rez-Villegas}
  {et~al.}(2017{\natexlab{a}}){P{\'e}rez-Villegas}, {Portail}, \&
  {Gerhard}}]{2017MNRAS.464L..80P}
{P{\'e}rez-Villegas}, A., {Portail}, M., \& {Gerhard}, O. 2017{\natexlab{a}},
  \mnras, 464, L80

\bibitem[{{P{\'e}rez-Villegas}
  {et~al.}(2017{\natexlab{b}}){P{\'e}rez-Villegas}, {Portail}, {Wegg}, \&
  {Gerhard}}]{2017ApJ...840L...2P}
{P{\'e}rez-Villegas}, A., {Portail}, M., {Wegg}, C., \& {Gerhard}, O.
  2017{\natexlab{b}}, \apjl, 840, L2

\bibitem[{{Pettitt} \& {Wadsley}(2018)}]{2018MNRAS.474.5645P}
{Pettitt}, A.~R. \& {Wadsley}, J.~W. 2018, \mnras, 474, 5645

\bibitem[{{Piskunov} \& {Valenti}(2017)}]{2017A&A...597A..16P}
{Piskunov}, N. \& {Valenti}, J.~A. 2017, \aap, 597, A16

\bibitem[{{Plummer}(1911)}]{1911MNRAS..71..460P}
{Plummer}, H.~C. 1911, \mnras, 71, 460

\bibitem[{{Poggio} {et~al.}(2021){Poggio}, {Drimmel}, {Cantat-Gaudin}, {Ramos},
  {Ripepi}, {Zari}, {Andrae}, {Blomme}, {Chemin}, {Clementini}, {Figueras},
  {Fouesneau}, {Fr{\'e}mat}, {Lobel}, {Marshall}, {Muraveva}, \&
  {Romero-G{\'o}mez}}]{2021A&A...651A.104P}
{Poggio}, E., {Drimmel}, R., {Cantat-Gaudin}, T., {et~al.} 2021, \aap, 651,
  A104

\bibitem[{{Portail} {et~al.}(2017){Portail}, {Gerhard}, {Wegg}, \&
  {Ness}}]{2017MNRAS.465.1621P}
{Portail}, M., {Gerhard}, O., {Wegg}, C., \& {Ness}, M. 2017, \mnras, 465, 1621

\bibitem[{{Purcell} {et~al.}(2011){Purcell}, {Bullock}, {Tollerud}, {Rocha}, \&
  {Chakrabarti}}]{2011Natur.477..301P}
{Purcell}, C.~W., {Bullock}, J.~S., {Tollerud}, E.~J., {Rocha}, M., \&
  {Chakrabarti}, S. 2011, \nat, 477, 301

\bibitem[{{Quillen} {et~al.}(2018){Quillen}, {Carrillo}, {Anders}, {McMillan},
  {Hilmi}, {Monari}, {Minchev}, {Chiappini}, {Khalatyan}, \&
  {Steinmetz}}]{2018MNRAS.480.3132Q}
{Quillen}, A.~C., {Carrillo}, I., {Anders}, F., {et~al.} 2018, \mnras, 480,
  3132

\bibitem[{{Quillen} {et~al.}(2011){Quillen}, {Dougherty}, {Bagley}, {Minchev},
  \& {Comparetta}}]{2011MNRAS.417..762Q}
{Quillen}, A.~C., {Dougherty}, J., {Bagley}, M.~B., {Minchev}, I., \&
  {Comparetta}, J. 2011, \mnras, 417, 762

\bibitem[{{Quillen} \& {Minchev}(2005)}]{2005AJ....130..576Q}
{Quillen}, A.~C. \& {Minchev}, I. 2005, \aj, 130, 576

\bibitem[{{Quinn} \& {Goodman}(1986)}]{1986ApJ...309..472Q}
{Quinn}, P.~J. \& {Goodman}, J. 1986, \apj, 309, 472

\bibitem[{{Ram{\'\i}rez} \& {Allende Prieto}(2011)}]{2011ApJ...743..135R}
{Ram{\'\i}rez}, I. \& {Allende Prieto}, C. 2011, \apj, 743, 135

\bibitem[{{Ramos} {et~al.}(2018){Ramos}, {Antoja}, \&
  {Figueras}}]{2018A&A...619A..72R}
{Ramos}, P., {Antoja}, T., \& {Figueras}, F. 2018, \aap, 619, A72

\bibitem[{{Ramya} {et~al.}(2012){Ramya}, {Reddy}, \&
  {Lambert}}]{2012MNRAS.425.3188R}
{Ramya}, P., {Reddy}, B.~E., \& {Lambert}, D.~L. 2012, \mnras, 425, 3188

\bibitem[{{Reid} {et~al.}(2019){Reid}, {Menten}, {Brunthaler}, {Zheng}, {Dame},
  {Xu}, {Li}, {Sakai}, {Wu}, {Immer}, {Zhang}, {Sanna}, {Moscadelli}, {Rygl},
  {Bartkiewicz}, {Hu}, {Quiroga-Nu{\~n}ez}, \& {van
  Langevelde}}]{2019ApJ...885..131R}
{Reid}, M.~J., {Menten}, K.~M., {Brunthaler}, A., {et~al.} 2019, \apj, 885, 131

\bibitem[{{Reid} {et~al.}(2014){Reid}, {Menten}, {Brunthaler}, {Zheng}, {Dame},
  {Xu}, {Wu}, {Zhang}, {Sanna}, {Sato}, {Hachisuka}, {Choi}, {Immer},
  {Moscadelli}, {Rygl}, \& {Bartkiewicz}}]{2014ApJ...783..130R}
{Reid}, M.~J., {Menten}, K.~M., {Brunthaler}, A., {et~al.} 2014, \apj, 783, 130

\bibitem[{{Russeil}(2003)}]{2003A&A...397..133R}
{Russeil}, D. 2003, \aap, 397, 133

\bibitem[{{Saburova} {et~al.}(2017){Saburova}, {Katkov}, {Khoperskov}, {Zasov},
  \& {Uklein}}]{2017MNRAS.470...20S}
{Saburova}, A.~S., {Katkov}, I.~Y., {Khoperskov}, S.~A., {Zasov}, A.~V., \&
  {Uklein}, R.~I. 2017, \mnras, 470, 20

\bibitem[{{Sanders} {et~al.}(2019{\natexlab{a}}){Sanders}, {Smith}, \&
  {Evans}}]{2019MNRAS.488.4552S}
{Sanders}, J.~L., {Smith}, L., \& {Evans}, N.~W. 2019{\natexlab{a}}, \mnras,
  488, 4552

\bibitem[{{Sanders} {et~al.}(2019{\natexlab{b}}){Sanders}, {Smith}, \&
  {Evans}}]{2019MNRAS.tmp.1855S}
{Sanders}, J.~L., {Smith}, L., \& {Evans}, N.~W. 2019{\natexlab{b}}, \mnras,
  1855

\bibitem[{{Sch{\"o}nrich} {et~al.}(2010){Sch{\"o}nrich}, {Binney}, \&
  {Dehnen}}]{2010MNRAS.403.1829S}
{Sch{\"o}nrich}, R., {Binney}, J., \& {Dehnen}, W. 2010, \mnras, 403, 1829

\bibitem[{{Sch{\"o}nrich} \& {Dehnen}(2018)}]{2018MNRAS.478.3809S}
{Sch{\"o}nrich}, R. \& {Dehnen}, W. 2018, \mnras, 478, 3809

\bibitem[{{Sellwood} \& {Binney}(2002)}]{2002MNRAS.336..785S}
{Sellwood}, J.~A. \& {Binney}, J.~J. 2002, \mnras, 336, 785

\bibitem[{{Sellwood} {et~al.}(2019){Sellwood}, {Trick}, {Carlberg}, {Coronado},
  \& {Rix}}]{2019MNRAS.484.3154S}
{Sellwood}, J.~A., {Trick}, W.~H., {Carlberg}, R.~G., {Coronado}, J., \& {Rix},
  H.-W. 2019, \mnras, 484, 3154

\bibitem[{{Siebert} {et~al.}(2012){Siebert}, {Famaey}, {Binney}, {Burnett},
  {Faure}, {Minchev}, {Williams}, {Bienaym{\'e}}, {Bland-Hawthorn}, {Boeche},
  {Gibson}, {Grebel}, {Helmi}, {Just}, {Munari}, {Navarro}, {Parker}, {Reid},
  {Seabroke}, {Siviero}, {Steinmetz}, \& {Zwitter}}]{2012MNRAS.425.2335S}
{Siebert}, A., {Famaey}, B., {Binney}, J., {et~al.} 2012, \mnras, 425, 2335

\bibitem[{{Siebert} {et~al.}(2011){Siebert}, {Famaey}, {Minchev}, {Seabroke},
  {Binney}, {Burnett}, {Freeman}, {Williams}, {Bienaym{\'e}}, {Bland-Hawthorn},
  {Campbell}, {Fulbright}, {Gibson}, {Gilmore}, {Grebel}, {Helmi}, {Munari},
  {Navarro}, {Parker}, {Reid}, {Siviero}, {Steinmetz}, {Watson}, {Wyse}, \&
  {Zwitter}}]{2011MNRAS.412.2026S}
{Siebert}, A., {Famaey}, B., {Minchev}, I., {et~al.} 2011, \mnras, 412, 2026

\bibitem[{{Skuljan} {et~al.}(1997){Skuljan}, {Cottrell}, \&
  {Hearnshaw}}]{1997ESASP.402..525S}
{Skuljan}, J., {Cottrell}, P.~L., \& {Hearnshaw}, J.~B. 1997, in ESA Special
  Publication, Vol. 402, Hipparcos - Venice '97, ed. R.~M. {Bonnet},
  E.~{H{\o}g}, P.~L. {Bernacca}, L.~{Emiliani}, A.~{Blaauw}, C.~{Turon},
  J.~{Kovalevsky}, L.~{Lindegren}, H.~{Hassan}, M.~{Bouffard}, B.~{Strim},
  D.~{Heger}, M.~A.~C. {Perryman}, \& L.~{Woltjer}, 525--530

\bibitem[{{Spitoni} {et~al.}(2019){Spitoni}, {Cescutti}, {Minchev},
  {Matteucci}, {Silva Aguirre}, {Martig}, {Bono}, \&
  {Chiappini}}]{2019A&A...628A..38S}
{Spitoni}, E., {Cescutti}, G., {Minchev}, I., {et~al.} 2019, \aap, 628, A38

\bibitem[{{Taylor} \& {Cordes}(1993)}]{1993ApJ...411..674T}
{Taylor}, J.~H. \& {Cordes}, J.~M. 1993, \apj, 411, 674

\bibitem[{{Taylor}(2005)}]{2005ASPC..347...29T}
{Taylor}, M.~B. 2005, Astronomical Society of the Pacific Conference Series,
  Vol. 347, {TOPCAT \&amp; STIL: Starlink Table/VOTable Processing Software},
  ed. P.~{Shopbell}, M.~{Britton}, \& R.~{Ebert}, 29

\bibitem[{{Ting} {et~al.}(2019){Ting}, {Conroy}, {Rix}, \&
  {Cargile}}]{2019ApJ...879...69T}
{Ting}, Y.-S., {Conroy}, C., {Rix}, H.-W., \& {Cargile}, P. 2019, \apj, 879, 69

\bibitem[{{Trick}(2020)}]{2020arXiv201101233T}
{Trick}, W.~H. 2020, arXiv e-prints, arXiv:2011.01233

\bibitem[{{van der Walt} {et~al.}(2011){van der Walt}, {Colbert}, \&
  {Varoquaux}}]{2011CSE....13b..22V}
{van der Walt}, S., {Colbert}, S.~C., \& {Varoquaux}, G. 2011, Computing in
  Science and Engineering, 13, 22

\bibitem[{{Vasiliev}(2019)}]{2019MNRAS.482.1525V}
{Vasiliev}, E. 2019, \mnras, 482, 1525

\bibitem[{{Virtanen} {et~al.}(2020){Virtanen}, {Gommers}, {Oliphant},
  {Haberland}, {Reddy}, {Cournapeau}, {Burovski}, {Peterson}, {Weckesser},
  {Bright}, {van der Walt}, {Brett}, {Wilson}, {Jarrod Millman}, {Mayorov},
  {Nelson}, {Jones}, {Kern}, {Larson}, {Carey}, {Polat}, {Feng}, {Moore}, {Vand
  erPlas}, {Laxalde}, {Perktold}, {Cimrman}, {Henriksen}, {Quintero}, {Harris},
  {Archibald}, {Ribeiro}, {Pedregosa}, {van Mulbregt}, \&
  {Contributors}}]{2020SciPy-NMeth}
{Virtanen}, P., {Gommers}, R., {Oliphant}, T.~E., {et~al.} 2020, Nature
  Methods, 17, 261

\bibitem[{{Wang} {et~al.}(2020{\natexlab{a}}){Wang}, {Huang}, {Zhang},
  {L{\'o}pez-Corredoira}, {Cui}, {Chen}, {Guo}, \&
  {Chang}}]{2020ApJ...902...70W}
{Wang}, H.~F., {Huang}, Y., {Zhang}, H.~W., {et~al.} 2020{\natexlab{a}}, \apj,
  902, 70

\bibitem[{{Wang} {et~al.}(2020{\natexlab{b}}){Wang}, {L{\'o}pez-Corredoira},
  {Huang}, {Carlin}, {Chen}, {Wang}, {Chang}, {Zhang}, {Xiang}, {Yuan}, {Sun},
  {Li}, {Yang}, \& {Deng}}]{2020MNRAS.491.2104W}
{Wang}, H.~F., {L{\'o}pez-Corredoira}, M., {Huang}, Y., {et~al.}
  2020{\natexlab{b}}, \mnras, 491, 2104

\bibitem[{{Wegg} {et~al.}(2015){Wegg}, {Gerhard}, \&
  {Portail}}]{2015MNRAS.450.4050W}
{Wegg}, C., {Gerhard}, O., \& {Portail}, M. 2015, \mnras, 450, 4050

\bibitem[{{Wegg} {et~al.}(2019){Wegg}, {Rojas-Arriagada}, {Schultheis}, \&
  {Gerhard}}]{2019A&A...632A.121W}
{Wegg}, C., {Rojas-Arriagada}, A., {Schultheis}, M., \& {Gerhard}, O. 2019,
  \aap, 632, A121

\bibitem[{{W}es {M}c{K}inney(2010)}]{mckinney-proc-scipy-2010}
{W}es {M}c{K}inney. 2010, in {P}roceedings of the 9th {P}ython in {S}cience
  {C}onference, ed. {S}t\'efan van~der {W}alt \& {J}arrod {M}illman, 56 -- 61

\bibitem[{{Wheeler} {et~al.}(2021){Wheeler}, {Abril-Cabezas}, {Trick},
  {Fragkoudi}, \& {Ness}}]{2021arXiv210505263W}
{Wheeler}, A., {Abril-Cabezas}, I., {Trick}, W.~H., {Fragkoudi}, F., \& {Ness},
  M. 2021, arXiv e-prints, arXiv:2105.05263

\bibitem[{{Wheeler} {et~al.}(2020){Wheeler}, {Ness}, {Buder}, {Bland
  -Hawthorn}, {De Silva}, {Hayden}, {Kos}, {Lewis}, {Martell}, {Sharma},
  {Simpson}, {Zucker}, \& {Zwitter}}]{2020arXiv200108227W}
{Wheeler}, A., {Ness}, M., {Buder}, S., {et~al.} 2020, arXiv e-prints,
  arXiv:2001.08227

\bibitem[{{Widrow} {et~al.}(2014){Widrow}, {Barber}, {Chequers}, \&
  {Cheng}}]{2014MNRAS.440.1971W}
{Widrow}, L.~M., {Barber}, J., {Chequers}, M.~H., \& {Cheng}, E. 2014, \mnras,
  440, 1971

\bibitem[{{Widrow} {et~al.}(2012){Widrow}, {Gardner}, {Yanny}, {Dodelson}, \&
  {Chen}}]{2012ApJ...750L..41W}
{Widrow}, L.~M., {Gardner}, S., {Yanny}, B., {Dodelson}, S., \& {Chen}, H.-Y.
  2012, \apjl, 750, L41

\bibitem[{{Williams} {et~al.}(2009){Williams}, {Freeman}, {Helmi}, \& {RAVE
  Collaboration}}]{2009IAUS..254..139W}
{Williams}, M. E.~K., {Freeman}, K.~C., {Helmi}, A., \& {RAVE Collaboration}.
  2009, in IAU Symposium, Vol. 254, The Galaxy Disk in Cosmological Context,
  ed. J.~{Andersen}, {Nordstr{\"o}ara}, B.~{m}, \& J.~{Bland -Hawthorn},
  139--144

\bibitem[{{Williams} {et~al.}(2013){Williams}, {Steinmetz}, {Binney},
  {Siebert}, {Enke}, {Famaey}, {Minchev}, {de Jong}, {Boeche}, \&
  {Freeman}}]{2013MNRAS.436..101W}
{Williams}, M.~E.~K., {Steinmetz}, M., {Binney}, J., {et~al.} 2013, \mnras,
  436, 101

\bibitem[{{Wozniak}(2020)}]{2020ApJ...889...81W}
{Wozniak}, H. 2020, \apj, 889, 81

\bibitem[{{Xiang} {et~al.}(2019){Xiang}, {Ting}, {Rix}, {Sand ford}, {Buder},
  {Lind}, {Liu}, {Shi}, \& {Zhang}}]{2019ApJS..245...34X}
{Xiang}, M., {Ting}, Y.-S., {Rix}, H.-W., {et~al.} 2019, \apjs, 245, 34

\bibitem[{{Xiang} {et~al.}(2017){Xiang}, {Liu}, {Yuan}, {Huo}, {Huang}, {Wang},
  {Chen}, {Ren}, {Zhang}, {Tian}, {Yang}, {Shi}, {Zhao}, {Li}, {Zhao}, {Cui},
  {Li}, {Hou}, {Zhang}, {Zhang}, {Wang}, {Wu}, {Cao}, {Yan}, {Yan}, {Luo},
  {Zhang}, {Bai}, {Yuan}, {Dong}, {Lei}, \& {Li}}]{2017MNRAS.467.1890X}
{Xiang}, M.~S., {Liu}, X.~W., {Yuan}, H.~B., {et~al.} 2017, \mnras, 467, 1890

\bibitem[{{Zhao} {et~al.}(2014){Zhao}, {Zhao}, {Chen}, {Oswalt}, {Tan}, \&
  {Zhang}}]{2014ApJ...787...31Z}
{Zhao}, J.~K., {Zhao}, G., {Chen}, Y.~Q., {et~al.} 2014, \apj, 787, 31

\end{thebibliography}

\begin{appendix}
\section{Linking chemical abundances with $R-R_g$ value}\label{sec::app1}
\begin{figure}[t]
\includegraphics[width=0.95\hsize]{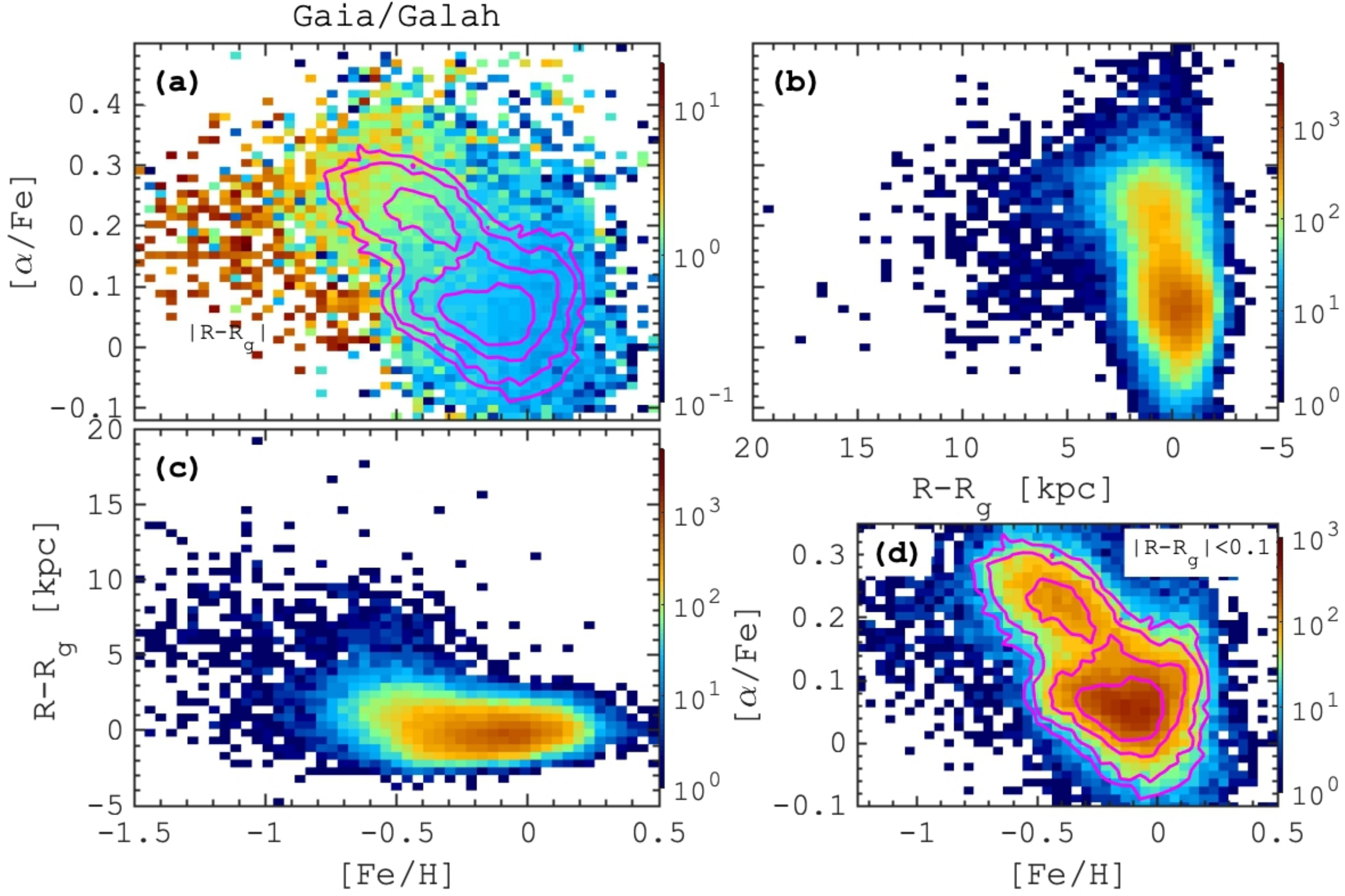}
\includegraphics[width=0.95\hsize]{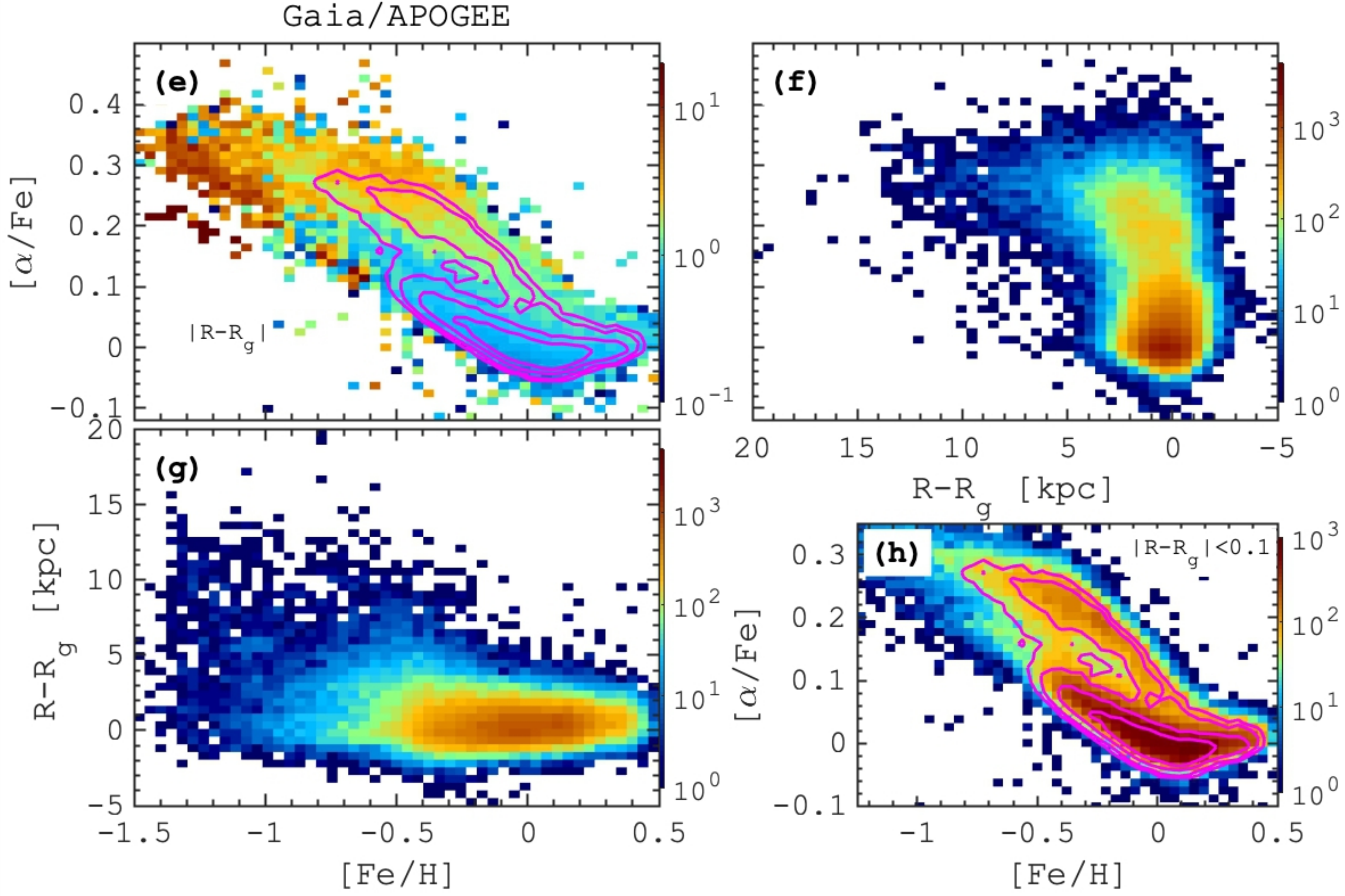}
\includegraphics[width=0.95\hsize]{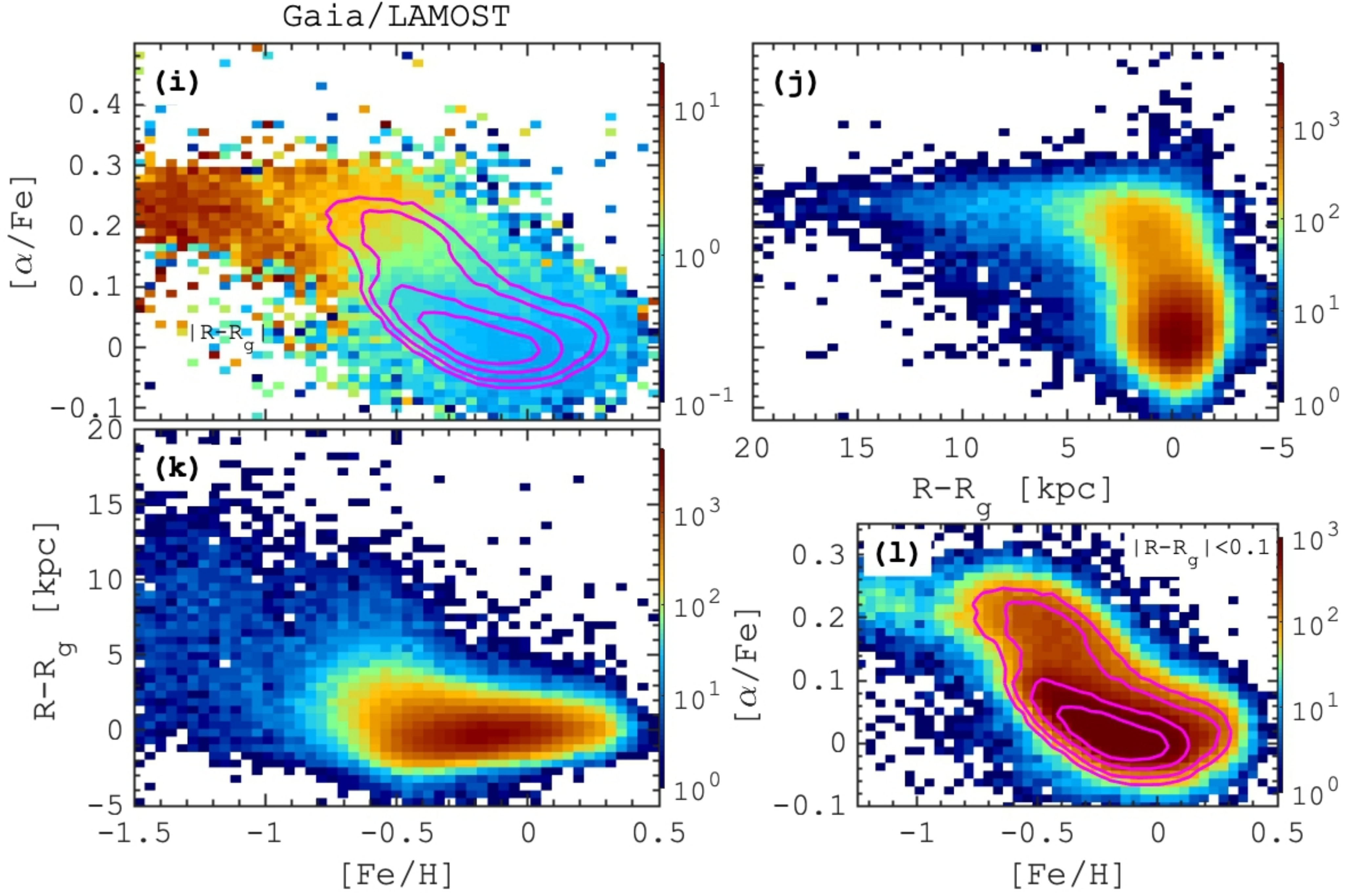}
\caption{Relation between chemical abundances as a function of $R-R_g$ for Galah~(a-d), APOGEE~(e-h) and LAMOST~(i-l) cross-matched with Gaia DR2~(G3RV2). Frames a, e and i show \aFe-\FeH relation colourcoded by the mean $|R-R_g|$ value in log scale. Frames b, f and j show the number of stars in \aFe-$|R-R_g|$ plane. Frames c, g and k show the number of stars in $|R-R_g|$-\FeH plane. Finally, frames d, h and l show the number of stars with $|R-R_g|<0.25$ in \aFe-\FeH plane. Magenta contours in \aFe-\FeH planes show the total number of stars distribution.}\label{fig::afe_fe_h}
\end{figure}

In Fig.~\ref{fig::afe_fe_h} we present the well-known $\aFe-\FeH$ plane together with $\aFe-|\RRg|$ and $|\RRg|-\FeH$ relations for G3RV2+G.A.L. subsamples providing an important chemo-kinematical information about stellar populations in the MW. $\aFe-\FeH$ relation for all adopted surveys shows well-known dichotomy implying the presence of two main populations, usually associated with thick~(high-$\alpha$) and thin~(low-$\alpha$) components~\citep[see, e.g.,][]{2014A&A...562A..71B,2013A&A...560A.109H}. Kinematical diversity of stellar populations of these components is clearly visible once we colour-coded \aFe-\FeH relation by the mean \RRg value~(Fig~\ref{fig::afe_fe_h}a, e, i) where thin disk~(low-$\alpha$) contains mostly cold stars with low radial oscillation amplitude which gradually increases towards high-$\alpha$ sequence. A third distinct component is also visible at low metallicities~(\FeH$<-0.5$) where \RRg$\gtrsim5$~kpc corresponding to the accreted stars~\citep[][]{2010A&A...511L..10N} which we also find as a separate component in \RRg distributions~(see Fig.~\ref{fig::r_rg}).  Interestingly that the \aFe-\FeH relation colour-coded by the mean \RRg looks very similar to one presented by \cite{2019MNRAS.482.3426M,2020MNRAS.493.2952H} for the SNd stars colour-coded by the orbital eccentricity. Presence of three main aforementioned components is also revealed in \aFe-|\RRg| relation. The decrease of the stellar radial oscillations with \FeH is clearly visible in frames c, g and k. Another feature which we also mention is the presence of very hot stars with \RRg$>5$~kpc up to solar metallicities thus understanding of such a population may provide some new constraints on the parameters of the last MW merger and the follow-up disk heating. Since the low-$\alpha$ stars mainly contain stars with low \RRg values~(see frames d, h and l) our results presented in Fig.~\ref{fig::Gaia_vs_masers} suggest that the MW spiral arms show a larger fraction of thin-disk stars.

\end{appendix}
\end{document}